\documentclass[aps, prd, superscriptaddress, nofootinbib, preprintnumbers]{revtex4}
\usepackage{amsmath}
\usepackage{graphicx}
\usepackage{color} 
\usepackage{amsmath}
\newcommand{\lesssim}{\mathrel{\mathpalette\vereq<}}
\newcommand{\gtrsim}{\mathrel{\mathpalette\vereq>}}

\newcommand{\chushi}[1]{}
\newcommand{\1}{\mbox{1}\hspace{-0.25em}\mbox{l}}

\begin{document}
 \title{{\bf Dynamical Gauge Boson of Hidden Local Symmetry \\
 within \\
 the Standard Model}
 \vspace{5mm}}
\author{{Koichi Yamawaki}} \thanks{
      {\tt yamawaki@kmi.nagoya-u.ac.jp}}
      \affiliation{ Kobayashi-Maskawa Institute for the Origin of Particles and 
the Universe (KMI), Nagoya University, Nagoya 464-8602, Japan.}
\date{\today}

\begin{abstract} 
The Standard Model (SM) Higgs Lagrangian is straightforwardly rewritten 
into the {\it scale-invariant}  nonlinear
sigma model 
$G/H=[SU(2)_L \times SU(2)_R]/SU(2)_{V}\simeq O(4)/O(3)$, with 
the (approximate) scale symmetry  
realized nonlinearly by the (pseudo) dilaton ($=$ SM Higgs).
 It is further gauge equivalent to that having the symmetry $O(4)_{\rm global}\times O(3)_{\rm local}$,
with $O(3)_{\rm local}$ being the Hidden Local Symmetry (HLS).
In the large $N$ limit of the scale-invariant version of the Grassmannian model $G/H=O(N)/[O(N-3)\times O(3)] \simeq 
O(N)_{\rm global}\times [O(N-3)\times O(3)]_{\rm local}$, identical to  
the SM 
 for $N\rightarrow 4$, we show  that 
 the kinetic term of the HLS gauge bosons (``SM rho'') $\rho_\mu$  of the  $O(3)_{\rm local}\simeq [SU(2)_V]_{\rm local}$ 
 are dynamically generated  by the nonperturbative dynamics of the  SM itself. 
The dynamical SM rho  stabilizes the skyrmion (``SM skyrmion'') $X_s$ as a dark matter candidate within the SM: 
The mass $M_{X_s} ={\cal O}(10\, {\rm GeV})$ consistent with the direct search experiments implies 
the induced HLS gauge coupling $g_{_{\rm HLS}}={\cal O}(10^3)$, which  realizes   the relic abundance, $\Omega_{X_s} h^2 ={\cal O}(0.1)$. If instead $g_{_{\rm HLS}}\lesssim 3.5$ ($M_\rho \lesssim 1.2 $ TeV), the SM rho  could be detected with ``narrow width'' $\lesssim 100 \,{\rm GeV}$ at LHC,
having all the
``$a=2$ results'' of the generic HLS Lagrangian ${\cal L}_A+ a {\cal L}_V$, i.e.,  
$\rho$-universality, KSRF relations 
and  the vector meson dominance,  
 independently of ``$a$''. There exists the second order  phase phase transition 
to the unbroken phase having massless  
$\rho_\mu$ and massive $\pi$ (no longer NG bosons), 
both becoming 
 massless free
particles just on the transition point
(scale-invariant ultraviolet fixed point). The results readily apply to the 2-flavored QCD as well. \end{abstract} 
\maketitle

  \section{Introduction}
\label{intro}
The Standard Model (SM)  Higgs Lagrangian is customarily written in the {\it linear sigma model}  
which is convenient for the perturbation theory. Although 
the perturbative SM (pSM)  has been very successful phenomenologically, 
there still remains the mystery of the origin of mass and the related Higgs particle itself within the pSM. 
Moreover,  there is some concrete tension between the conventional (perturbative) understanding of the SM and the reality: 
apparent absence of the dark matter candidate, $\theta$ vacuum parameters due to instantons (strong CP problem, etc.),  and absence of the first order phase transition for finite temperature
and  of large enough CP violation, both required by the baryogenesis, and so on. Although these problems might hint the possible new physics beyond the SM (BSM),
 they may simply indicate our ignorance of the full SM  including {\it the nonperturbative dynamics}.

A possibility was in fact pointed out \cite{Matsuzaki:2016iyq}
 that the {\it dark matter candidate already exists in the SM} as a soliton (``SM skyrmion''): The SM Higgs Lagrangian is  rewritten \cite{Fukano:2015zua} (see also \cite{Yamawaki:2015tmu,Yamawaki:2016kdz,Yamawaki:2016qux}) into the equivalent  {\it nonlinear sigma model} having the {\it hidden local symmetry (HLS)} \cite{Bando:1984ej,Bando:1985rf,Bando:1984pw,
 Bando:1987br,Harada:2003jx} as well as {\it (approximately) scale-invariance},
which then generates the kinetic term of the HLS gauge boson (``SM rho'') $\rho_\mu$ through  {\it nonperturbative physics} at quantum level, thereby  stabilizes  a soliton as a skyrmion in the SM itself, 
similarly to the hadronic skyrmion (nucleon) stabilized by the $\rho$ meson as an HLS gauge boson within the same nonlinear sigma model
(except for the scale symmetry).  It was found \cite{Matsuzaki:2016iyq}
that the predicted skyrmion dark matter in the SM,
$X_{_s}$, with the mass  $M_{X_{_s}}={\cal O} (10\,  {\rm GeV})$ consistent with the direct detection experiments, gives the relic abundance $\Omega_{X_s} h^2 \simeq 0.1$ in rough agreement with the observation.
\\

In this paper, as a follow-up of Ref.\cite{Matsuzaki:2016iyq}, we establish  the {\it quantum dynamical generation of the SM rho, $\rho_\mu$, the gauge boson 
of the HLS hidden within the SM itself} as a notable nonperturbative dynamics within the SM Higgs Lagrangian,
based on an approach more reliable than that in Ref.\cite{Matsuzaki:2016iyq}.  
\\

In fact,  
{\it pSM is not a whole story of the SM}: Even within the perturbation the SM Higgs self-coupling $\lambda$ grows indefinitely 
to hit the {\it Landau  pole} in the ultraviolet region thus eventually {\it invalidating 
the perturbation itself}, resulting in   some 
nonperturbative effects such as the 
bound state within the SM.
Even if the perturbative Landau pole happens to be removed or pushed far above the weak scale so that the pSM would be logically consistent   all the way up to the
high scale, there arises ironically  a 
notorious {\it naturalness problem} in turn. An immediate solution to this would be the nonperturbative quantum dynamics within the SM to show up with the ``cutoff'' $\Lambda$  acting as the {\it nonperturbative Landau pole} not far from the weak scale {\it without affecting the successful pSM at lower energy.}   Such a nonperturbative physics  within the SM at lower scale may be a signal of the  BSM having such a scale as a dual theory, similarly to the hadron-quark duality:  With ``cutoff'' or ``Landau pole'' ${\cal O}(\Lambda_\chi) ={\cal O}( 4\pi f_\pi)$,  the nonlinear sigma model with HLS $\rho$ meson  and skyrmion nucleon is dual to the underlying QCD. \footnote{
 In the case of QCD, the
 perturbation in the   {\it linear} sigma model already breaks down at physical point with $\lambda \gg 1$, i.e., the ``perturbative'' Landau pole is very close to the nonperturbative one ${\cal O}(\Lambda_\chi)$. This is due to {\it absence of the scale symmetry} \cite{Yamawaki:2016qux},  in sharp contrast to the SM Higgs Lagrangian $\lambda\ll 1$, having the perturbative Landau pole far away from the weak scale (physical point).  
 }

Note also that the nonperturbative effects may not necessarily require the ``strong coupling'':  Sphaleron and instanton 
are well-known nonperturbative objects not to be described by the pSM but certainly {\it exist in the SM as nonperturbative objects} even for the weakest coupling. 
Also in the  
Georgi-Glashow model which is perturbatively renormalizable similarly to the SM Higgs Lagrangian, there exists a nonperturbative
object, the 't Hooft-Polyakov monopole,  
even in the vanishing quartic coupling, known as a ``Bogomol'nyi-Prasad-Sommerfield (BPS) limit'',   similarly to the SUSY flat direction limit~\cite{Harvey:1996ur}.  So even for the region of a small Higgs self-coupling, nonperturbative physics could be operative.
\\

The nonperturbative quantum physics can often be better described 
by a different parameterization of the same Lagrangian at the classical level. 
In fact it was shown~\cite{Fukano:2015zua}  that the SM Higgs Lagrangian written in the linear sigma model on the broken vacuum 
can be straightforwardly cast through the polar decomposition into an (approximately) {\it scale-invariant} version of the {\it nonlinear sigma model} 
based on the manifold $G/H=SU(2)_L\times SU(2)_R/SU(2)_V\simeq O(4)/O(3)$. Namely, 
{\it both the (approximate) scale symmetry and internal symmetry $G$ are realized nonlinearly}, with 
{\it the SM Higgs being nothing but a (pseudo) dilaton}, $\varphi$, a (pseudo) Nambu-Goldstone (NG) boson of the spontaneously broken scale symmetry, 
in addition to  the NG bosons $\pi$ living on $G/H$. 
(Since the (pseudo) dilaton parts in such a parameterization are $G-$invariant,
the discussions hereafter are confined, unless otherwise mentioned, to the nonlinear realization of the internal symmetry $G$.)

Once written in the form of nonlinear sigma model, 
one readily sees~\cite{Fukano:2015zua}
that it has the HLS,  since it is known~\cite{Bando:1984pw,Bando:1987br} that 
any nonlinear sigma model based on $G/H$ $\grave{a}$ la Callan-Coleman-Wess-Zumino (CCWZ)~\cite{Coleman:1969sm,Callan:1969sn} is gauge equivalent to another model 
(HLS Lagrangian) 
having a symmetry $G_{\rm global}\times H_{\rm local}$, with the Lagrangian consisting of two invariants : ${\cal L}= {\cal L}_A + a {\cal L}_V$, $a$ being a free parameter (See Appendix \ref{CCWZ}).  While ${\cal L}_A$ is reduced after gauge fixing  to the original $G/H$ model, $ {\cal L}_V$ term yields the mass of the HLS gauge boson $\rho_\mu$,  in such a way that 
the gauge symmetry (HLS) $H_{\rm local}$ and $G_{\rm global}$ are both spontaneously broken down to the diagonal
group $H =H_{\rm local} \oplus H_{\rm global}$ ($H_{\rm global} \subset G_{\rm global}$) through the Higgs mechanism. In the case at hand, the SM Higgs Lagrangian
rewritten into the (approximately) scale-invariant version of the $G/H=[SU(2)_L\times SU(2)_R]/SU(2)_V\simeq O(4)/O(3)$ nonlinear sigma model 
has the HLS $H_{\rm local}=[SU(2)_V]_{\rm local}\simeq O(3)_{\rm local}$\cite{Bando:1984ej,Bando:1985rf,Bando:1984pw,
 Bando:1987br,Harada:2003jx}.   Here the HLS gauge boson   is an auxiliary field (without kinetic term) as 
a static massive composite of the NG bosons  and can be solved/gauged away at classical level: ${\cal L}_V=0$. 
\footnote{
A similar s-HLS model was studied \cite{Kurachi:2014qma,Fukano:2015zua,
Fukano:2015hga,Fukano:2015uga} as the effective theory of the walking technicolor~\cite{Yamawaki:1985zg,Bando:1986bg} having the (approximate) scale symmetry and a pseudo-dilaton (``technidilaton'') as a light composite Higgs.
The s-HLS model was also discussed in a different context, the ordinary QCD in medium~\cite{Lee:2015qsa}. Note that 
the pseudo-dilaton $\varphi$ in the present paper is of course the SM Higgs itself (see a simple re-parameterization in Eqs.(\ref{NLscale}) and (\ref{Higgs2})), with decay constant $F_\varphi=F_\pi=246\, {\rm GeV}$, which {\it should not be confused with  the technidilaton} having a different decay constant and hence  couplings to the SM particles different from those of the SM Higgs
 \cite{Bando:1986bg} (see \cite{Matsuzaki:2015sya} for a recent review of the technidilaton  consistent with the LHC experiments in spite of the different couplings).
} 
\\
  
However, it is well known (see Appendix \ref{CPNApp}) that the HLS gauge bosons in many nonlinear sigma models, such as the $CP^{N-1}$ model with $G/H=U(N)/[U(N-1)\times U(1)]\simeq SU(N)/[SU(N-1)\times U(1)]\simeq SU(N)_{\rm global} \times U(1)_{\rm local}$, do acquire kinetic term at quantum level through nonperturbative dynamics like the  large $N$ limit~\cite{Eichenherr:1978qa,Golo:1978de,DAdda:1978vbw,DAdda:1978dle,Witten:1978bc,Arefeva:1980ms,Haber:1980uy,Kugo:1985jc,Bando:1987br,Weinberg:1997rv,Harada:2003jx}
~\footnote{
The $CP^{N-1}$ model 
minimally written in terms of ($2N-2$) NG bosons is usually 
parameterized having the symmetry $SU(N)_{\rm global} \times$ $U(1)_{\rm local}$,  including redundancy: one constraint with Lagrange multiplier and 
the $U(1)_{\rm local}$ as an HLS whose gauge boson (having mass by the Higgs mechanism) is
an auxiliary field to be solved away at classical level. It is well established~\cite{Eichenherr:1978qa,Golo:1978de,DAdda:1978vbw,DAdda:1978dle,Witten:1978bc,Arefeva:1980ms,Haber:1980uy, Kugo:1985jc, Bando:1987br,Weinberg:1997rv,Harada:2003jx} that  in the large $N$ limit, there exists a phase transition from the perturbative (broken) phase to nonperturbative (unbroken) phase and the HLS gauge boson in both phases necessarily 
acquires the kinetic term, becoming the propagating gauge boson, massive (broken phase) or massless (unbroken phase).
In the unbroken phase the NG bosons at classical  level are no longer the NG bosons but have a mass  
given by the Lagrange multiplier. 
The model in $D=4$ dimensions  has a cutoff, an {\it extra free parameter} to define the quantum theory, acting as a Landau pole where the induced kinetic term of HLS gauge boson vanishes (``compositeness condition'') to  return to the auxiliary field as at classical level \cite{Bando:1987br,Harada:2003jx} (See also \cite{Weinberg:1997rv} for different formulation (renormalizable in the sense of effective theory) in $D=4$ dimensions, which also needs extra free parameters as counter terms from the onset). 
See Appendix \ref{CPNApp} for details. 
\label{CPN} 
}. In the unbroken phase a minimal parameterization of the {\it classical Lagrangian without gauge symmetry redundancy  
is ill-defined at quantum level}~\cite{Bando:1987br}, 
thus the HLS parameterization is crucial to  the nonperturbative quantum physics (Exactly the same applies to the present case as we shall explain later in details).
This is in sharp contrast to the perturbation, the pSM, which  is known \cite{Alonso:2016oah} to be independent of the parameterization for a generic metric for the {\it CCWZ nonlinear realization}
 not just the original linear sigma model parameterization.

The same dynamical generation of the HLS gauge bosons in the large $N$ limit is also known in the nonlinear sigma model on the Grassmannian manifold,
$G/H=U(N)/[U(N-p)\times U(p)]$~\cite{Brezin:1980ms,Bando:1996pg}  as an extension of $CP^{N-1}$ model ($p=1$), and also on 
$G/H=O(N)/[O(N-p)\times O(p)]$~\cite{Brezin:1980ms}.  Similarly, in the Nambu-Jonal-Lasinio (NJL) model~\cite{Nambu:1961tp}, such a dynamical generation at quantum level of the kinetic term of the  auxiliary field in the large $N$ limit is very well known~\cite{Eguchi:1976iz,Kugo:1976tq,Bardeen:1989ds}, see Appendix \ref{NJL} for details. 
{\it Thus the dynamical generation of the kinetic term of the auxiliary fields is a very common nonperturbative phenomenon.}

Further in the case of the SM written in the form of a scale-invariant nonlinear sigma model, it was shown  \cite{Matsuzaki:2016iyq}  that  the massive (Higgsed) $[SU(2)_V]_{\rm local}$ HLS gauge boson in the SM, SM rho,  acquires kinetic term by the {\it nonperturbative dynamics} of the SM  itself at order ${\cal O}(p^4)$ 
in the systematic derivative expansion
within the framework of the chiral perturbation theory  
\cite{Gasser:1983yg,Gasser:1984gg} in a version
extended to include the HLS~\cite{Harada:1992np,Harada:1999zj,
Harada:2003jx}.
 \\

Here we show  the key dynamical issue of Ref.\cite{Matsuzaki:2016iyq}, the {\it quantum dynamical generation of the SM rho, the gauge boson 
of the HLS hidden within the SM itself}, in a more transparent and well-established nonperturbative method than that in Ref.\cite{Matsuzaki:2016iyq} \footnote{
The stability of the skyrmion crucially depends on the short distance dynamics, thereby the low momentum expansion in Ref.\cite{Matsuzaki:2016iyq} may not be ideal particularly for the skyrmion calculation. 
The present paper overcomes this ``weak point'' of Ref.\cite{Matsuzaki:2016iyq}.},  namely the conventional large $N$ 
expansion widely used for many nonperturbative dynamics. As an $N$ extension of the SM with $G/H= SU(2)_L\times SU(2)_R/SU(2)_V\simeq O(4)/O(3)$, we take a Grassmannian manifold $G/H=O(N)/[O(N-p)\times O(p)]$, with $p=3=$ fixed, which, combined with the pseudo-dilaton parts, is reduced to precisely the SM Higgs Lagrangian for $N\rightarrow 4$ and $p=3$.

According to the generic arguments~\cite{Bando:1984pw,Bando:1987br},  this Grassmannian model is 
gauge equivalent to that having an HLS with the symmetry $G_{\rm global}\times H_{\rm local}=O(N)_{\rm global} \times [O(N-p) \times O(p)]_{\rm local}$; 
this time  the Lagrangian besides the dilatonic parts consists of three invariants: ${\cal L}={\cal L}_A + a^{(p)} {\cal L}^{(p)}_V + a^{(N-p)}{\cal L}^{(N-p)}_V$, with  $a^{(p)}$ and $a^{(N-p)}$ being free parameters corresponding to the $O(p)_{\rm local}$ and $O(N-p)_{\rm local}$ gauge boson mass terms, respectively.  
\\

Similarly to the $CP^{N-1}$ model (see Appendix \ref{CPNApp}), a popular parameterization to study the large $N$ limit of this model~\cite{Brezin:1980ms}   is to use $p\times N$ real scalar field $\phi_{i,\alpha}$ ($i=1,\cdots, p; \alpha= 1,\cdots, N$) and introduce the HLS gauge boson, SM rho,  $\rho_\mu$ by the covariant derivative $D_\mu \phi = (\partial_\mu -i \rho_\mu) \phi$.
The $\phi$ consists of  the $p (N-p)$ NG bosons ($\pi$) and  the $p(p-1)/2$ HLS gauge degrees of freedom (would-be NG bosons $\check \rho$ absorbed into $\rho_\mu$)\footnote{In the HLS papers~\cite{Bando:1984ej,Bando:1985rf,Bando:1984pw,Bando:1987br,Harada:2003jx}, 
$\check{\rho}$ was denoted by $\sigma$. In order to avoid confusion in the present paper using $\sigma$ for a different object, we will use $\check{\rho}$ in this paper.}
, plus $p(p+1)/2$ redundant massive components corresponding to the
constraints (through the Lagrange multipliers $\eta_{i,j}(x)$),  besides the (pseudo-)dilaton (SM Higgs) $\varphi$ to make the theory equivalent to the SM in the $N\rightarrow 4$ limit with $p=3$.

Curiously, this parameterization is equivalent to a specific  ``golden point'' $a (=a^{(p)}) =2$  realizing all the successful results in the QCD for  the generic HLS Lagrangian, 
${\cal L}_A + a  {\cal L}_V$ when the kinetic term  is simply assumed~\cite{Bando:1984ej,Bando:1985rf,Bando:1984pw,
Bando:1987br,Harada:2003jx};
The $\rho-$universality, 
Kawarabayashi-Suzuki-Riazuddin-Fayyazuddin  (KSRF) relation II 
and  the vector meson dominance (VMD) for $\pi$ form factor (See Eqs.(\ref{rhouniversality}), (\ref{KSRFII}) and (\ref{VMD2}) in Appendix \ref{CCWZ}.)
\\

 We first discuss 
 the phase structure of this model in the large $N$ limit, which is essentially the same as that of the $CP^{N-1}$ model: (i)  the broken phase $\langle \phi_{i,\alpha}(x) \rangle= \delta_{i,j} \sqrt{N} v\ne 0\, ,
 \langle \eta_{i,j}(x)\rangle=0$, with 
 the NG bosons $\pi$ (with the decay constant $F_\pi=\sqrt{N} v$) living on the coset $G/H=O(N)/[O(N-p)\times O(p)]$ as in the classical theory, and 
(ii) the unbroken phase $\langle \phi_{i,\alpha}(x) \rangle= 0\, ,
 \langle \eta_{i,j}(x)\rangle=\delta_{i,j} \eta; \eta\ne 0$, which exists only at the quantum level, with $\pi$ being no longer the NG bosons but massive. 
 
In generic  $D$ dimensions ($2\leq D \leq 4$),  the gap equation takes a form very similar to that of the $CP^{N-1}$ model, and also to that of the NJL model in $D$ dimensions though with opposite direction (weak coupling for the broken, and strong coupling for the unbroken). The SM Higgs boson $\varphi$, sitting in the theory both through the 
 Higgs potential and the dilatonic factor, plays only a minor role for the phase structure. 
  \\
 
 Then we discuss the dynamical generation of the HLS gauge boson $\rho_\mu$.
It was in fact already shown~\cite{Brezin:1980ms} 
  in (a conventional non-scale invariant version of) this model that the kinetic term of the HLS gauge boson is dynamically generated in the  large $N$ limit
  in both phases.
Although the pseudo-dilaton $\varphi$ ($=$ SM Higgs)  in our case  additionally
is coupled to the HLS gauge boson, it does not affect the large $N$ counting of the  2-point function of the 
HLS gauge boson and  hence is irrelevant to the kinetic term generation.

For concrete case $p=3$ relevant to the SM (extension to $p\ne 3$ is trivial), we show that the  {\it  $O(p)_{\rm local}$ gauge boson $\rho_\mu$ becomes dynamical in the large $N$ limit, in both phases :
massive (broken phase) or massless (unbroken phase)}, while not the $O(N-p)_{\rm local}$ gauge boson, carrying index running $1,\cdots, N-p$ thus subject to all the planar diagrams contributions in the large $N$ limit, which stays as an auxiliary field (i.e., ${\cal L}^{(N-p)}_V=0$) in either phase. This is similar to the $SU(N-1)_{\rm local}$ gauge boson in the $CP^{N-1}$ model with $G/H=SU(N)/[SU(N-1)\times U(1)]$, 
which,  carrying the index running 
through $1,\cdots, N-1$, is not dynamically generated in the large $N$ limit, in contrast to the dynamical generation of the $U(1)_{\rm local}$ part. 
  This is also contrasted to
a popular $N$ extension $G/H=O(N)/O(N-1)\simeq O(N)_{\rm global} \times O(N-1)_{\rm local}$ whose $O(N-1)_{\rm local}$ HLS will not be dynamical in the large $N$ limit for the same reason as for  the $O(N-p)_{\rm local}$ gauge boson in our case. (In the limit to  the SM, with $N \rightarrow 4\, (p=3)$,  $O(N-p)$ does not exist, anyway.)

We thus find that the kinetic term of the $O(3)_{\rm local}$ HLS gauge boson $\rho_\mu$ in the large $N$ limit is indeed generated, with the $N-$independent induced HLS coupling ('t Hooft coupling) $\lambda_{_{\rm HLS}}=N g^2_{_{\rm HLS}}$ given as:
\begin{eqnarray}
\frac{1}{\lambda_{_{\rm HLS}}(\mu^2)}= \frac{1}{N g_{_{\rm HLS}}^2(\mu^2)} &=&\frac{1}{3(4 \pi)^2} \ln \left(\frac{{\tilde \Lambda}^2}{\mu^2}\right) \quad \quad  \rightarrow 0\quad \left(\mu \rightarrow \tilde \Lambda\right)\,,
\label{kinetic1}
\end{eqnarray}
where $\tilde \Lambda= e^{4/3}\cdot \Lambda\simeq 3.8 \, \Lambda$ (broken phase) ($= \Lambda$ (unbroken phase))  is  identified with the Landau pole and $\mu^2$ the ``renormalization scale'' traded for $q^2$,  the  $({\rm momentum})^2$ of $\rho_\mu$.
The cutoff $\Lambda$ is needed to define the nonperturbative quantum theory by regularizing the divergence of the kinetic term which is absent as a counter term in the tree-level SM Lagrangian and thus 
cannot be renormalized in the ordinary sense, in sharp contrast to the pSM producing no such an extra kinetic term. Then the HLS gauge coupling $g_{_{\rm HLS}}$ (or $\Lambda$) is an {\it extra free parameter} of the nonperturbative dynamics within the SM, similarly to the $CP^{N-1}$ model in $D=4$ (see footnote \ref{CPN}) and other nonperturbative dynamics.
In the broken phase,  we have the ``on-shell'' $\rho_\mu$ mass $M_\rho^2=M_\rho^2(\mu^2=M_\rho^2)$ for the mass function $M_\rho(\mu^2)= g_{_{\rm HLS}}^2(\mu^2)\cdot F_\rho^2$ as: 
\begin{eqnarray}
M_\rho^2 &=& 
g_{_{\rm HLS}}^2(M_\rho^2) \cdot F_\rho^2= \lambda_{_{\rm HLS}}(M_\rho^2) \cdot 2 v^2\, ,\nonumber\\
F_\rho^2&\equiv&2\cdot  N v^2 =2\cdot F_\pi^2\simeq 2\cdot  (246\, {\rm GeV})^2\simeq (350 \, {\rm GeV})^2\,,
\label{rhomassresult1}
\end{eqnarray}
where  $v$ is the $N-$independent VEV $v^2 =F^2_\pi/N $, 
and 
the factor 2 for $F_\rho^2/F_\pi^2$ reflects  the covariant derivative parameterization which corresponds to $a=2$.

Thus  {\it we find the dynamical gauge boson, SM rho $\rho_\mu$, of $O(3)_{\rm local}\simeq [SU(2)_V]_{\rm local}$ in the SM as the extrapolation  $N\rightarrow 4$ with $p=3 (=$ fixed) of the large $N$ limit}.\footnote{
It is known that the large $N$ results remain qualitatively true even for the smallest  value of $N=2$ in the $CP^{N-1}$ model,  which is checked by  the equivalent  $O(3)$ model exactly solvable in 2 dimension. See e.g., Ref.\cite{Witten:1978bc}. Not to mention that the large $N_c$ QCD  also well describes the reality with a small $N_c=3$.
} 
Note that explicit $N-$dependence enters only in the relation Eq.(\ref{kinetic1}) between $g_{_{\rm HLS}}=g_{_{\rm HLS}}(M_\rho^2)$ and the Landau pole $\tilde \Lambda$, while the phenomenologically relevant relation Eq.(\ref{rhomassresult1}) has no explicit $N-$dependence, given the SM value $F_\pi\simeq 246$ GeV ($F_\rho\simeq 350$ GeV).

Needless to say,   the SM fermions 
 carries no $H_{\rm local}$ charges, so that the dynamically generated 
 {\it new gauge symmetry HLS  is trivially anomaly-free} by construction.\footnote{Even in the generic case (not  the present SM case)  having gauged-$G_{\rm global}$ anomaly (Wess-Zumino-Witten anomaly) as in QCD, the dynamical HLS such as the one associated with $\rho/\omega$ mesons in QCD is still anomaly-free, see Ref.\cite{Fujiwara:1984mp,Bando:1987br}. }
\\

Although the above results are obtained in a specific parameterization corresponding to $a=2$, we further show that the large $N$ dynamics yields, {\it   independently of} ``$a$'',  not just Eqs.(\ref{kinetic1}) and ({\ref{rhomassresult1})  but all the on-shell relations, analogues of 
{\it the successful
 results of $a=2$ choice in the conventional HLS treatment} (simply assuming the kinetic term) in QCD, i.e., the $\rho-$ universality ($g_{\rho\pi\pi}= g_{_{\rm HLS}}$) and KSRF II, 
and some of the off-shell physics like analogue of the  VMD of the $\pi$ form factor ($W/Z/\gamma-W_L W_L/Z_L$ dominated by $W/Z/\gamma\rightarrow \rho_\mu\rightarrow W_L W_L/Z_L$ in the present case). 

 On the other hand, the {\it dynamically generated $\rho_\mu$ propagator does depend on $a$} in the large $N$ limit, in such a way that the equation of motion of $\rho_\mu$ at classical level (in the absence of the kinetic term)  is violated at quantum level by the term depending on $a$ as proportional to $1/a$. Thus the off-shell physics depends on $a$ in
 principle, with a curious exception of  the
 vector meson dominance for $\pi$ form factor as mentioned above.  As a result,  {\it at $a\rightarrow \infty$ the equation of motion of $\rho_\mu$ at classical level (in the absence of the kinetic term)  remains intact at quantum level, so that the $\rho_\mu$ kinetic term is
totally replaced by the Skyrme term}, which was the case explicitly shown by Ref.\cite{Matsuzaki:2016iyq}  that 
the SM skyrmion as a dark matter is stabilized by the dynamical HLS gauge boson $\rho_\mu$. We thus establish the key ingredient of Ref.\cite{Matsuzaki:2016iyq}. 
\\

Phenomenological implications of our result  for the SM rho would be two different scenarios depending on the possible value of a single free parameter existing in the theory,  $M_\rho=g_{\rho\pi\pi}\cdot F_\rho$ (or $g_{\rho\pi\pi}=g_{_{\rm HLS}}\equiv g_{_{\rm HLS}}(M_\rho^2)
=M_\rho/F_\rho=M_\rho/(350 {\rm GeV})$or the cutoff $\Lambda$ (or the Landau pole $\tilde \Lambda$) in view of  Eqs.(\ref{rhomassresult1}) and (\ref{kinetic1}). The cutoff $\Lambda= 
e^{-4/3}\cdot  M_\rho  \cdot \exp{[\frac{3}{8}(4\pi  F_\rho)^2/M_\rho^2]} $ implies that  $\Lambda < M_\rho\, (g_{_{\rm HLS}}>6.7,\,  M_\rho> 2.3\,  {\rm TeV})$ and  $\Lambda > M_\rho\, ( g_{_{\rm HLS}}< 6.7, \, M_\rho< 2.3\,  {\rm TeV})$.  

1) ``Low $M_\rho$ scenario'' ($M_\rho<2.3 \, {\rm TeV},\,  \Lambda> M_\rho$):

The SM rho $\rho_\mu$ at the collider experiments may be produced through Drell-Yang processes  $q \bar q  \rightarrow W/Z/\gamma\rightarrow \rho_\mu$ with the coupling $\sim \alpha_{\rm em} F_\rho/M_\rho=\alpha_{\rm em} /g_{\rho\pi\pi}$. 
Given a reference value $M_\rho=2$ TeV for instance, we would have  $g_{\rho\pi\pi}\simeq 5.7$ and $\Lambda\simeq 3.3\, {\rm TeV}\simeq 4\pi F_\pi$ (simple scale-up of  the QCD $\rho$ meson). This  yields the width $\Gamma_\rho \simeq \Gamma_{\rho\rightarrow WW}\simeq g_{\rho\pi\pi}^2 M_\rho/(48\pi) 
\simeq  433 $ GeV, so broad as barely detectable at
 LHC.   For larger (smaller) $M_\rho$ the width gets larger (smaller) as $\sim M_\rho^3$, and the  production cross section gets smaller (larger) as $\sim 1/M_\rho^2$, thus more difficult for $M_\rho>2$ TeV to be seen at LHC. 
 The SM rho with narrow resonance $\Gamma_\rho\lesssim100$ GeV could be detected at LHC for  $M_\rho\lesssim1.2$ TeV, which corresponds to $g_{_{\rm HLS}}\lesssim 3.5$ and $\Lambda \gtrsim 50\, {\rm TeV}$.

  2) ``High $M_\rho$  scenario''  ($M_\rho={\cal O} (10^2-10^3) {\rm TeV},\,   \Lambda < M_\rho$,  as a stabilizer of the skyrmion dark matter $X_s$)\cite{Matsuzaki:2016iyq}:
  
 Even if no direct evidence were seen at the collider experiments,  physical effects of the dynamical $\rho_\mu$ are still observable through the skyrmion dark matter $X_s$ in the SM.  In fact the SM skyrmion is stabilized by the
  {\it off-shell} $\rho_\mu$ in the {\it short distance} physics as shown in Ref.\cite{Matsuzaki:2016iyq}, the result of which corresponds to $a\rightarrow \infty$ calculation, while the results are numerically similar even for $a\sim 2 
  $ \cite{MOY2018}. The HLS coupling is extremely large $g_{_{\rm HLS}}={\cal O}(10^3) $ to realize   $M_{X_s}\lesssim {\cal O} (10)$ GeV  consistent with  the direct detection of the dark matter,  in rough agreement with  
 the relic abundance of the dark matter: $\Omega_{X_s} h^2 \simeq 0.1$ \cite{Matsuzaki:2016iyq,MOY2018}. Note the cutoff is  $\Lambda =e^{-4/3} \tilde \Lambda \simeq e^{-4/3}\cdot M_\rho ={\cal O} (10^2\, {\rm TeV})$, where
$M_\rho = g_{_{\rm HLS}} \cdot F_\rho $ is  a typical mass scale (no longer  the ``on-shell'' mass, since the SM rho  is deeply off-shell).  
In either scenario, the phenomenologically interesting nonperturbative SM physics has  typical strong SM rho gauge coupling
$ g_{_{\rm HLS}}\simeq  1/3 - 10^3$, which will have the cutoff $\Lambda ={\cal O} (10^0 -10^2)\, {\rm TeV}$ 
close to the weak scale in sharp contrast to the pSM, thus {\it resolving the naturalness problem within the full SM} including the nonperturbative effects, even without recourse to the BSM.\footnote{This indicates that the quadratic divergence  corrections  to the  weak scale $\delta F_\pi^2\sim 4\cdot \Lambda^2/(4\pi)^2\sim (0.1\, {\rm TeV})^2 - (10\,  {\rm TeV})^2$ (see the gap equation Eq.(\ref{quadraticdiv})).   This also suggests a possibility that the SM in the full nonperturbative
formulation  eventually reveals itself as  
 a  ``dual'' to a possible BSM underlying theory with  such a scale,
  similarly to the hadron-quark duality (nonlinear sigma model/chiral Lagrangian vs QCD). 
   }
   \\

We further show that the theory of this type has a salient phase transition, though at this moment  it is purely formal discussion. 
As in many nonlinear sigma models such as the $CP^{N-1}$ model, 
 the  dynamical HLS  $O(p)_{\rm local}$  gauge bosons $\rho_\mu$ in the genuine nonperturbative unbroken phase 
  of
the $O(N)/[O(N-p)\times O(p)]$ Grassmannian nonlinear sigma model in the large $N$ limit
are {\it massless}~\cite{Brezin:1980ms}. In this respect, we demonstrate that the gauge symmetry, HLS, is mandatory to keep the theory well-defined at quantum level not just in the broken phase but also in the unbroken  phase, similarly to the $CP^{N-1}$ model.
In particular,  in  2 dimensions the model having vanishing critical coupling has only the unbroken 
phase in accord with the Mermin-Wagner-Coleman theorem, thus the well-defined quantum theory exists only at presence of the HLS.

  The (zero temperature) {\it phase transition between the broken and the unbroken phase takes place independently of $a$  as the second order phase transition} at a critical point (nontrivial ultraviolet fixed point)  of a dimensionless ``coupling'' related to the condensate (decay constant). 
  The phase change goes through, with the (induced) HLS gauge coupling tending to zero continuously 
   from both sides of the phases (second order phase transition), where the massless and massive spectrum interchanged between the HLS gauge boson $\rho_\mu$  and the $\pi$  modes (corresponding to
 NG bosons in the broken phase).  This $a-$independent phase transition is compared with a similar symmetry restoration  ``Vector Manifestation'' \cite{Harada:2000kb,Harada:2003jx} proposed in the non-scale-invariant nonlinear sigma model at $a=1$ (a fixed point) at one loop (of ${\cal O} (p^4)$ in the sense of the chiral perturbation theory). 
 \\

Finally,  the results in this paper  are of direct relevance to the 2-flavored QCD, not just the SM Higgs Lagrangian. 
  In fact, the dynamical results obtained here for the SM in the large $N$ limit are quite independent of the presence of the (pseudo) dilaton $\varphi=$ SM Higgs.   Thus they apply most directly to the 2-flavored QCD,
which is described by  the same nonlinear sigma model (without scale invariance/pseudo-dilaton) having the $\rho$ meson as the dynamical gauge boson of HLS, thereby proving all the otherwise mysterious ``$a=2$ relations'' as the reality of QCD, such as the $\rho-$ universality, KSRF relation II, and the VMD,  be realized $a-$ independently, together with the skyrmion (nucleon) stabilized by the dynamical $\rho$ meson simply as  nonperturbative dynamical effects in the large $N$ limit,  {\it without recourse to the underlying QCD}. 
The dynamically generated kinetic term has a new free parameter, the $\rho$ coupling $g_{\rho\pi\pi}=g_{_{\rm HLS}}  \simeq 5.9$ corresponding to
 $m_\rho \simeq 770\, {\rm MeV}$ and  through  Eq.(\ref{kinetic1}) we have  $\Lambda \simeq 1.1\, {\rm GeV}(\simeq 4 \pi f_\pi)$.
Then this establish that 
  the nonperturbative dynamics of the nonlinear sigma model having dynamical $\rho$ meson  is certainly dual to the underlying QCD, matched each other at $\Lambda \sim \Lambda_\chi \sim \Lambda _{\rm QCD}$. The Grassmannian manifold, thus being the right macroscopic theory dual to the underlying theory, QCD, gives us  an unmistakable evidence for the observation~\cite{Harada:1999zj,Harada:2003jx,Komargodski:2010mc}  that the HLS is a ``magnetic gauge symmetry'' $\grave{a}$ la Seiberg duality~\cite{Seiberg:1994bz} even in the non-SUSY QCD.
\\

The paper is organized as follows:
\\

 In the next section we recapitulate the re-parameterization of the SM Higgs Lagrangian in terms of the (approximately) scale-invariant version of the
 nonlinear sigma model 
 $G/H=SU(2)_L\times SU(2)_R/SU(2)_V\simeq O(4)/O(3)$.

 In Section III,
 as an $N$ extension of the model $G/H=O(4)/O(3)\simeq O(4)_{\rm global} \times O(3)_{\rm local}$, we introduce the HLS model  $G_{\rm global}\times H_{\rm local}=O(N)_{\rm global} \times [O(N-p) \times O(p)]_{\rm local}$ which is gauge equivalent to the
 ``CCWZ  representation'' of
 the Grassmannian manifold $G/H=O(N)/[O(N-p)\times O(p)]$. 
 
 In Section IV we  first introduce an alternative parameterization of the model in terms of the covariant derivative of the $p\times N$ component real field $\phi$ with constraints through 
 Lagrange multiplier. Solving the constraints we show that this parameterization is equivalent to a specific 
 parameter choice $a=2$ in the generic HLS Lagrangian.
We then show that the effective action and gap equation in the large $N$ limit and identify the two phases, broken and unbroken $O(N)$ symmetry,
 with the scale symmetry spontaneously (and explicitly) broken  in each phase by a different order parameter.

Section V is the main part of the paper where we demonstrate the dynamical generation of the $O(p)_{\rm local}$ HLS gauge boson $\rho_\mu$, while  not  of the $O(N-p)_{\rm local}$ in the 
 large $N$ limit for a concrete case $p=3$, and hence the dynamical generation of $\rho_\mu$ in the SM. The second order phase transition
 with the vanishing induced HLS gauge coupling at the transition point is also discussed.
 
 In section VI,  the $a-$independence of the physical quantities of the $\rho_\mu$ is shown, while the $a-$dependent part
 of the $\rho_\mu$ propagator.  Its physical implications are further discussed. In particular, the $\rho-$universality, KSRF relations, I,II, and the VMD are shown to be realized independently of the parameter $a$, while the skyrmion dynamics is shown to  depend on $a$ in such a away that $a\rightarrow \infty$ limit realizes the pure Skyrme term.  Phenomenological implications of the SM rho for the collider physics and the skyrmion dark matter are further discussed, 
 both acting as solution to the naturalness problem within the SM. 
 
 SectionVII is devoted to  Summary and Discussions. 
  
 In Appendix \ref{CCWZ}, we summarize the basic formalism of the CCWZ nonlinear sigma model based on the manifold $G/H$,
 and its gauge-equivalent HLS model in the scale-invariant version. 
 In Appendix \ref{CPNApp}, we recapitulate the well-known  dynamical generation of the auxiliary field (HLS gauge boson) in $CP^{N-1}$ model
 which is the same dynamical phenomenon in the large $N$ limit as the SM discussed in the present paper. Appendix \ref{NJL} is
 for a review of the dynamical generation of the auxiliary fields  in the NJL model, which is also the same large $N$ dynamical phenomenon
 as that in the present paper. Appendix \ref{universality} is for a direct calculation to prove the $\rho-$universality in the large $N$ limit.

\section{SM Higgs Lagrangian as a nonlinear sigma model} 
Let us first recapitulate the fact~\cite{Fukano:2015zua} that the SM Higgs Lagrangian is re-parameterized into a scale-invariant nonlinear sigma model. 

\subsection{$G/H=SU(2)_L \times SU(2)_R/SU(2)_V$ parameterization}

As is well-known, the SM Higgs Lagrangian 
takes the form of the $SU(2)_L\times SU(2)_R$ linear sigma model: 
\begin{eqnarray}
{\cal L}_{\rm SM}&=&
 |\partial_\mu h|^2 -m 
 ^2 |h|^2 -\lambda|h|^4 
\nonumber
\\
&=& \frac{1}{2} \left[
\left(\partial_\mu {\hat \sigma}\right)^2 +\left(\partial_\mu {\hat \pi_a}\right)^2
\right]
-\frac{m 
^2 }{2}
 \left[ 
{\hat \sigma}^2+{\hat \pi_a}^2 
\right]-\frac{\lambda}{4} \left[ 
{\hat \sigma}^2+{\hat \pi_a}^2 
\right]^2 
\nonumber
\\
&=&
\frac{1}{2} {\rm tr} \left( \partial_\mu M\partial^\mu M^\dagger \right)
 - \frac{m 
 ^2}{2} {\rm tr}\left(M M^\dagger\right)-\frac{\lambda}{4}  \left({\rm tr}\left(M M^\dagger\right)\right)^2 \,,
\label{Higgs1} 
  \end{eqnarray}
  where 
  \begin{equation}
h=\left(\begin{array}{c}
\phi^+\\
\phi^0\end{array}  \right)=\frac{1}{\sqrt{2}} \left(\begin{array}{c}
i{\hat \pi}_1+{\hat \pi}_2\\
\hat{\sigma}-i {\hat \pi}_3\end{array}\right)\,,
\end{equation}   
  and
  the $2\times 2$ matrix $M$ reads:
\begin{equation}
M=(i \tau_2 h^*, h) = \frac{1}{\sqrt{2}}\left({\hat \sigma}\cdot  1_{2\times 2} +2i {\hat \pi}\right)\,, \quad \left({\hat \pi} \equiv {\hat \pi}_a \frac{\tau_a}{2}\right)\,,
\end{equation}  
which transforms under $SU(2)_L\times SU(2)_R$ as
\begin{equation}
M \rightarrow g_L \, M\, g_R^\dagger \,,\quad \left(g_{R,L} \in SU(2)_{R,L}\right)\,.
\end{equation} 
The potential term,
\begin{eqnarray}
V(\hat \pi, \hat \sigma) 
&=& m 
 ^2 |h|^2 +\lambda|h|^4 \nonumber \\
&=&\frac{m 
^2 }{2}
 \left[ 
{\hat \sigma}^2+{\hat \pi_a}^2 
\right]+\frac{\lambda}{4} \left[ 
{\hat \sigma}^2+{\hat \pi_a}^2 
\right]^2 =
 \frac{m^2}{2} {\rm tr}\left(M M^\dagger\right)+\frac{\lambda}{4}  \left({\rm tr}\left(M M^\dagger\right)\right)^2 \nonumber\\
&=& \frac{m^2 }{2}
\sigma^2 + \frac{\lambda}{4} \sigma^4= V(\sigma)\,,
\end{eqnarray}
with
$ \sigma(x) \equiv \sqrt{ {\hat \sigma}^2(x) +{\hat \pi_a}^2(x)}$, 
  has a minimum  for $m 
  ^2<0$ at the chiral-invariant circle:
\begin{equation}
\langle \sigma(x)\rangle = \sqrt{\frac{-m^2}{\lambda}}\equiv v=246\,{\rm GeV} \,.
\label{chicircle}
\end{equation}

On this vacuum the 
complex matrix $M$ can be decomposed into a positive Hermitian (diagonalizable) matrix $H$  and a unitary matrix $U$ as $M=HU$ 
(``polar decomposition'') \cite{Bando:1987br} : 
 \begin{equation}
 M(x) = H(x)\cdot U(x)\,, \quad H(x)=\frac{1}{\sqrt{2}} \left(\begin{array}{cc}
 \sigma(x) & 0\\
 0  &\sigma(x)
 \end{array}\right)
 \,, \quad U(x)= \exp\left(i \frac{2\pi(x)}{F_\pi}\right)  \,,\,  F_\pi=v=\langle  \sigma(x) \rangle\,,
  \label{Polar}
 \end{equation}
 with $\pi(x)=\pi^a(x) \frac{\tau^a}{2} \,(a=1,2,3)$ and the $\pi$ decay constant $F_\pi$. 
 The chiral transformation of $M$ is 
 carried by $U$,
 while $H$ is a chiral singlet such that:
 \begin{equation}
 U \rightarrow g_L \, U\, g_R^\dagger\,,\quad H \rightarrow H\,,
 \label{transformation}
 \end{equation}
where $g_{L/R} \in SU(2)_{L/R}$.
 Note that the {\it physical Higgs is the
 radial mode $\sigma$ which is a chiral-singlet} (electroweak gauge singlet when electroweak coupling switched on),  while {\it $\hat \sigma$ is a chiral non-singlet} (electroweak gauge non-singlet) transforming 
 to the chiral partner ${\hat \pi}$ by the chiral rotation, both being tachyons with ${\rm mass}^2 =m^2 <0$, in contrast to the physical modes $\sigma$  and $\pi$
 (angular/phase modes or gauge parameters totally absorbed into $W/Z$ in the unitary gauge). 
In fact $\sigma$ is a chiral singlet and thus  $\langle  \sigma(x) \rangle=v\ne 0$ breaks spontaneously the scale symmetry, but not the chiral symmetry which is actually spontaneously broken by $\langle U(x)\rangle =1 \ne 0\, (\langle \pi(x)\rangle =0)$.

  We thus may parametrize $\sigma(x)$  as the nonlinear base of the scale transformation:
 \begin{equation} 
 \sigma(x) =v \cdot \chi(\varphi)\,,\quad \chi(\varphi)=\exp\left(\frac{\varphi(x)}{F_\varphi}\right)\,, \,\, F_\varphi=v \,,
 \label{NLscale}
 \end{equation}
such that  $\langle \chi(\varphi) \rangle = 1\ne 0$ ($\langle\varphi(x)\rangle=0$): Now the {\it physical Higgs is $\varphi(x)$ which is a dilaton}, NG boson of the spontaneously broken scale symmetry,    
with the decay constant $F_\varphi=v$~\footnote{
 The scale (dilatation) transformations for these fields are 
 \begin{equation}
 \delta_D \sigma =(1 +x^\mu \partial_\mu) \sigma \,, \qquad  
\delta_D \chi=(1+x^\mu \partial_\mu) \chi\,, \qquad 
\delta_D \varphi= v +x^\mu \partial_\mu\varphi\,. 
\nonumber
 \end{equation}
Although $\chi$ is a dimensionless field,
it transforms as that of dimension 1, while $\varphi$ having dimension 1 transforms as the dimension 0, instead.
}. 
 Eq.(\ref{Higgs1})  is then straightforwardly rewritten into the form~\cite{Fukano:2015zua}: 
    \begin{eqnarray}
{\cal L}_{\rm SM} 
&= &
\frac{1}{2} \left(\partial_\mu \sigma \right)^2  +\frac{\sigma^2}{4}{\rm tr} \left(\partial_\mu U \partial^\mu U^\dagger\right) -V(\sigma) \nonumber \\
&=&\chi^2(\varphi) \cdot \left[ \frac{1}{2} \left(\partial_\mu \varphi\right)^2  +\frac{v^2}{4}{\rm tr} \left(\partial_\mu U \partial^\mu U^\dagger\right)\right] -V(\varphi)\,, 
  \label{Higgs2}  \\
  V(\varphi) &=&V(\sigma) 
=\frac{\lambda}{4} v^4 \left[\left(\chi^2(\varphi) -1\right)^2-1\right]\,.
\label{potential}
      \end{eqnarray}

Thus the SM Higgs Lagrangian Eq.(\ref{Higgs1})  
is trivially  identical to Eq.(\ref{Higgs2}). Note that the kinetic term of the latter, 
$\chi^2(\varphi) \cdot \left[ \frac{1}{2} \left(\partial_\mu \varphi\right)^2  +\frac{v^2}{4}{\rm tr} \left(\partial_\mu U \partial^\mu U^\dagger\right)\right]$, 
 contains the usual nonlinear sigma model 
 $\frac{v^2}{4}{\rm tr} \left(\partial_\mu U \partial^\mu U^\dagger\right) $ 
 which transforms as dimension 2 to make the action not scale-invariant.
However, the extra dilaton factor 
$\chi^2(\varphi)=e^{2\varphi(x)/v}$,  
transforming as dimension 2, makes the whole kinetic term to be dimension 4. Hence 
the action becomes scale-invariant as it should, since it is  just a rewriting of the original kinetic term in Eq.(\ref{Higgs1}) which is scale-invariant (dimension 4).

Actually the kinetic term coincides with the 
scale-invariant nonlinear chiral Lagrangian based on the coset $G/H=SU(2)_L\times SU(2)_R/SU(2)_{L+R}$, 
with  the scale symmetry as well as the chiral symmetry being realized nonlinearly. 
The {\it scale $v$} is in fact a measure of the {\it spontaneous} breaking  but {\it not the explicit breaking} of the scale symmetry.
Thus  
{\it the SM Higgs 
$\varphi$ is nothing but a pseudo dilaton}, 
with 
the {\it explicit breaking of the scale symmetry} from the potential term $V(\varphi)$ 
characterized by the {\it dimensionless parameter}:
\begin{equation}
\lambda=\frac{M_\varphi^2}{
2 v^2}
\simeq 
\frac{(125 \,{\rm GeV})^2}
{
2\times (246\, {\rm GeV})^2
} \simeq \frac{1}{
8} \ll 1\,,
\end{equation}
which is very close to the {\it  ``conformal limit''}, 
\begin{equation}
\lambda\rightarrow 0\,\,\, {\rm with} \,\,\,v= {\rm fixed} \,,
\label{BPSlimit}
\end{equation}
 where $V(\varphi)\propto  \lambda v^4 \rightarrow 0$ for {\it any} $\chi(x)=e^{\varphi(x)/v}
 $~\footnote{
The opposite  limit, $\lambda$ $\rightarrow$$\infty$ with $v$$=$ fixed, leads to the 
ordinary nonlinear sigma model, where we also have $V(\varphi) \rightarrow 0$ but 
with $\chi(x) \equiv 1$, so that the scale symmetry compensated by $\chi^2$ factor 
in Eq.(\ref{Higgs2}) is completely lost. See Ref.\cite{Fukano:2015zua,Yamawaki:2015tmu,Yamawaki:2016kdz,Yamawaki:2016qux}. 
Either limit has no
 $\lambda$ coupling, but has derivative couplings instead, which are ``weak'' in the low energy
 $p^2/(4\pi v)^2 \ll 1$, so that the perturbation according to the derivative expansion (``chiral perturbation theory'') makes sense.
}. 
The limit  actually
corresponds to the ``Bogomol'nyi-Prasad-Sommerfield (BPS) limit'' of 
't Hooft-Polyakov monopole in the Georgi-Glashow model~\cite{Harvey:1996ur}, 
similarly to the SUSY flat direction~\footnote{
Even if we take such a conformal/BPS limit, 
the theory is still an interacting theory with {\it derivative} {\it coupling} as in the usual chiral Lagrangian,
 and thus the quantum corrections will produce the trace anomaly of dimension 4, 
$\sim$ $v^4$ $ \chi^4 \ln \chi$, as a new source of the
SM Higgs mass as a pseudo-dilaton, which, however, 
would do not affect the dynamical generation of the HLS gauge boson discussed here, similarly to the tiny explicit 
scale-symmetry breaking 
 in the tree-level potential $V(\varphi)$ with 
$\lambda$ $\simeq $ $1/8$ $\ll$ $1$. 
}.

By the electro-weak gauging as usual; $\partial_\mu U\Rightarrow {\cal D}_\mu U= \partial_\mu U -i g_2 W_\mu U +i g_1 U B_\mu$
 in Eq.(\ref{Higgs2}), we see the followings: first, the physical Higgs field $\varphi$ as a pseudo-dilaton is a gauge singlet and hence {\it manifestly gauge invariant} object, in sharp contrast to the conventional shifting $h_0=h_0^\prime +v/\sqrt{2}
$ or $\hat \sigma ={\hat \sigma}^\prime +v$, where  $h_0^\prime$ or $ {\hat \sigma}^\prime$ is gauge variant.  Second,
 the mass term of $W/Z$ is scale-invariant thanks to the dilaton factor $\chi$, and so is the mass term of the SM fermions $f$: $g_Y  \bar f h f
 =(g_Y v/\sqrt{2}) (\chi \bar f f)$, all with the scale dimension 4.  
 This implies that {\it the couplings of the SM Higgs as a pseudo dilaton to all the SM particles are written in the scale-invariant form and thus 
 obey the low energy theorem of the scale symmetry} in perfect agreement with the experiments: 
 
 The low energy theorem for the pseudo dilaton $\varphi(q_\mu)$ coupling to the canonical matter filed $X$ at  $q_\mu$$\rightarrow 0$ reads 
 \begin{eqnarray}
 g_{\varphi X^\dagger X}=\frac{2M_X^2}{F_\varphi},\quad g_{\varphi \bar X X}= \frac{M_X}{F_\varphi} \quad \left(F_\varphi=v\right)\,,
  \label{LET}
    \end{eqnarray}
   for complex scalar and spin $1/2$ fermion, respectively \cite{Carruthers:1971vz},  which can also be read from
    the scale invariance of the mass term;
 \begin{eqnarray}
 M_X^2 \cdot \chi^2\, X^\dagger  X&=&M_X^2 X^\dagger X + \frac{2 M_X^2}{v} \varphi X^\dagger X +\cdots\,,\nonumber \\
  M_X \cdot\chi\,  \bar X  X&=&M_X \bar X X + \frac{M_X}{v} \varphi {\bar X} X +\cdots\,,
 \end{eqnarray}
 for the respective canonical field with 
 the canonical dimension.  
\footnote{
For the general form of the low energy theorem of the scale symmetry including the anomalous dimension such as in the walking technicolor with large anomalous $\gamma_m=1$~\cite{Yamawaki:1985zg}, see Ref.~\cite{Bando:1986bg,Matsuzaki:2012vc}.
}

  \subsection{$G/H=O(4)/O(3)$ Parameterization}
 
Since the nonlinear sigma model on the manifold  $G/H=[SU(2)_L\times SU(2)_R]/SU(2)_V$ is 
equivalent to another model  $G/H=O(4)/O(3)$, we here  present an explicit form of the scale-invariant form of the
latter, based on the generic CCWZ formalism reviewed in Appendix \ref{CCWZ2}, where that for $G/H=[SU(2)_L\times SU(2)_R]/SU(2)_V$
is given, see Eq.(\ref{Higgs4}). The resultant form  is equivalent to Eq.({\ref{Higgs2}) and hence to the SM Higgs Lagrangian Eq.(\ref{Higgs1}).
 This is further regard as an extrapolation $N\rightarrow 4$ with $p=3$ of the  Grassmannian manifold $G/H=O(N)/[O(N-p)\times O(p)]$ ($p=3$) as will be discussed  
 in the next section.
 \\
 
Here we take  a  vector representation of $O(4)$ with different normalization $ {\rm tr} (T_A
T_B
)= 2 \delta_{AB}$,  $T^t_A=-T_A$, with   $\pi=\pi_a X_a\,,  {\rm tr} (\pi^2)\, = 2 \pi_a^2$.
 we have 
\begin{eqnarray}
\xi^{(4)}
(\pi)_{\alpha\beta} 
&=&
\left(
\begin{array}{c}
\phi(\pi)_{i\beta} \\
\Phi(\pi)_{4 \beta}\end{array}
\right) =\exp \left(
i \frac{\pi_a\cdot X_a}{F_\pi}\right) 
=\exp\left[\frac{1}{F_\pi} \left(
  \begin{array}{cccc}
   & & &\pi_1\\
   &  \text{{\huge{0}}} & &\pi_2\\
      &  &  &\pi_3\\
   -\pi_1&-\pi_2 & -\pi_3&0  
  \end{array}
\right)\right] \,,\nonumber\\
&&\left[\xi^{(4)}(\pi)\right]^t \cdot
\,\xi^{(4)}(\pi)= \phi^t(\pi)\, \phi(\pi)  + \Phi^t(\pi) \Phi(\pi)=\1
=\xi^{(4)}(\pi)\cdot \left[\xi^{(4)}(\pi)\right]^t \,,\nonumber\\
&&\alpha,\beta= i,4\,;\quad  i=1,2,3\,, 
\label{O4}
\end{eqnarray}
which transforms under $G$ as $\xi^{(4)}
(\pi) \rightarrow h(g,\pi) \cdot \xi^{(4)}(\pi)\cdot  g^t$ ($g\in G$).

The Maurer-Cartan one-form reads: 
\begin{eqnarray}
\alpha_\mu^{(4)}(\pi)&=&\frac{1}{i} \partial_\mu \xi^{(4)}
(\pi)  \cdot \left(\xi^{(4)}(\pi)\right)^t 
= \frac{1}{i} \left[\left(\frac{i}{F_\pi}\right)\partial_\mu \pi + \frac{1}{2!}\left(\frac{i}{F_\pi}\right)^2  \left[\pi,\partial_\mu \pi\right]  +\frac{1}{3!}
\left(\frac{i}{F_\pi}\right)^3\left[\pi,\left[\pi,\partial_\mu \pi \right] \right]+ \cdots\right]\,
\nonumber\\
&=&\frac{1}{i} \left(
\begin{array}{c}
\partial_\mu \phi(\pi)\\
\partial_\mu \Phi(\pi) 
\end{array}
\right)
\cdot \left(
\begin{array}{cc}
\phi^t(\pi) &\Phi^t(\pi)
\end{array}
\right)
=\frac{1}{i}\left(
\begin{array}{cc}
\partial_\mu \phi(\pi) \cdot \phi^t(\pi)& \partial_\mu \phi(\pi) \cdot\Phi^t(\pi)\\
\partial_\mu \Phi(\pi) \cdot \phi^t(\pi)& \partial_\mu \Phi(\pi) \cdot\Phi^t(\pi)
\end{array}
\right)=\alpha_{\mu,\perp}(\pi) + \alpha_{\mu,||}(\pi)\,,\nonumber \\
\alpha_{\mu,\perp}^{(4)}(\pi)
&=&\frac{1}{2} {\rm tr}^{(4)} \left(\alpha_{\mu}^{(4)}(\pi) X_a \right) \cdot X_a
= \frac{1}{i}
\left(
\begin{array}{cc}
 {\bf 0}_{3\times 3}& \partial_\mu \phi(\pi) \cdot\Phi^t(\pi)\\
\partial_\mu \Phi(\pi) \cdot \phi^t(\pi)&  0
\end{array}
\right)=\frac{1}{F_\pi} \partial_\mu \pi  + \cdots
\,, \nonumber\\
\alpha_{\mu,||}^{(4)}(\pi)
&=&\frac{1}{2} {\rm tr}\left(\alpha_{\mu}^{(4)}(\pi) S_a \right) \cdot S_a
= \frac{1}{i}
\left(
\begin{array}{cc}
\partial_\mu \phi(\pi) \cdot\phi^t(\pi)& {\bf 0}_{3\times 1} \\
{\bf 0}_{1\times 3}& \partial_\mu \Phi(\pi) \cdot \Phi^t(\pi)
\end{array}
\right)= \frac{i}{2F_\pi^2}  \left[\pi,\partial_\mu \pi\right] +\cdots\,, 
\end{eqnarray}
and the CCWZ Lagrangian reads:
\begin{eqnarray}
{\cal L}_{_{\rm CCWZ}}&=&
\frac{F_\pi^2}{4} {\rm tr} \left(
\left(\alpha_{\mu,\perp}^{(4)}(\pi)\right)^2 \right)
= - \frac{F_\pi^2}{2} {\rm tr}\left(\partial_\mu \phi(\pi) \Phi^t(\pi)\partial^\mu \Phi(\pi)
\phi^t(\pi) \right)\nonumber\\
&=& \frac{F_\pi^2}{2}   {\rm tr}_{_{3\times 3}} \left(\partial_\mu \phi(\pi)\cdot \Phi^t(\pi) \,\, \Phi(\pi) \cdot  \partial^\mu \phi^t(\pi)
\right)\nonumber\\
&=&\frac{F_\pi^2}{2} {\rm tr}_{_{3\times 3}} \left(\partial_\mu \phi(\pi)\cdot \partial^\mu \phi^t(\pi)+ \phi(\pi)\partial_\mu \phi^t(\pi)\cdot \phi(\pi)\partial^\mu \phi^t(\pi)
\right)\nonumber\\
&=&\frac{1}{2 } \left(\partial_\mu \pi_a\right)^2 + \cdots\,,
\end{eqnarray}
where we have used $\Phi \phi^t=0$ and $\Phi^t \Phi  = 1 -\phi^t \phi$. 

  Then the SM Lagrangian Eq.(\ref{Higgs2}) with $F_\pi=v$ is further rewritten in the form of the CCWZ Lagrangian similarly to Eq.(\ref{Uform}) as:
   \begin{eqnarray}
   {\cal L}_{\rm SM} 
&=&\chi^2(\varphi) \cdot 
 \left[ 
\frac{1}{2} 
\left(
\partial_\mu \varphi 
\right)^2  
+ \frac{v^2}{4} {\rm tr}\left((\alpha^{(4)}_{\mu,\perp}(\pi))^2\right)
\right] 
-V(\varphi)\,.
\label{SMO4}
  \end{eqnarray}
  \\
  
It is now straightforward to  introduce the HLS in the SM written in the form of the nonlinear realization $G/H$ besides the nonlinear realization of the scale symmetry. In the Appendix \ref{sHLS} the generic HLS formalism~\cite{Bando:1984pw,Bando:1987br} for the CCWZ representation
$G/H$ is reviewed 
   and the explicit (scale-invariant) HLS form of the SM is given 
  for $G_{\rm global} \times H_{\rm local} =[SU(2)_L\times SU(2)_R]_{\rm global} \times [SU(2)_V]_{\rm local}$,
   and
  in particular for 
  $G_{\rm global} \times H_{\rm local} =O(4)_{\rm global} \times O(3)_{\rm local}$ in Eq.(\ref{SM-HLSO4}) which is 
  the base for the Grassmannian $N-$extension to be discussed in the followings. For reader's convenience the physical implications
  of the standard HLS formalism (with the HLS kinetic term put by hand) are also  reviewed in Appendix \ref{DGHLS}, which 
  are for comparison with the genuine dynamical generation of the HLS kinetic term in the large $N$ limit to be given in the present paper.

 \section{Grassmannian $N$-Extension of the SM} 
 
In order to discuss the dynamical generation of the HLS gauge bosons by the nonperturbative dynamics within the SM in terms of the $1/N$ expansion, in this section\footnote{This section largely depends on the explicit Grassmannian coset parameterization given by Taichiro Kugo (private communication). We thank him for his generous offer.}  we here write down 
the scale-invariant HLS model with $G_{\rm global} \times H_{\rm local}=O(N)_{\rm global} \times [O(N-p)\times O(p)]_{\rm local}$.
This is gauge equivalent to a scale-invariant version of the Grassmannian model on the manifold $G/H=O(N)/[O(N-p)\times O(p)]$, according to the generic HLS formalism~\cite{Bando:1984pw,Bando:1987br} (see Appendix \ref{sHLS}). By taking $N=4\,,p=3$ the model is reduced into the  scale-invariant HLS Lagrangian  of the $G_{\rm global}\times H_{\rm local}=O(4)_{\rm global}\times O(3)_{\rm local}$, Eq.(\ref{SM-HLSO4}), which is gauge equivalent to that of the
$G/H= O(4)/O(3)$ model, Eq.(\ref{SMO4}),
and thus is equivalent to the SM Higgs Lagrangian Eq.(\ref{Higgs2}) and Eq.(\ref{Higgs1}).\footnote{
 Extension to $G/H=[SU(N)\times SU(N)]/SU(N)$ and scale symmetry (not $[SU(N)\times SU(N)]$ linear sigma model!) is
    precisely the same form as Eq.(\ref{Higgs2}) with the $N\times N$ matrix
    \begin{eqnarray}
   U(x)= \exp\left(2 i \frac{\pi}{F_\pi}\right)\,,\quad \pi(x)= \pi_a(x) T^a\,,\quad \chi(x)=\exp\left(\frac{\varphi}{F_\varphi}\right)\,, \quad
  \frac{F_\varphi}{\sqrt{N/2}}= F_\pi = v \,,   
  \end{eqnarray}
    $T^a$ being the generator of $SU(N)$. In contrast,  the $SU(N)_L\times SU(N)_R$ linear sigma model has two
    independent quartic couplings, $\left({\rm tr} M M^\dagger\right)^2$ and ${\rm tr} [\left(M M^\dagger\right)^2]$,
 and $M$ has $N^2-1$  scalars in addition to  $\sigma$.    However this extension is not suitable for the study of the nonperturbative dynamics
 based on the large $N$ limit, because   
 all planar graphs of the induced HLS gauge bosons loops are equally on the leading order of $1/N$ expansion.
 On the other hand, a popular $N$ extension of the $O(4)/O(3)$ model for the large $N$ limit
   is $G/H=O(N)/O(N-1)$. One might consider a gauge equivalent HLS model $O(N)_{\rm global} \times O(N-1)_{\rm local}$, which however does not
   give the dynamical gauge boson of $O(N-1)$, since again all planar graphs do contribute at the leading order. 
 } 
 \\
   
Following the generic HLS formalism in Appendix \ref{sHLS},   let us define the HLS version of the $N-$ extension of the  CCWZ base, an  $N\times N$ real matrix field $\xi(x)$ which transforms under $G_{\rm global}\times H_{\rm local}$ as $\xi(x) \rightarrow h(x) \cdot \xi(x) \cdot g^{-1}$ with $h(x) \in H_{\rm local}\,, g\in G_{\rm global}$,  
    \begin{eqnarray}
  \xi (x)_{\alpha \beta} &=&\left[\exp\left(i \frac{1}{2} \theta_{\gamma\delta} T_{\gamma\delta} \right)\right]_{\alpha \beta} 
  =\exp
  \left[
  \left(
  \begin{array}{cccc}
 \left( \theta_{ij}\right)_{p\times p}&\left(\theta_{ik}\right)_{p\times N-p}\\
 \left(\theta_{ki}\right)_{N-p\times p}& \left( \theta_{kl}\right)_{N-p\times N-p}
    \end{array}
     \right)
  \right]\nonumber\\
  &=& \xi({\check \rho}^{(p)})\cdot \xi({\check \rho}^{(N-p)}) \cdot \xi(\pi)= \exp
  \left(
  i \frac{
  {\check \rho}_a^{(p)} S_a^{(p)}
  }{
  F_\rho^{(p)}
  }
  \right)
  \cdot 
  \exp\left(
 i \frac{
  {\check \rho}_a^{(N-p)} S_a^{(N-p)}
  }{
  F_\rho^{(N-p)}
  }  
  \right)\cdot \exp \left(i \frac{\pi_aX_a}{F_\pi }
  \right) \nonumber\\
&=&\left(\begin{array}{c}
   \phi_{i,\beta}(x)\\
    \Phi_{k,\beta}(x)
   \end{array}
   \right)\,,
     \label{ONpNp}  
     \\  
  T_{\alpha \beta}&=&-T_{ \beta\alpha}\,,\left(T_{\alpha \beta}\right)_{\gamma\delta}  =-
  i \left(\delta_{\alpha\gamma}\delta_{\beta\delta}-\delta_{\alpha\delta}\delta_{ \beta\gamma} \right)\,, \quad \left(\alpha, \beta=1,2,\cdots N\right)\,,
  \,\, {\rm tr} \left( T_A T_B  \right)=2 \, \delta_{AB}\,, \nonumber\\ 
   \theta_{\alpha \beta}&=&-\theta_{ \beta\alpha}\,;  \theta_{ij}={\check \rho}^{(p)}_{ij}/F_\rho^{(p)}\,, \theta_{kl}={\check \rho}^{(N-p)}_{kl}/F_\rho^{(N-p)}\,,
   \theta_{ik}=\pi_{ik}/F_\pi\,, \quad\left(i,j=1\cdots p;\, k,l=p+1\cdots N    \right)\,, \nonumber\\   
    \xi^t(x)\cdot \xi(x) &=&\phi^t(x) \phi(x) + \Phi^t(x) \Phi(x) =\1\,,\nonumber\\
  \xi(x)\cdot \xi^t(x) &=&  \left(\begin{array}{cc}
   \phi(x) \phi^t(x) &\phi(x) \Phi^t(x)\\
  \Phi(x) \phi^t(x)& \Phi(x) \Phi^t(x)\end{array}\right)
  =\left(\begin{array}{cc}
  \1_{p\times p} & {\huge 0} \\
  {\huge 0}& \1_{(N-p)\times (N-p)}
  \end{array}\right) =\1\,,
    \label{ONpNpconstraint}
     \end{eqnarray}
where $ {\check \rho}^{(p)}= {\check \rho}^{(p)}_a S_a^{(p)}$ and $ {\check \rho}^{(N-p)} ={\check \rho}^{(N-p)}_a S_a^{(N-p)}$ are the would-be NG bosons to be absorbed into the HLS gauge bosons of $O(p)_{\rm local}$ and $O(N-p)_{\rm local}$,
respectively. 

The degrees of freedom of $\phi$ is $N\times p$ which is divided into  $(N-p)\times p$ (NG bosons $\pi$) plus 
 $p\times (p-1)/2$ (would-be NG bosons $ {\check \rho}_a^{(p)}$), plus $p\times (p+1)/2$ modes (i.e., constraints $\phi \phi^t=\1_{p\times p}$
 corresponding to ``massive'' scalars in the broken phase, while the diagonal component out of them is a pseudo-dilaton in the unbroken phase), 
 \begin{eqnarray}
 N\times p\Big|_\phi = (N-p) \times p \Big|_\pi + \frac{p\times (p-1)}{2}\Big|_{\check \rho}
  + \frac{p\times (p+1)}{2}\Big|_{\rm constraint}\,,
  \label{components}
 \end{eqnarray}
 and similarly for $\Phi$. 
 
 In the case of $N=4, p=3$ which corresponds to the SM, there exists an HLS only for the $O(p)_{\rm local}$, since $O(N-p)$ does not exist. Moreover, as we shall see later, dynamical generation of the $O(N-p)_{\rm local}$ does not take place in the large $N$
limit. Needless to say, when the gauge fixed ${\check \rho}_a^{(p)}={\check \rho}_a^{(N-p)}=0$, Eq.(\ref{ONpNp}) is reduced  to the CCWZ base in the original 
$O(N)/[O(p)\times O(N-p)]$ model, which for $N=4, p=3$ is nothing but the $O(4)/O(3)$ parameterization Eq.(\ref{O4})
 in  the SM Lagrangian Eq.(\ref{SMO4}).  
\\

   The Maurer-Cartan one-form reads:
   \begin{eqnarray}
   \alpha_\mu(x)\equiv \frac{1}{i} \partial_\mu \xi(x)  \cdot \xi^t(x)  &=&\alpha_{\mu,\perp}(x) + \alpha_{\mu,||}(x)\,,\nonumber\\
    \alpha_{\mu,\perp} (x)&=&
  \frac{1}{i}   \left(\begin{array}{cc}
 0& \partial_\mu \phi  \cdot \Phi^t\\
  \partial_\mu \Phi  \cdot \phi^t & 0\end{array}
  \right)\,,\,\,
   \alpha_{\mu,||} (x)  = \frac{1}{i}   \left(\begin{array}{cc}
 \partial_\mu \phi  \cdot \phi^t&0\\
  0&\partial_\mu \Phi  \cdot \Phi^t\end{array}
  \right)\,,   \end{eqnarray}
  and its covariantized one:
     \begin{eqnarray}
  \hat  \alpha_\mu(x)&\equiv& \frac{1}{i} D_\mu \xi(x)  \cdot \xi^t(x) =\frac{1}{i} 
 \left( \begin{array}{c}
 \partial_\mu \phi- i\rho^{(p)}_\mu \phi\\
   \partial_\mu \Phi -i \rho^{(N-p)}_\mu \Phi
   \end{array}  
    \right) \cdot \left(\phi^t\,\, \Phi^t\right)\,, \nonumber\\
     {\hat \alpha}_{\mu,\perp}(x) &=& \alpha_{\mu,\perp}(x)      
    = \frac{1}{i}   \left(\begin{array}{cc}
 0& \partial_\mu \phi  \cdot \Phi^t\\
  \partial_\mu \Phi  \cdot \phi^t & 0\end{array}
  \right)\,, \nonumber\\ 
  {\hat  \alpha}_{\mu,||}(x)   &=&  \left(\begin{array}{cc}\alpha^{(p)}_{\mu,||} -  \rho^{(p)}_\mu &0\\
  0&\alpha^{(N-p)}_{\mu,||} -  \rho^{(N-p)}_\mu 
  \end{array}\right)
  =
   \frac{1}{i}   \left(\begin{array}{cc}
\partial_\mu \phi \cdot \phi^t - i\rho^{(p)}_\mu & 0\\
  0&\partial_\mu \Phi \cdot \Phi^t - i\rho^{(N-p)}_\mu 
  \end{array}
  \right)
        \,    ,   
    \end{eqnarray}
   where $\rho^{(p)}_\mu$ and $\rho^{(N-p)}_\mu$ are the HLS gauge bosons of $O(p)_{\rm local}$ and $O(N-p)_{\rm local}$,
   respectively. 
 ${\hat \alpha}_{\mu,\perp} (x)$ and ${\hat \alpha}_{\mu,||}(x)$ transform as
 \begin{eqnarray}
 {\hat \alpha}_{\mu,\perp}(x) &\rightarrow& h(x) \cdot {\hat \alpha}_\mu(x) \cdot h^{-1}(x)\,,\,\,\nonumber \\
  {\hat \alpha}^{(p)}_{\mu,|| } (x)&\rightarrow& h_{(p)}(x) \cdot {\hat \alpha}^{(p)}_{\mu,||}(x)  \cdot h_{(p)}^{-1}(x)\,,\,\,
   {\hat \alpha}^{(N-p)}_{\mu,|| }(x) \rightarrow h_{(N-p)}(x) \cdot {\hat \alpha}^{(N-p)}_{\mu,||}(x)  \cdot h_{(N-p)}^{-1}(x)\,,
   \end{eqnarray}
  where $h_{(p)}(x) \in O(p)_{\rm local}\,,\,
    h_{(N-p)}(x)\in O(N-p)_{\rm local}$.      
  \\
    
    Thus we arrive at an s-HLS model with  $G_{\rm global} \times H_{\rm local}
    =O(N)_{\rm global}\times [O(N-p)\times O(p)]_{\rm local}$,     as an $N-$ extension of the SM with HLS (see Eq.(\ref{SM-HLS})),
    which consists of  three independent invariants at the lowest derivative
 \begin{eqnarray}
  {\cal L}^{(N,p)}_{\rm SM-HLS} 
  &=&\chi^2(\varphi) \cdot 
 \left[ 
\frac{1}{2} 
\left(
\partial_\mu \varphi 
\right)^2  
+ {\cal L}_A + a^{(p)} {\cal L}_V^{(p)}+  a^{(N-p)} {\cal L}_V^{(N-p)} \right] - V(\varphi)
\,,
\label{Model}
\end{eqnarray}
where
\begin{eqnarray}
 {\cal L}_A&=&
 \frac{F_\pi^2}{4} {\rm tr} \left(
     {\hat \alpha}^2_{\mu,\perp}(x) \right)
     =\frac{F_\pi^2}{4} {\rm tr} \left({\alpha}^2_{\mu,\perp}(x) \right)^2        
  = 
     - \frac{F_\pi^2}{4}  {\rm tr} 
    \left( \begin{array}{cc}
    \partial_\mu \phi \cdot \Phi^t \partial^\mu \Phi\cdot\phi^t&0\\
    0&  \partial_\mu \Phi \cdot \phi^t \partial^\mu \phi\cdot\Phi^t    
     \end{array}  
     \right)\nonumber\\
     &=&- \frac{F_\pi^2}{2}   {\rm tr} \left(\phi^t \partial_\mu \phi \cdot \Phi^t \partial^\mu \Phi  
     \right)
    = \frac{F_\pi^2}{2}  {\rm tr}_{_{p\times p}}\left(\partial_\mu \phi   \partial^\mu \phi^t  
     +\left(\phi \partial_\mu  \phi^t \right)^2   \right) =\frac{F_\pi^2}{2} {\rm tr}_{_{N-p\times N-p}}\left(\partial_\mu \Phi   \partial_\mu \Phi^t  
     +\left(\Phi \partial_\mu  \Phi^t \right)^2   \right) \nonumber\\
     &=&\frac{F_\pi^2}{4} {\rm tr} \left({\alpha}^2_{\mu,\perp}(\pi) \right)^2=
      \frac{F_\pi^2}{4} {\rm tr}  \left(\frac{1}{F_\pi}\partial_\mu \pi+\cdots\right)^2     
    = \frac{1}{2} \left(\partial_\mu \pi_a\right)^2  +\cdots 
    \,, \label{gaugefixedLA}
   \end{eqnarray}
   (see Eq.(\ref{ONpNpconstraint}) for the second line), and
\begin{eqnarray}
 a^{(p)} {\cal L}_V^{(p)}&=&\frac{\left(F^{(p)}_\rho\right)^2}{4}{\rm tr}_{_{p\times p}} \left(
     \left[
     {\hat \alpha}^{(p)}_{\mu,||}(x)
     \right]^2 
     \right)
=\frac{\left(F^{(p)}_\rho\right)^2}{4}{\rm tr}_{_{p\times p}}\left[
     \left(\frac{1}{i} \partial_\mu \phi\cdot\phi^t -\rho^{(p)}_\mu
     \right)^2
     \right]
     \label{LVp}
    \nonumber \\
&=&   \frac{\left(F^{(p)}_\rho\right)^2}{4}\, 
   {\rm tr}_{_{p\times p}}   \left[
  \left(\rho^{(p)}_\mu-
  \frac{1}{F^{(p)}_\rho} \partial_\mu {\check \rho}^{(p)}
  \right) -
 \frac{i}{2 (F^{(p)}_\rho)^2}
 \left[
 \partial_\mu {\check \rho}^{(p)}, \check \rho^{(p)} 
 \right]- 
 \frac{i}{2F_\pi^2}
\left[
 \partial_\mu \pi,\pi\right]
 +\cdots
  \right]^2\,,  \label{LVp} 
   \end{eqnarray}
 \begin{eqnarray}
  a^{(N-p)} {\cal L}_V^{(N-p)}&=&
 \frac{\left(F^{(N-p)}_\rho\right)^2}{4}{\rm tr}_{_{N-p\times N-p}} \left(
     \left[
     {\hat \alpha}^{(N-p)}_{\mu,||}(x)
     \right]^2 
     \right)=  \frac{\left(F^{(N-p)}_\rho\right)^2}{4} {\rm tr}_{_{N-p\times N-p}}  \left[
     \left(\frac{1}{i} \partial_\mu \Phi\cdot\Phi^t -\rho^{(N-p)}_\mu
     \right)^2
     \right]\nonumber\\
     &=&
 \frac{\left(F^{(N-p)}_\rho\right)^2}{4} 
 \nonumber \\
&\times&  {\rm tr}_{_{N-p\times N-p}} 
   \left[
  \left(\rho^{(N-p)}_\mu-
  \frac{\partial_\mu {\check \rho}^{(N-p)}}{F^{(N-p)}_\rho} 
  \right) -
i \frac{\left[
 \partial_\mu {\check \rho}^{(N-p)}, \check \rho^{(N-p)} 
 \right] }{2 (F^{(N-p)}_\rho)^2}
 - \frac{i}{2F_\pi^2}
\left[
 \partial_\mu \pi,\pi\right]
  +\cdots
 \right]^2\,.   \label{LVNp}
  \end{eqnarray}  

Here
we have two arbitrary parameters $a^{(p)}$ and $a^{(N-p)}$ instead of a single one, $a$,  in the standard HLS model (Eq.(\ref{SM-HLS})), with 
 $[F_\rho^{(p)}]^2= a^{(p)} F_\pi^2= a^{(p)} v^2$, $[F_\rho^{(N-p)}]^2=a^{(N-p)} F_\pi^2= a^{(N-p)} v^2$, as the normalization of the kinetic term of  ${\check \rho}^{(p)}$ and ${\check \rho}^{(N-p)}$, 
   respectively. Note that the third  line of Eq.(\ref{gaugefixedLA}) is a gauge-fixed form (unitary gauge $\xi(x)=\xi(\pi)$, or $\check \rho^{(p)}=\check \rho^{(N-p)}=0$)  as a consequence of the parameterization of the second line of Eq.(\ref{ONpNp}) where $\xi(\check \rho^{(p)})$ and  $\xi(\check \rho^{(N-p)})$ are automatically dropped in the trace, as noted for the generic case in  Appendix \ref{sHLS}.     
 At the classical level without the kinetic term  of the HLS gauge bosons, $\rho^{(N-p)}_\mu$ and $\rho^{(p)}_\mu$, 
 we can solve away  these auxiliary fields through the respective equation of motion, which give  ${\cal L}_V^{(N-p)} = {\cal L}_V^{(p)} =0$, and we are left with ${\cal L}_A$ which is identical to the   genuine nonlinear sigma model based on $G/H=O(N)/[O(N-p)\times O(p)]$ as an $N-$extension of $G/H=O(4)/O(3)$ in Eq.(\ref{SMO4}):
 
  When we take $N=4$ with $p=3$, the model is reduced to that of $O(4)_{\rm global} \times O(3)_{\rm local}$ having a single parameter $a= a^{(p=3)}$ for
 ${\cal L}_V= {\cal L}_V^{(p=3)}$ and without 
  $ {\cal L}^{(N-p)}$ term,  which    is   identical to the standard  s-HLS model (see Eq.(\ref{SM-HLS})).
   Thus the model is gauge equivalent to the scale-invariant version of 
 the $O(4)/O(3)$ model Eq.(\ref{SMO4}) and hence to   the SM itself in the form of Eq.(\ref{Higgs2}) and eventually to the standard SM Higgs Lagrangian Eq.(\ref{Higgs1}):
  \begin{eqnarray}
\displaystyle \lim_{N \to 4}   {\cal L}^{(N,3)}_{\rm SM-HLS}= 
  \left[    \chi^2(\varphi) \cdot 
 \left( 
\frac{1}{2} 
\left(
\partial_\mu \varphi 
\right)^2  
+ {\cal L}_A + a {\cal L}_V \right)  
- V(\varphi)\right]_{_{N=4,p=3}}={\cal L}_{\rm SM-HLS}\simeq {\cal L}_{\rm SM}\,.
\label{SMHLSO4}
     \end{eqnarray}
 \\
 
  The model Eq.(\ref{Model})  is our starting Lagrangian.

 \section{Phase structure and phase transition}

  Now we expect nonperturbative dynamics of SM can be realized in the large $N$ limit of our Lagrangian Eq.(\ref{Model}). 
 We disregard the  part $a^{(N-p)} {\cal L}^{(N-p)}$, since it does not give rise to the dynamical generation of the
 kinetic term of $\rho^{(N-p)}$ in the large $N$ limit: Namely, 
  in order to generate its kinetic term, all the planar graphs having the index $\alpha$ running to $1-N$ are involved, which   not controllable even in that limit. The situation is  the same as  that  in the $CP^{N-1} = U(N)/[U(N-1) \times U(1)]$ model
where  only the $U(1)_{\rm local}$ part among the whole HLS $[U(N-1) \times U(1)]_{\rm local}$ can generate the dynamical gauge boson in the large $N$ limit (see Appendix \ref{CPNApp}).

 \subsection{Covariant derivative parameterization and multiplier}
 
 The most convenient parametrization to study the large $N$ dynamics is the $p\times N$ matrix $\phi$ in Eq.(\ref{ONpNp}).
 To calculate the effective action in the large $N$ limit  we use rescaling the quantities and new abbreviations in notation:
 \begin{eqnarray}
\phi &\rightarrow &  
F_\pi\,\phi\,, \quad \alpha^{(p)}_{\mu,||}\rightarrow   \alpha_{\mu,||}=\frac{1}{i  F_\pi^2} \partial_\mu \phi \phi^t =i \frac{G}{N} \phi \partial_\mu \phi^t \,, \quad \rho^{(p)}_\mu \rightarrow \rho_\mu\,,\nonumber\\
{\rm s.t.} \,\,\,\,  \phi\phi^t &=& F_\pi^2 \cdot \1 \equiv \frac{N}{G}\cdot \1 \,, 
 \end{eqnarray}
 in terms of which the Lagrangian consisting of the terms in 
  Eq.(\ref{gaugefixedLA}) and (\ref{LVp}) takes the form:
 \begin{eqnarray}
 {\cal L}&=&  {\cal L}_A + a {\cal L}_V\,,\nonumber\\
 {\cal L}_A &=&\frac{1}{2} {\rm tr}_{_{p\times p}}\left(\partial_\mu \phi  \partial^\mu \phi^t + \frac{1}{F_\pi^2} \left(\phi \partial_\mu \phi^t\right)^2
 \right)= \frac{1}{2} {\rm tr}_{_{p\times p}}\left(\partial_\mu \phi  \partial^\mu \phi^t + \frac{G}{N} \left(\phi \partial_\mu \phi^t\right)^2
 \right)\,,\nonumber\\
  a {\cal L}_V&=& \frac{F_\rho^2}{4} {\rm tr}_{_{p\times p}} 
  \left(\rho_\mu- \alpha_{\mu,||} 
  \right)^2=\frac{F_\rho^2}{4} {\rm tr}_{_{p\times p}} 
  \left(\rho_\mu- i \frac{G}{N} \phi \partial_\mu \phi^t 
  \right)^2=\frac{a}{2}\cdot \frac{N}{G} \frac{1}{2}{\rm tr}_{_{p\times p}} 
  \left(\rho_\mu- i \frac{G}{N} \phi \partial_\mu \phi^t 
  \right)^2\,.
  \label{GrassmannHLS}
   \end{eqnarray}
This suggests that $G=N/F_\pi^2$ may be regarded as the coupling and  the large $N$ limit is taken as $N \rightarrow \infty$ with $G=N/F_\pi^2$ fixed (See Appendix \ref{CPNApp} and \cite{Bando:1985rf} for a similar definition of the coupling in $CP^{N-1}$ model, and Ref.\cite{Harada:2003jx} for the chiral Lagrangian).  Then our $N-$ extension of the
SM takes the form
\begin{eqnarray}
{\cal L}^{(N,p)}_{\rm SM-HLS:a} =   \chi^2(\varphi) \cdot 
 \left[ 
\frac{1}{2} 
\left(
\partial_\mu \varphi 
\right)^2  +  {\cal L}_A + a {\cal L}_V\right] - V(\varphi)\,,
\label{N-SMa}
\end{eqnarray}
which appears to depend on the arbitrary parameter $a$.

However, as far as $\rho_\mu$ is the auxiliary field without kinetic term, we may use the equation of motion 
 $\rho_\mu=  i \frac{G}{N} \phi \partial_\mu \phi^t 
$ (or simply add $ a {\cal L}_V (\equiv 0)$ with arbitrary weight $a$) and 
always rewrite the Lagrangian  independently of the parameter $a$ as:~\footnote{
 It is amusing that  for a special value $a=2$,  Eq.(\ref{covariantform})  still remains true as a simple  identity,  even  without using the equation of motion for $\rho_\mu$, namely the term $\frac{G}{N} \left(\phi \partial_\mu \phi^t\right)^2$ gets automatically cancelled between those from 
 ${\cal L}_A$ and $a {\cal L}_V=2  {\cal L}_V$ to yields the form Eq.(\ref{covariantform}), even when the $\rho_\mu$ 
 acquires the kinetic term. This is also the case in the auxiliary field formulation of the NJL model, see Appendix \ref{NJL}. We shall return to this point later, Section \ref{a-dependence}. 
} 
:\begin{eqnarray}
 {\cal L}_A + a {\cal L}_V =  \frac{1}{2} {\rm tr}_{_{p\times p}}\left(D_\mu \phi \cdot (D^\mu \phi)^t\right)  \,,\quad D_\mu \phi=\left(\partial_\mu- i \rho_\mu\right) \phi\,,\quad \rho_\mu^t=- \rho_\mu\,.
 \label{covariantform}
\end{eqnarray}
{\it This simply reflects the trivial fact that the classical theory without kinetic term of the $\rho_\mu$ is independent of $a$},
although the quantum theory does in general depend on $a$ as we shall 
discuss later.

\subsection{Gap equation in the large $N$ limit}

We now consider the large $N$ limit nonperturbative dynamics of the $N$-extension of the SM as:
\begin{eqnarray}
 {\cal L}^{(N,p)}_{\rm SM-HLS} =    \chi^2(\varphi) \cdot 
 \left[ 
\frac{1}{2} 
\left(
\partial_\mu \varphi 
\right)^2  + 
\frac{1}{2} {\rm tr}_{_{p\times p}}
\left\{
D_\mu \phi \cdot (D^\mu \phi)^t 
\right\}
\right] 
- V(\varphi) \,.
 \label{Model3}
 \end{eqnarray}
 As we repeatedly mentioned, for $N=4\,, p=3$ this is equivalent to Eq.(\ref{SM-HLS}) and hence gauge equivalent to the SM Lagrangian Eq.(\ref{Higgs1}) in the form of Eq.(\ref{Higgs2}).

The Lagrangian can be further rewritten in terms of the rescaled fields:
\begin{eqnarray}
\hbox{\boldmath$\phi$}(x)=\chi(\varphi) \cdot \phi(x)\,, \quad 
\hbox{\boldmath$\sigma$}(x) =\frac{1}{G^{1/2}}\cdot\chi(\varphi)=\frac{1}{\sqrt{N}} \sigma(x)\,,
\end{eqnarray}
having a canonical scale dimension 
$d_{\hbox{\boldmath$\phi$}}=d_{\hbox{\boldmath$\sigma$}} =1$ for $D=4$ 
($d_{\hbox{\boldmath$\phi$}}=d_{\hbox{\boldmath$\sigma$}} =D/2-1$ for $2\leq D\leq 4$ dimensions):
\begin{eqnarray}
 {\cal L}^{(N,p)}_{\rm SM-HLS} &=&    
\frac{1}{2} {\rm tr}_{_{p\times p}}
\left\{
D_\mu \hbox{\boldmath$\phi$} \cdot (D^\mu \hbox{\boldmath$\phi$})^t 
-\eta(x) 
\left(
\hbox{\boldmath$\phi$}\, \hbox{\boldmath$\phi$}^t - N \hbox{\boldmath$\sigma$}^2 \1
\right)\right\}
- V(\hbox{\boldmath$\sigma$}) \,,\nonumber\\
V(\hbox{\boldmath$\sigma$}) &=& V(\varphi)= V(\sigma)= N \cdot \frac{\hat \lambda}{4} \left[\left(  {\hbox{\boldmath$\sigma$}}^2-
\frac{1}{G}\right)^2-\frac{1}{G^2}\right] \,,
\label{Model4}
 \end{eqnarray}
with $\hat \lambda= N \lambda=$ constant ('t Hooft coupling) in the $N\rightarrow \infty$ limit as usual.
We have used Lagrange multiplier $p\times p$ matrix $\eta(x)
$ (with the scale dimension $d_\eta
=2$) instead of the constraint $\phi\phi^t 
= F_\pi^2 \cdot \1 \equiv \frac{N}{G}\cdot \1$ (the dimensionful field $\phi$ has the scale dimension $d_\phi=0$), and also rescaled the dilaton decay constant $F_{\varphi}=v \rightarrow p^{1/2} v$ such that
$\chi(\varphi)=e^{\varphi/v} \rightarrow e^{\varphi/(p^{1/2} v)}$. 
 Note that Eq.(\ref{Model4}) is equivalent to Eq.(\ref{Model3}) which is scale-invariant except for $V(\varphi)$.

 Indeed  the constraint originates from the potential term $V(\varphi)$ itself, which breaks spontaneously the internal symmetry and the scale symmetry by the very existence of the explicit scale breaking term (SM Higgs mass $m^2$ in Eq.(\ref{Higgs1})). Thus once we imposed the constraint (or, Lagrange multiplier) , the role of the potential $V(\varphi)$ is only limited, irrelevant to the phase structure (determined by the derivative couplings even for $V(\varphi)=0$), thereby giving only the perturbative self interactions of $\phi$ itself, if any. Indeed,  in the form of Eq.(\ref{Model4}) 
 the conformal limit (BPS limit), Eq.(\ref{BPSlimit}), i.e.,  $\lambda \rightarrow 0$ ($V(\varphi)\rightarrow 0$) 
 with  $\langle \sigma(x)\rangle$ ($\ne 0$) fixed, can be realized automatically by simply taking $\lambda (\hat \lambda) \rightarrow 0 $ (See the gap equations, Eq.(\ref{sigmagap})). \footnote{
 As we noted in the footnote to Eq.(\ref{BPSlimit}), even in the conformal limit, to define the quantum corrections we need the regularization scale $\Lambda$, and hence the scale symmetry is explicitly broken so as to give rise to the trace anomaly as the induced potential for $\varphi$, such  as $\sim v^4 \chi^4 \ln \chi$ for $D=4$. This also yields a similar (but numerically different) self couplings of $\varphi$. Direct test of the SM Higgs self couplings in future experiments will test the precise form of the $V(\varphi)$.
 }    
 \\
 
 We now study the nonperturbative dynamics in the large $N$ limit of Eq.(\ref{Model4}). 
 The $F_\pi$ (and hence $G$) in the classical Lagrangian
 Eq.(\ref{Model4}) should be regarded as the bare quantity and receives quantum corrections in the large $N$ limit.
 The effective action at leading order of $1/N$ expansion reads:
 \begin{eqnarray}
\Gamma_{\rm eff}\left[ \hbox{\boldmath$\phi$} , \eta, \hbox{\boldmath$\sigma$},  \rho_\mu \right]
&=& \int d^D x \,\,\frac{1}{2} {\rm tr}_{_{p\times p}}
\left[
  D_\mu \hbox{\boldmath$\phi$} (D^\mu  \hbox{\boldmath$\phi$})^t 
  - \eta(x)
  \left( 
    \hbox{\boldmath$\phi$} \hbox{\boldmath$\phi$}^t - N \hbox{\boldmath$\sigma$}^2 \1
  \right) 
\right]
- V(\hbox{\boldmath$\sigma$}) \nonumber\\
&&+ \frac{i}{2} N \,\mbox{Tr} \mbox{Ln} \left(
  - D_\mu D^\mu - \eta
\right)
\,, \quad \left( 2\leq D \leq 4\right) \,,
\label{effaction}
\end{eqnarray} 
where  in $D$ dimensions $\hbox{\boldmath$\phi$}(x)$ and $\hbox{\boldmath$\sigma$}(x) $ and $\eta(x)$ have a canonical dimension $d_{\hbox{\boldmath$\phi$}/\hbox{\boldmath$\sigma$} }=D/2-1$, and $d_\eta=2$, respectively, while $\rho_\mu$
scales in the same way as the derivative in the covariant derivative,  $d_{\rho_\mu}=1$.

 The effective potential for $\langle \hbox{\boldmath$\phi$}_{i,\beta}(x) \rangle=\sqrt{N} v (\delta_{i,j},0)$ and $\langle \eta_{i,j}(x) \rangle=\eta\, \delta_{i,j}$, $\langle \hbox{\boldmath$\sigma$}(x)\rangle = 
 \hbox{\boldmath$\sigma$} $ takes the form: 
 \begin{eqnarray}
 \frac{1}{Np} V_{\rm eff}\left(v,\eta, \hbox{\boldmath$\sigma$}\right)=\eta \left(v^2 -{\hbox{\boldmath$\sigma$}}^2\right) 
 +\frac{1}{N p} V( \hbox{\boldmath$\sigma$}
 ) + \int\frac{d^D k}{i (2\pi)^4} \ln \left(k^2-\eta\right)\,.
 \end{eqnarray}
  This yields  the gap equation:
 \begin{eqnarray}
\frac{1}{Np}  \frac{\partial V_{\rm eff}}{\partial v}&=& 2 \eta v=0\,,
\label{sta1}
\\
\frac{1}{Np} \frac{\partial V_{\rm eff}}{\partial \hbox{\boldmath$\sigma$}} &=& - 2 
\eta
 \hbox{\boldmath$\sigma$}
+ \frac{\hat \lambda}{p} \hbox{\boldmath$\sigma$} \left({\hbox{\boldmath$\sigma$}}^2 -\frac{1}{G}\right)=0\,,\label{sta2}\\
\frac{1}{Np} \frac{\partial V_{\rm eff}}{\partial \eta} &=& v^2-
{\hbox{\boldmath$\sigma$}}^2 
+ \int \frac{d^D k}{i (2\pi)^D} \frac{1}{\eta-k^2}=0 \,.\label{sta3}
\end{eqnarray}

Eq.(\ref{sta3}) together with (\ref{sta1}) is the same form  as that of $CP^{N-1}$ in $D$ dimensions (see e.g., \cite{Bando:1985rf,Harada:2003jx}), and 
implies either of the two
cases:
\begin{equation}
\left\{ \begin{array}{ccc}
  \eta
  =0 \,, & v \neq 0 \, ; & \mbox{case (i)} \, \\
  v=0 \,, \, & \eta 
   \neq 0 \, ; &  \mbox{case (ii)} \, .
\end{array}\right.
\label{two phases}
\end{equation}
Eq.(\ref{sta2}) yields two cases:
\begin{eqnarray}
\begin{array}{cc}
{\hbox{\boldmath$\sigma$}}=0\,,&\\
{\hbox{\boldmath$\sigma$}}\ne 0\,,& 
- 2 
\eta
+ \frac{\hat \lambda}{p}\left({\hbox{\boldmath$\sigma$}}^2 -\frac{1}{G}\right)=0\,.
\end{array}\,
\label{sigmagap}
\end{eqnarray}
where the first solution ${\hbox{\boldmath$\sigma$}}=0$ in Eq.(\ref{sigmagap}) contradicts Eqs.(\ref{sta3}) and (\ref{sta1}), and hence we are left with the second one,
 which implies $\eta=0$ for $\hat \lambda \rightarrow 0$, the BPS limit in the broken phase, case (i), while for 
 $\hat \lambda\ne 0$ we have:
 \begin{eqnarray}
{\hbox{\boldmath$\sigma$}}^2=\frac{1}{G} +\frac{2 p\, \eta }{\hat \lambda}. 
\end{eqnarray} 

The stationary condition in Eq.(\ref{sta3}) gives a relation
between $\eta
$ and $v$.
By putting $\eta
=v=0$ in Eq.~(\ref{sta3}), the critical point
$G(\equiv G(\Lambda)) =G_{\rm crit}(\equiv G_{\rm crit}(\Lambda))$ separating the two phases in Eq.~(\ref{two phases})
is determined as
\begin{equation}
\frac{1}{G_{\rm crit}} = 
\int \frac{d^Dk}{i(2\pi)^D} \, \frac{1}{- k^2}
= \frac{1}{\left(\frac{D}{2} - 1 \right) \Gamma(\frac{D}{2}) }
  \frac{\Lambda^{D-2}}{(4\pi)^{\frac{D}{2}}}
\,,
\label{Gcr}
\end{equation}
by which  the integral in Eq.(\ref{sta3}) reads: 
\begin{eqnarray}
 \int \frac{d^D k}{i (2\pi)^4} \frac{1}{\eta-k^2}=\frac{1}{G_{\rm crit}} -
 \frac{\Gamma(2- D/2)}{\left(D/2 - 1 \right)}
 \cdot 
\frac{\eta^{D/2-1}}{(4\pi)^{D/2}} \,.
 \end{eqnarray}
\\

Hence the gap equation takes the form: 
\begin{eqnarray}
v^2 - \left(\frac{1}{
G}-\frac{1}{G_{\rm crit}}\right)=
\frac{\Gamma(2- D/2)}{\left(D/2 - 1 \right)}
 \cdot 
\frac{\eta^{D/2-1}}{(4\pi)^{D/2}} + \frac{2 p\, \eta }{\hat \lambda}
\equiv v_\eta^2\,.
\label{gapa2}
\end{eqnarray}
We may define
the renormalized coupling at renormalization point $\mu^2$ as:
\begin{eqnarray}
\frac{1}{ 
G} - \frac{1}{G_{\rm crit}}&=& \frac{1}{{ 
G}^{(R)}} - \frac{1}{G_{\rm crit}^{(R)}}\ ,
\label{renormalized}\\
\frac{1}{{
G}^{(R)}} 
&\equiv& 
\frac{1}{{
G}^{(R)}(\mu)}= 
\frac{1}{
G} 
-\int \frac{d^Dk}{i(2\pi)^D} \, \frac{1}{\mu^2 - k^2}
\,, \notag \\ 
 \frac{1}{G^{(R)}_{\rm crit}} 
 &\equiv& 
 \frac{1}{G^{(R)}_{\rm crit}(\mu)}
 = 
 \int \frac{d^Dk}{i(2\pi)^D} 
 \left(\frac{1}{- k^2}-\frac{1}{\mu^2 - k^2} \right)
\notag \\ 
&=& 
\frac{\Gamma(2- D/2)}{\left(D/2 - 1 \right)}
\cdot \frac{\mu^{D-2}}{(4\pi)^{D/2}}
\,.  
\end{eqnarray}

Now the gap equation Eq.(\ref{gapa2}) takes the form, depending on the phase (i) and (ii) as:
\begin{eqnarray}
&&\mbox{(i)} \quad G < G_{\rm cr} :
 v \neq 0 \ , \ 
 \eta
 =0\,  
\nonumber\\
&&
\frac{1}{G(\Lambda)} - \frac{1}{G_{\rm crit}(\Lambda)} 
=
 \frac{1}{G^{(R)}(\mu)} - \frac{1}{G_{\rm crit}^{(R)}(\mu)}
=v^2 \,
>0 \,\, ,
\label{broken}\\
&&
\mbox{(ii)} \quad G > G_{\rm cr} :
  v = 0 \ , \ 
  \eta
  \ne 0
\nonumber\\
&& 
\frac{1}{G(\Lambda)} - \frac{1}{G_{\rm crit}(\Lambda)} 
=
 \frac{1}{G^{(R)}(\mu)} - \frac{1}{G_{\rm crit}^{(R)}(\mu)} =
-v_\eta^2\, <0\,.
\label{symmetric}
\end{eqnarray}
\\

\subsection{Phase Structure}

The form of the gap equation, Eq.(\ref{broken}) and (\ref{symmetric}),  is the same as that in NJL model up to the opposite sign, i.e.,  opposite phase for strong coupling $G>G_{\rm crit}$ vs weak coupling $G<G_{\rm crit}$, since the classical (bare)  theory is formulated in the opposite phase (Wigner realization for the NJL model versus
NG (nonlinear) realization for the present case).  
See e.g., Eq.(\ref{NJLgap4D}) in Appendix \ref{NJL}.
\\

The case (i) is the {\it perturbative phase} where the classical theory structure remains. 
  Eq.(\ref{broken})  is the gap equation for the spontaneous breaking of the symmetry
 $G_{\rm global} \times H_{\rm local}$, 
with the
 Higgs mechanism of $H_{\rm local}$ yielding the ``mass''  of $\rho_\mu$: 
 \begin{eqnarray}
 (M_\rho)_0^2 = 2\cdot N  v^2 \ne 0 \,,
  \end{eqnarray}
 (with mass dimension $D-2$), 
 as read from Eq.(\ref{effaction}), with $\langle \hbox{\boldmath$\phi$}_{i,\beta}(x) \rangle=\sqrt{N} v (\delta_{i,j},0)$. \footnote{The factor 2 is an ``accidental value'' $a=2$ in the particular representation in Eq.(\ref{Model3}), which can actually be shifted to arbitrary value as far as the equation of motion of $\rho_\mu$ as the auxiliary field is used. See discussions for Eq.(\ref{covariantform}) and the related footnote. 
 }
 This mass already differs from the ``bare mass'' $2 N/G$ at classical level  by the power divergent corrections $1/G_{\rm crit}$ as seen in Eq.(\ref{broken}), but still
 receives additional quantum effects arising from the kinetic term after rescaled to the canonical form, see later discussions. The scale symmetry is also spontaneously broken by the same $\langle \hbox{\boldmath$\phi$}_{i,\beta}(x)\rangle =\langle \chi(\varphi)\cdot \phi_{i,j} \rangle =\sqrt{N} v \delta_{i,j}$, $v\ne 0$,  with $\varphi$ in the $\chi(\varphi)=
 e^{\varphi/F_{\varphi}}$ being the 
 (pseudo-)dilaton with  the decay constant $F_{\varphi}=\sqrt{N} v$ at the quantum level (not the value at classical level $\sqrt{p\, N/G}$).

The case (ii) is a genuine {\it nonperturbative phase} in strong coupling $G>G_{\rm crit}$. It implies that
{\it the quantum theory  is actually in the unbroken phase} of $G_{\rm global} \times H_{\rm local}$,
although the theory at classical level 
is written in terms of the NG boson variables as parts of  $\mbox{\boldmath$\phi$}$ living in the coset $G/H$ as if it were in the broken phase.
The HLS gauge symmetry $H_{\rm local}$ thus is never spontaneously broken and the gauge boson if exists as 
a particle should 
be massless.  
In fact,  the originally the NG bosons $\pi$ in the $\hbox{\boldmath$\phi$}$ (and would-be NG bosons $\check \rho$)
at classical level (see Eq.(\ref{components}) ) are {\it no longer
 the NG bosons (would-be NG bosons)  at quantum level}  by the nonperturbative dynamics (at large $N$) and acquire dynamically the  mass
  \begin{equation}
 M^2_\pi=M^2_{\check \rho}
 =\eta \ne 0\,, \quad \left(G>G_{\rm crit}\,, \quad v=0 \right)\,,
 \end{equation}
as readily seen from Eq.(\ref{Model4}).  
 Note that
$ \langle \eta(x)\rangle
=\eta \, \1 \ne 0$  {\it breaks no internal symmetry
 but the scale symmetry} due to VEV of the field $\eta(x) 
 $ carrying the scale dimension 2. Writing the $p$-flavor-singlet (trace part)
 $\eta(x)=\eta e^{\varphi_\eta(x)/\eta} \cdot \1$, we may regard $\varphi_\eta(x)$ as a pseudo-dilaton in this phase (its mass from the trace anomaly due to the regularization with $\Lambda$, or the renormalization with $\mu$). 
 \\

  Eq.(\ref{broken}) and Eq.(\ref{symmetric}) imply that the dynamical  phase transition from the case (i) ($\eta
 =0\,, v^2=1/G-1/G_{\rm crit}>0$) to the case (ii) ($v=0\,, -v_\eta^2= 1/G-1/G_{\rm crit}<0)$ is induced {\it continuously} at the transition point $v=v_\eta=0$ by the power divergent (quadratic divergent for $D=4$) loop contributions 
 $1/G_{\rm crit}$ to the classical $(F_\pi^2)_0=N/G$. In $D=4$ it reads from the broken side ($v\rightarrow 0$)
 \begin{eqnarray}
v^2=\frac{1}{G} -\frac{1}{G} _{\rm crit} = \frac{F_\pi^2}{N} =\frac{\left(F_\pi^2\right)_0}{N} -\frac{\Lambda^2}{(4\pi)^2}\,\, \quad \longrightarrow 0\,,
\quad  \left(G\rightarrow G_{\rm crit}-0\right)\,,
\label{quadraticdiv}
\end{eqnarray}
(see, e.g., Ref.\cite{Harada:2003jx}), while from the side of the unbroken phase ($\eta\rightarrow 0$)
\begin {eqnarray}
- v_\eta^2 = \frac{1}{G} -\frac{1}{G} _{\rm crit} =\frac{\left(F_\pi^2\right)_0}{N} -\frac{\Lambda^2}{(4\pi)^2}\, \quad\longrightarrow 0\,,
\quad  \left(G\rightarrow G_{\rm crit}+ 0\right)\,.
\end {eqnarray}
 The phase transition is the second order,
analogously to the well-known gap equation in the NJL model (See Appendix \ref{NJL}).
In $D=4$ this is the same as the pSM tuning the weak scale $F_\pi^2=(246\,{\rm GeV})^2$ against the quadratically divergent corrections to the $({\rm Higgs} \, {\rm mass})^2$, except  that the cutoff $\Lambda$ (to be shown as related to the Landau pole) in the present case will be shown to be close to the weak scale, in contrast to  the pSM whose Landau pole is much higher to be plagued with the naturalness problem.
\\

Note also that the gap equations Eq.(\ref{broken}) and Eq.(\ref{symmetric}) are finite relations 
 for $2 \le D< 4$  
 as they should, since the theory is renormalizable. 
Indeed, the gap equations  take the same form as  that of the $D$-dimensional NJL model
which is also renormalizable for $2\leq D<4$~\cite{Kikukawa:1989fw,Kondo:1992sq}.  
Then the full quantum theory has a beta function with a nontrivial ultraviolet fixed point for the dimensionless coupling $g=G\,\Lambda^{D-2}$: 
\begin{eqnarray}
\beta (g)=\Lambda \frac{\partial g}{\partial \Lambda}\Big|_{v\,,
\eta={\rm fixed}} =
- (D-2)\, \frac{g}{g_{\rm crit}} \, \left( g - g_{\rm crit}\right)\,,\quad g_{\rm crit}=G_{\rm crit}\, \Lambda^{D-2} =(4\pi)^{\frac{D}{2}} \left(\frac{D}{2} - 1 \right) \Gamma\left(\frac{D}{2}\right) \,, 
 \label{beta}
  \end{eqnarray}
  (the same form of $\beta(g^{(R)})$ for $g^{(R)}
  (\mu)=G^{(R)}(\mu) \cdot \mu^{D-2}$ with $g^{(R)}_{\rm crit}=G^{(R)}_{\rm crit}\, \mu^{D-2}$),  where
 $g_{\rm crit}$ 
  is the nontrivial (non-Gaussian)  ultraviolet fixed point  at which the interacting quantum theory is defined. This is in fact similar to that of the $D$-dimensional NJL model~\cite{Kikukawa:1989fw,Kondo:1992sq}: 
 \\
 
Special attention should be paid to {\it $D=2$ dimensions, where $g_{\rm crit}=0$ and hence  the {\it case (i) (the classical/perturbative phase, broken phase with $v\ne 0$) does not exist at all}, in accord with the Mermin-Wagner-Coleman theorem on 
absence of the spontaneous symmetry breaking in $D=2$ dimensions}. On the other hand, the gap equation Eq.(\ref{symmetric}) with $D=2$  
takes the form $\frac{1}{g}=- \frac{1}{4\pi} \ln \frac{\eta}{\Lambda^2}$, or:
\begin{equation}
\langle \eta(x)\rangle =\Lambda^2 \cdot \exp\left(-\frac{4\pi}{g(\Lambda)}\right) = \mu^2 \cdot \exp\left(-\frac{4\pi}{g^{(R)}(\mu)}\right)\,,
\end{equation}
where the scale symmetry appears to be spontaneously broken by $\langle \eta(x)\rangle\ne 0$ in the same sense as $D>2$ (up to explicit breaking due to the trace anomaly), but actually undergoes  the Berezinskii-Kosterlitz-Thouless (BKT) phase transition similarly to the $D=2$ NJL model (Gross-Neveu model).\footnote{
 The
$D=2$ NJL model (Gross-Neveu model) also has $G_{\rm crit}=0$, which would imply the broken phase of the chiral symmetry for all $G>G_{\rm crit}=0$, opposite to the present model. This would be in apparent contradiction to the Mermin-Wagner-Coleman theorem, but actually undergoes the BKT phase transition at $G=G_{\rm crit}=0$, which is a typical example of the ``conformal phase transition''~\cite{Miransky:1996pd}.
\label{MWCtheorem}
}  
\\
 
 For $D=4$, on the other hand, the logarithmic divergence remains in $g^{(R)}_{\rm crit}(\mu)=G^{(R)}_{\rm crit}(\mu)\cdot \mu^2
 = (4\pi)^{-2} \ln (\Lambda^2/\mu^2)$ due to  the factor $\Gamma(2-D/2)$ in Eq.(\ref{broken}) and Eq.(\ref{symmetric}) in the large $N$ theory, in such a way that in the vicinity of the phase transition point $v/\Lambda, v_\eta/\Lambda\simeq 0$, the renormalized coupling behaves as 
 \begin{eqnarray}
 g^{(R)}(\mu) = \left[\frac{1}{(4\pi)^2} \ln \left(\frac{\Lambda^2}{\mu^2} \right) \right]^{-1} \quad 
 &\rightarrow& 0 \quad \left(\frac{\Lambda^2}{\mu^2} \rightarrow \infty\right)\,, \quad \left({\rm triviality} \right)\,,
 \label{triviality}\\
&\rightarrow& \infty \quad \left(\mu^2 \rightarrow \Lambda^2\right)\,,\quad \left({\rm Landau}\,\,{\rm pole}\right)\, ,
  \end{eqnarray}
 even though the bare coupling $g(\Lambda)$ has a nontrivial UV fixed point $g_{\rm crit}=(4\pi)^2$ (Gaussian fixed point). Namely the ``renormalized'' coupling  $g^{(R)}(\mu)$ for $D=4$ is infrared (IR) zero, having a trivial IR fixed point:
\begin{eqnarray}
\beta(g^{(R)}(\mu)) = \mu \frac{\partial g^{(R)}(\mu)}{\partial \mu}= \frac{2}{(4\pi)^2} \left(g^{(R)}(\mu)\right)^2\quad >0 \,,
\end{eqnarray}
and hence  $g^{(R)}(\mu)\rightarrow 0$ as $\mu^2 \rightarrow 0$. This  is  the same as in  the NJL model in four dimensions. 
\\
 
 This appears different from 
 the original SM which is  perturbatively renormalizable. However, the perturbative SM in the original parameterization  is also plagued with
  the Landau pole  at a certain scale $\Lambda$ indicating that the theory is a trivial theory; the coupling $\lambda(\mu) \rightarrow 0$ ($\mu<\Lambda$) for the limit  $\Lambda \rightarrow \infty$. So the situation is essentially the same.
 We here define the full quantum theory as a cutoff theory for $D=4$ where the cutoff $\Lambda$  is regarded as the Landau pole, which will be explicitly related to the dynamically generated kinetic term of the HLS gauge boson, having no counter term and thus characterized by a  new extra parameter, the HLS gauge coupling $g_{_{\rm HLS}}$.\footnote{This corresponds to the extra free parameters in the ``renormalized formulation'' in the effective field theory approach  for the nonlinear sigma model such as $CP^{N-1}$ model in $D=4$~\cite{Weinberg:1997rv}. Such an extra free parameter 
 also appears in the $D=4$ NJL model, i.e., dynamically generated kinetic term and quartic coupling of the auxiliary scalar field (analogue of the kinetic term of the HLS gauge boson in the present case), see  Appendix \ref{NJL}, and also in the  chiral perturbation theory at loop orders (extra counter terms).
 }
   For $\Lambda\rightarrow \infty$ limit there still remains derivative coupling, which however  vanishes at the infrared 
 (low energy) limit, corresponding to the triviality of the conventional formulation of the SM.

 \section{Dynamical Generation of the HLS gauge bosons in the SM (SM rho) }

\subsection{Large $N$ result of the Grassmannian model 
}

We now discuss the dynamical generation of the HLS gauge boson $\rho^{ij}_\mu=\rho^a_\mu (S^a)^{ij}$.
We are interested in the large $N$ limit of the SM as a scale-invariant Grassmannian model
$O(N)_{\rm global} \times O(p)_{\rm local}$ with $N=4$ and $p=3$. Hereafter we confine ourselves to  $O(N)_{\rm global} \times
O(3)_{\rm local}$ model,
in which case  $O(3)$ generator takes the form $S^a_{ij}= i \epsilon_{iaj}$ such that:
\begin{eqnarray}
 \sum_{ij} \rho_\mu^{ij} \rho_\nu^{ji} =
 \sum_{ij}\rho_\mu^a \rho_\nu^b \cdot \left(i\epsilon_{iaj} i\epsilon_{jbi}\right)= \sum_{ab} \rho_\mu^a \rho_\nu^b\cdot 2 \delta^{ab} 
=2\cdot   \sum_{a} \rho_\mu^a \rho_\nu^a\,, \quad \left(p=3,\,\, S^a_{ij}= i \epsilon_{iaj}\right)\,.
\label{rhonormalization}
\end{eqnarray}

From the effective action Eq.(\ref{effaction}) the (amputated) two-point function of $\rho_\mu$ reads:
\begin{eqnarray}
\frac{1}{2}\int \frac{d^D q}{(2\pi)^D} \sum_{ij}\rho^\mu_{ij} (-q)\cdot  {\tilde \Gamma}^{(\rho)}_{\mu\nu}(q)\cdot \rho^\nu_{ji}(q) =\frac{1}{2}\int \frac{d^D q}{(2\pi)^D} \sum_{a}\rho^\mu_a (-q)\cdot  \Gamma^{(\rho)}_{\mu\nu}(q)\cdot \rho^\nu_a(q)\end{eqnarray}
where 
\begin{eqnarray}
\Gamma^{(\rho)}_{\mu\nu}(q)= 2\cdot  {\tilde \Gamma}^{(\rho)}_{\mu\nu}(q)
&=&
\frac{N}{2} \cdot 2 \cdot 
\left[
g_{\mu\nu} \cdot 2 v^2 +  
\int\frac{d^D k}{i (2\pi)^D} 
\left[
\frac{(q+2k)_\mu\cdot (q+2 k)_\nu}{(k^2-\eta) ((q+k)^2-\eta)}
-  2 \frac{g_{\mu\nu}}{k^2-\eta}
\right] 
\right]
\nonumber\\
&=&
2 N \left[
g_{\mu\nu}  \cdot v^2 + \left(g_{\mu\nu} -\frac{q_\mu q_\nu}{q^2}\right)\cdot q^2 \cdot  f(q^2, \eta
)
\right] \,,
\label{twopointfunc}
\end{eqnarray}
\begin{eqnarray}
f(q^2,\eta)= - \frac{1}{2}\frac{\Gamma(2-\frac{D}{2})}
{\left(4\pi\right)^{\frac{D}{2}}\Gamma(2)}
\int_0^1 dx \frac{\left(1- 2x\right)^2}{
\left[x\left(1-x\right) q^2+\eta\right]^{2-\frac{D}{2}
}}\,.
\label{f-function}
\end{eqnarray}
It reads (see e.g.,\cite{Leibbrandt:1975dj} for massless integral with $\eta=0$):\footnote{We thank  Taichiro Kugo (private communication)
for giving us explicit calculations  in the broken phase for $D=4$.
}
\begin{eqnarray}
f(q^2,0)&=&
- \frac{1}{D-1} 
\frac{
\Gamma(2-\frac{D}{2}) \left[\Gamma(D/2-1)\right]^2
}{2 
(4\pi)^{\frac{D}{2}}\, \Gamma(D-2)
} 
 (q^2)^{D/2-2} \, \nonumber\\
 &\Longrightarrow& - \frac{1}{2} \cdot \frac{1}{3 \left(4\pi\right)^2}
\cdot \left[
\ln \left(\frac{\Lambda^2}{q^2}\right) +\frac{8}{3}
\right]\quad \left(D\longrightarrow 4\right)\,,
\quad \left({\rm case}\,\, {\rm (i)} : v\ne 0\,\,, \eta=0\right)\,, \label{f-function4Dbroken}\\
f(0,\eta)&=&
-\frac{1}{3} 
\frac{
\Gamma(2-\frac{D}{2})
}{2
(4\pi)^{\frac{D}{2}}\, \Gamma(2)}
  \eta ^{D-4} \nonumber\\  &\Longrightarrow&
- \frac{1}{2} \cdot \frac{1}{3 \left(4\pi\right)^2}\cdot \left[
\ln \left(\frac{\Lambda^2}{\eta}\right)\right]\quad \left(D\longrightarrow 4\right) \,, \quad \left({\rm case}\,\, {\rm (ii)} : v= 0\,\,, \eta\ne 0\right)\,,
\label{f-function4Symmetric}
\end{eqnarray}
where for $D=4$ we have retained the explicit $\ln q^2$-dependence   in the broken phase (case (i)) which is relevant in  the later discussions on the VMD.\footnote{
The finite part $8/3$ may be absorbed into the redefinition of the cutoff $\Lambda^2\Rightarrow \tilde \Lambda^2=(e^{4/3} \cdot \Lambda)^2 $ (Landau pole) and ignored hereafter unless otherwise mentioned.\label{rescaled}
 }, while only took the pole residue as a divergent part in the unbroken phase (case (ii)) where the
$\ln q^2$-dependence will not be discussed in this paper.
Similar results have been given in Ref.\cite{Brezin:1980ms}.\footnote{
There is a caveat about the coefficient $N$ for the loop integral part, which is from the loop of the $p\times (N-p)$
components $\pi$'s out of the full $p\times N$ components of $\hbox{\boldmath$\phi$}_{i,\alpha}$ ($i=1,\cdots, p(=3)$  for $(\rho_\mu)_{i,j}$ is fixed, while $\alpha=1,2,\cdots,N$ is running in the loop) and thus the coefficient factor $N$ might be replaced by  $N-p$. However, difference between $N$ and $N-p$ is only the ambiguity of the $1/N$ sub-leading effects which are on the same order of other (uncontrollable) sub-leading  effects. There is no justification for keeping only a part of the sub-leading effects, ignoring others.
  In the standard sense of the $1/N$ expansion we here took only  
the leading order.  
}
\\

It  then yields the propagator of the $O(3)_{\rm local}$ HLS gauge boson $\rho_\mu$:
\begin{eqnarray}
\langle \rho_\mu \rho_\nu\rangle (q) = {\cal F}.\, {\cal T}.  \langle T\left(\rho_\mu(x) \rho_\nu(0)
\right)\rangle
 &=&- {\tilde \Gamma}^{(\rho)}_{\mu\nu}(q)^{-1}= - 2 \Gamma^{(\rho)}_{\mu\nu}(q)^{-1} = 2\cdot \langle \rho^a_\mu \rho^a_\nu\rangle (q)  \nonumber\\
 &=&
 \frac{1}{N}
  \frac{\left(- f^{-1}(q^2,\eta)\right)}
{q^2 -  v^2(- f^{-1}(q^2,\eta))}\left[
g_{\mu\nu} -
\frac{
q_\mu q_\nu
}
{
 v^2(- f^{-1}(q^2,\eta))
}
\right]\,.
\label{propagator}
\end{eqnarray}

In the $\rho^a_\mu$ basis (see Eq.(\ref{rhonormalization})) this takes the form in each phase in $D=4$:
\begin{eqnarray}
\left({\rm case}{\rm (i)}; \quad v\ne 0\,, \quad M^2_\pi
= \eta=0\right) \nonumber\\
 \langle \rho_\mu^a \rho_\nu^b\rangle (q)&=& - \delta^{ab} \Gamma^{(\rho)}_{\mu\nu}(q)^{-1} = -\delta^{ab}\frac{1}{2} {\tilde \Gamma}^{(\rho)}_{\mu\nu}(q)^{-1}\nonumber\\
  &\approx&\delta^{ab} \frac{1}{N}
  \frac{\left( -2 f(M_\rho^2,0)\right)^{-1}}
{q^2 - M_\rho^2} \left[
g_{\mu\nu} - 
\frac{
q_\mu q_\nu
}
{M_\rho^2}
\right]
 \,\,  \quad {\rm near} \,\,q^2 = M_\rho^2 \,,
\label{propagatorbroken}\\  
M_\rho^2 &=&- f^{-1}(M_\rho^2,0) \cdot v^2 =2 \lambda_{_{\rm HLS}} v^2= 2\cdot  g_{_{\rm HLS}}^2 \cdot Nv^2 =2\cdot  g_{_{\rm HLS}}^2 F_\pi^2
\,,
 \label{rhomassa2}\\
\frac{1}{\lambda_{_{\rm HLS}}} &=&\frac{1}{ Ng_{_{\rm HLS}}^2} = -   2 f(M_\rho^2,0)  =\frac{ 
 1}{3 (4\pi)^2} \ln \left(\frac{\Lambda^2}{M_\rho^2}\right)   \,,
 \label{gaugecouplingbroken}
 \end{eqnarray}
 
 \begin{eqnarray}
 \left({\rm case}  {\rm (ii)}\,; \quad v=0\,, \quad M^2_\pi
 =\eta\ne0\right) \nonumber\\
 \langle \rho_\mu^a \rho_\nu^b\rangle (q) &=&\delta^{ab}\frac{1}{N} 
 \cdot
  \frac{\left( -2 f(0,\eta)\right)^{-1}}
 {q^2} \cdot g_{\mu\nu}  + {\rm gauge}\,\,{\rm terms}  \,,
  \label{propagatorsymmetric}
  \\
  M_\rho^2&=&0\,,\\
 \frac{1}{\lambda_{_{\rm HLS}}} &=&\frac{1}{ N g_{_{\rm HLS}}^2}=
 - 2f(0,M_\pi^2 
)
=\frac{ 
 1}{3 (4\pi)^2} \ln \left(\frac{\Lambda^2}{M_\pi^2}\right) 
\,.
 \label{gaugecouplingsymmetric}
 \end{eqnarray}
where ``gauge terms'' depend on the gauge fixing as usual, and $\lambda_{_{\rm HLS}}=N g_{_{\rm HLS}}^2$ is the 't Hooft coupling to be fixed in the large $N$ limit.
In either phase we may introduce ``running'' coupling $g_{_{\rm HLS}}^2(\mu^2)$ with $\mu$ representing a typical mass scale
$\mu=M_\rho$ for the case (i) and $\mu=M_\pi$ for (ii).
 \\

Note the transversality of the loop contribution (second term of Eq.(\ref{twopointfunc})), which as a consequence of the gauge symmetry of HLS does imply the  masslessness of the HLS gauge boson when $v=0$, i.e., the  unbroken phase (case (ii)). Were it not for the gauge symmetry, namely, the HLS,  
we could not have taken the
inverse of $\Gamma^{(\rho)}_{\mu\nu}(q)$, so that the theory in the unbroken phase would be inconsistent, the same situation as the $CP^{N-1}$ model in the parameterization without gauge symmetry (see Appendix \ref{CPNApp}). (This would be  a serious problem particularly for $D=2$ where there exists only the unbroken phase in both the present case and the $CP^{N-1}$ model.)
Thanks to the HLS as a gauge symmetry in the present case, we have a freedom to fix the gauge  by adding the gauge-fixing term as usual
 and  can take the inverse.

In the broken phase $v\ne 0$ (case (i)), on the other hand,
the first term of Eq.(\ref{twopointfunc}) is from the $\rho_\mu$ mass term, also receiving the loop contributions via the gap equation Eq.(\ref{broken}), and plays a role of the gauge-fixing term
(unitary gauge), since the gap equation solution $\langle \hbox{\boldmath$\phi$}_{i,\beta}(x) \rangle=\sqrt{N} v (\delta_{i,j},0)$
with $v\ne 0$ already fixed the gauge. These results are precisely the same as those
of the $CP^{N-1}$ model~\cite{Eichenherr:1978qa,Golo:1978de,DAdda:1978vbw,DAdda:1978dle,Witten:1978bc,Arefeva:1980ms,Haber:1980uy,Bando:1987br,Weinberg:1997rv,Harada:2003jx}.
\\

In the case (ii), unbroken phase ($v=0\,, \eta\ne 0$), 
the $\pi
$  fields (and $\check \rho$) in $\hbox{\boldmath$\phi$}$ are no longer the NG bosons (would-be NG bosons) and are massive with mass 
\begin{eqnarray}
M^2_\pi =M^2_{\check \rho}
=\eta \ne 0\,,
\label{pimass}
\end{eqnarray}
and hence the only singularity of $f(q^2,M_\pi^2)$ arises from the two-particle threshold $q^2=4 M^2_\pi$ and beyond. Thus $f(q^2,M_\pi^2)$ has no singularity at $q^2=0$ in the $q^2$ plane, $ - f(0,M_\pi^2) \ne 0$,   as seen from the explicit calculation in Eq.(\ref{propagatorsymmetric}), and
we see that the two-point Green function  develops a genuine massless pole. 

We then find the {\it massless} HLS gauge boson $\rho_\mu$ acquires the kinetic term:
\begin{eqnarray}
{\cal L}_\rho &=&-  \frac{1}{4 g_{_{\rm HLS}}^2}  \frac{1}{2}{\rm tr} \left(\rho_{\mu\nu}^2\right)=-  \frac{1}{4 g_{\rm HLS}^2} 
\left({\rho^a}_{\mu\nu}\right)^2\,,  \label{kineticsymmetric}
 \end{eqnarray}  
 with $g_{_{\rm HLS}}^2$ given in Eq.(\ref{gaugecouplingsymmetric}). 
 {\it Hence the kinetic term of the massless HLS gauge boson 
 indeed has been dynamically generated by the
  nonperturbative dynamics at $1/N$ leading order!! }

 As already noted~\cite{Kugo:1985jc,Bando:1987br} for $CP^{N-1}$ model (see Appendix \ref{CPNApp}), the result is  in perfect conformity with the Weinberg-Witten theorem~\cite{Weinberg:1980kq} which forbids the
 dynamical generation of the massless  particle with spin $J\geq 1$. The theorem was proved  in the Hilbert space with positive definite metric and hence without gauge symmetry. This is in  
 sharp contrast to our HLS Lagrangian Eq.(\ref{Model4}) which does have
 a gauge symmetry thus is quantized with indefinite metric Hilbert space, and hence generates a massless composite gauge boson without conflict to the Weinberg-Witten theorem.
\\

 As to the case (i), broken phase ($v\ne 0,\, \eta=0$), 
  the $\rho_\mu$ have a mass and thereby decay into massless NG bosons $\pi$  in  $\hbox{\boldmath$\phi$}$
and has no pole in the physical (time-like) momentum plane. Still the kinetic term can be generated as in Eq.(\ref{propagatorbroken}):
\begin{eqnarray}
{\cal L}_\rho &=&-  \frac{1}{4 g_{_{\rm HLS}}^2}\frac{1}{2}{\rm tr} \left(\rho_{\mu\nu}^2\right)=-  \frac{1}{ 4 g_{_{\rm HLS}}^2}\left({\rho^a}_{\mu\nu}\right)^2\,,  \label{kineticbroken}
\end{eqnarray}
with $g_{_{\rm HLS}}^2$ given in Eq.(\ref{gaugecouplingbroken}).

The  kinetic term may be rescaled into the canonical form in
\begin{eqnarray}
-  \frac{1}{4 g_{_{\rm HLS}}^2}\left({\rho^a_{\mu\nu}}\right)^2
\rightarrow -\frac{1}{4} \left({\rho^a_{\mu\nu}}\right)^2 \,,
 \end{eqnarray}
in such a way that the $\rho_\mu$ mass reads the typical type of the Higgs mechanism in Eq.(\ref{rhomassa2}), 
which is independent of $N$. 
We may define $F_\rho^2\equiv M_\rho^2/g_{_{\rm HLS}}^2$ at quantum level,  which then coincides with the $a=2$ relation at classical level
\begin{eqnarray}
F_\rho^2=2 \cdot F_\pi^2 .
\label{FrhoFpi}
\end{eqnarray}

 Then the kinetic term is also dynamically generated in the $N\rightarrow \infty$ limit, in exactly the same way as the $CP^{N-1}$ model in $D=4$ including the factor 2 in Eqs.(\ref{rhomassa2}) and (\ref{FrhoFpi})~\cite{Bando:1987br,Weinberg:1997rv}.
 \\
 
 In $D=4$ the kinetic term operator has scale dimension 4 (hence scale-invariant)  but the coefficient of the kinetic term $1/g_{_{\rm HLS}}^2(\mu^2)$ necessarily depends on the cutoff, thus also violating the 
scale symmetry in a sense different from $2\leq D <4$.\footnote{
   In $D\ne 4$ dimensions the scale symmetry existing at classical level (in the conformal/BPS limit $V(\varphi)\rightarrow 0$) has been broken  by the dynamical generation of the 
  kinetic term of the HLS gauge boson, 
  having the scale dimension 4 (not $D$), which is traced back to the spontaneous scale-symmetry breaking due to $\eta=\langle \eta(x) \rangle\ne 0$ (unbroken phase) and $v \ne 0$ (broken phase).
 However,  the HLS gauge coupling vanishes just on the ultraviolet fixed point $g=g_{\rm crit}$ (see Eq.(\ref{beta}))
 approaching from both sides of the phases:
 \begin{eqnarray}
 g_{_{\rm HLS}}^2(\mu^2) &\rightarrow& 0 \,, \quad M_\rho\equiv 0\,, \quad \mu^2=M^2_\pi=(M_{\check \rho})^2 = \eta\rightarrow 0\,,\quad \left(G\rightarrow G_{\rm crit}+0 \right)\,,\nonumber\\
g_{_{\rm HLS}}^2(\mu^2) &\rightarrow& 0\,,\quad M^2_\pi
=\eta\equiv 0\,, \quad \mu^2= M_\rho^2 \rightarrow 0\,, \quad \left( G\rightarrow G_{\rm crit} -0 \right)\,.\nonumber
\end{eqnarray} 
  and hence the dynamically generated massless HLS gauge bosons  get decoupled to be  free particles, in conformity with the exact scale invariance at the fixed point.  
   }
 Nevertheless the HLS gauge coupling at $D=4$ also vanishes on the fixed point where the scale symmetry is realized in a trivial sense:
\begin{eqnarray}
g^2_{_{\rm HLS}}(\mu^2)\rightarrow 0\quad  {\rm as}\quad \frac{\mu^2}{\Lambda^2}  \rightarrow 0 \quad  \left(g\rightarrow g_{\rm crit}\right)\,.
 \label{HLStrivial}
\end{eqnarray}
This implies that {\it in $D=4$  dynamically generated  HLS gauge coupling $\alpha_{_{\rm HLS}}(\mu^2)=g^2_{_{\rm HLS}}(\mu^2)/(4\pi)$ has an trivial IR fixed point  $\alpha_{_{\rm HLS}}(\mu^2)\rightarrow 0$ ($\mu^2 \rightarrow 0$), i.e.,  asymptotically non-free. 
\begin{eqnarray}
\beta(\alpha_{_{\rm HLS}}(\mu^2))= \mu^2 \frac{\partial \alpha_{_{\rm HLS}}(\mu^2)}{\partial \mu^2}= \frac{1}{3 \pi} \alpha^2(\mu^2) >0\,.
\end{eqnarray}
Thus the phase transition point $g=g_{\rm crit}$ is identified with 
the trivial IR fixed point for both $g^{(R)}(\mu^2)$ and $g^2_{_{\rm HLS}}(\mu^2)$ which vanish just on the
IR fixed point.} 

In fact, as we noted before (Eq.(\ref{triviality})), although the original  (bare)  coupling $g=G \Lambda^2$ of the model has a nontrivial UV fixed point, the ``renormalized'' one $
g^{(R)}(\mu^2)=G^{(R)}(\mu^2) \mu^2$ has an IR zero (trivial IR fixed point) $g^{(R)}(\mu^2)\rightarrow 0$ ($\mu^2 \rightarrow 0$) corresponding to the Landau pole $g^{(R)}(\mu^2)\rightarrow \infty$ ($\mu^2\rightarrow \Lambda^2$). 
Hence 
the HLS gauge bosons simply get decoupled at the critical point, both $\rho_\mu$ and $\pi$ becoming free massless particles, in accordance with the scale symmetry at the ultraviolet fixed point $g=g_{\rm crit}$ in Eq.(\ref{beta}).

 On the same token, the $D=4$ case  in Eq.(\ref{gaugecouplingbroken}) and (\ref{gaugecouplingsymmetric}) indicates the Landau pole $g_{_{\rm HLS}}^2(\mu^2)\rightarrow 0$ as $\Lambda^2 \rightarrow \infty$, in the same way as the 
   original coupling $g^{(R)}=G^{(R)} \mu^2$ in Eq.(\ref{triviality}).   Another view of this result is 
  \begin{eqnarray}
  \frac{1}{g_{_{\rm HLS}}^2(\mu^2)} \rightarrow 0 \quad \left(\mu^2 \rightarrow \Lambda^2\right)\,,
  \label{Landaupole}
      \end{eqnarray}
that is,  the kinetic term vanishes at the Landau pole in such a way that the HLS gauge boson $\rho_\mu$ returned to the
auxiliary filed as a static composite of $\pi$, 
the situation sometimes referred to as ``compositeness condition'' \cite{Bardeen:1989ds} advocated in a reformulation of the top quark condensate model~\cite{Miransky:1988xi} for the composite Higgs.

In this viewpoint the HLS gauge bosons as bound states of $\pi$'s develop the kinetic term as we integrate the higher frequency modes
in the large $N$ limit from $\Lambda^2$ down to the scale $\mu^2$ in the sense of the Wilsonian renormalization group\cite{Harada:2003jx}
\\

Thus either from the unbroken phase  $\eta \rightarrow 0$ in Eq.(\ref{kineticsymmetric}) or broken phase $v\rightarrow 0$ in Eq.(\ref{kineticbroken}), {\it the HLS gauge coupling and hence  $M_\rho^2 $ in Eq,(\ref{rhomassa2})  do vanish continuously across the phase transition point. The phase transition is the second order}, similarly to the $CP^{N-1}$ model~\cite{Weinberg:1997rv}.  
Then the HLS gauge boson gets degenerate with the massless $\pi
$ but actually decoupled, both $\rho_\mu$ and $\pi$ being massless free particles exactly on the phase transition point $g=g_{\rm crit}$ which is the UV fixed point of the original coupling $g=G\Lambda^2=N\Lambda^2/F_\pi^2 $ where the scale symmetry is realized. 
This looks similar to the ''vector manifestation'' \cite{Harada:2000kb,Harada:2003jx}  of the Wigner realization of symmetry, with $g_{\rm HLS}^2 \rightarrow 0$ and $M_\rho^2 \rightarrow 0$, $a\rightarrow 1$,  while in the present case we shall show later that {\it these quantities of $\rho_\mu$ are independent of $a$ and hence the phase transition is also independent of
$a$, accordingly.
} 
\\

\subsection{SM rho  $\rho_\mu$ in the extrapolation $N\rightarrow 4$ with $p=3$}

Finally, the  SM Higgs Lagrangian is equivalent to this model in  an extrapolation $N\rightarrow 4$ with $p=3$, Eq.(\ref{SMHLSO4}),
and hence Eqs.(\ref{kineticbroken}) and (\ref{kineticsymmetric})
 clearly indicate the dynamical generation of the SM rho, the HLS gauge boson $\rho_\mu$ in the SM, with $
 N\rightarrow 
 4$.
 
We then conclude that as a simple extrapolation $N\rightarrow 4$ of the large $N$ result, Eq.(\ref{propagator}), or Eqs.(\ref{propagatorbroken}) - (\ref{gaugecouplingsymmetric}), the kinetic term of the HLS gauge boson $\rho_\mu$ in the SM is generated with a mass 
as
\begin{eqnarray}
\frac{1}{\lambda_{_{\rm HLS}}(\mu^2)}
&=& \frac{1}{N g_{_{\rm HLS}}^2(\mu^2)}\Bigg|_{N \rightarrow 4}=\frac{1}{3} \frac{1}{(4 \pi)^2} \ln \left(\frac{\tilde \Lambda^2}{\mu^2}\right)\,,
 \label{SMkinetic}\\
M_\rho^2 (\mu^2)
&=& - v^2  f^{-1} (\mu^2,0) = 
2 \cdot \lambda_{_{\rm HLS}} (\mu^2) v^{2}
=g_{_{\rm HLS}}^2(\mu^2) \cdot F_\rho^2\,, 
\label{SMrhomass}\\
F_\rho^2 &=&2\cdot F_\pi^2
=2\cdot \left(246\, {\rm GeV}\right)^2\simeq \left(350 \,{\rm GeV}\right)^2\,, 
\label{SMFrho}
\end{eqnarray}
where  $\mu^2=M_\rho^2$ in the broken phase  
and $\mu^2= M^2_\pi=\eta$  in the unbroken phase is understood,  respectively. The Landau pole  in the broken (unbroken) phase  
is related to the cutoff $\Lambda$ 
as $\tilde \Lambda=e^{4/3}\cdot \Lambda\quad  (=\Lambda)$ (footnote{\ref{rescaled}).
Note that the resultant expression in Eqs.(\ref{SMrhomass}) and (\ref{SMFrho}) relevant to the phenomenology remains the same as that in $N\rightarrow \infty$, having no explicit dependence on $N$. 
 
 We shall later discuss  possible phenomenological consequences of the dynamical generation of the SM rho, which will indicate
 that the cutoff or the Landau pole $\Lambda ={\cal O} ({\rm TeV}) - {\cal O}(10^2\, {\rm TeV})$ for the coupling strength $g_{_{\rm HLS}}=
 {\cal O} (10^0 - 10^3)$, depending on the collider physics or dark matter physics, respectively, either case being much close to the weak scale as a solution to  the naturalness problem of the SM through the nonperturbative physics
 within the SM.  
\\
 
 If the large $N$ results persist at least qualitatively  for $N=4$, then the (zero temperature) phase transition is of the send order, both $\rho_\mu$ and $\pi$ becoming degenerate as free massless particles
 just on the transition point $g=g_{\rm crit}$ which is the scale-invariant ultraviolet fixed point. The  unbroken phase has massless HLS gauge bosons $\rho_\mu$ and massive $\pi$ and $\check \rho$, while broken phase does massive $\rho_\mu$ (absorbing $\check \rho$) and massless  $\pi$ (NG bosons).
 This phase transition is quite different from the  second order phase transition in the conventional view based on the linear sigma model parameterization of the SM, where the
unbroken phase is realized by the degenerate massive scalars, $\hat \pi$ and $\hat \sigma$ in Eq.(\ref{Higgs1}), which are interacting 
even at the transition point.

\section{$a-$ dependence}
\label{a-dependence}

So far we have discussed nonperturbative dynamics in the large $N$ limit in the particular form of the Lagrangian Eq.(\ref{Model3}), which corresponds to
$a=2$ of the $a {\cal L}_V$ term in the classical Lagrangian
Eq.(\ref{N-SMa}). At classical level the Lagrangian Eq.(\ref{Model3}) is independent of the parameter $a$, since it is equivalent to Eq.(\ref{N-SMa}) for any $a$, as far as we use the classical equation of motion for $\rho_\mu$ (by which $a {\cal L}_V\equiv 0$).
However, as noted below Eq.(\ref{N-SMa}), the value $a=2$ is special in that  this equivalence holds without using the classical equation of motion.
It will be seen more explicitly in Eq.(\ref{a-depaction}) where the only different term between them  is  $\left(1-\frac{a}{2}\right) 
 \frac{G}{N} 
  \left(\phi \partial_\mu \phi^t \right)^2$ which identically vanishes for $a=2$. 
 \\

Here we discuss  $a-$dependence at the quantum level. 
 For $a\ne 2$, not only the one-loop diagram, an infinite set of the bubble sum due to the additional term should be included  in the large $N$ limit. 
We shall show that {\it on-shell quantities, such as the $\rho_\mu$ pole position and the residue are independent of the parameter $a$} in a peculiar way.  On the other hand,  {\it classical equation of motion for $\rho_\mu$ is violated  at quantum level in the large $N$ limit    in an $a-$dependent way}: The (non-propagating)  contact term proportional to $1- \frac{2}{a}$  appears in the two-point Green function of the $\rho_\mu$ besides the propagating part of the unitary gauge form.  Thus the off-shell physics can in principle depend on $a$.

In the $a\rightarrow \infty$ limit, however,  
such an $a-$dependence vanishes so that  the classical equation of motion is recovered, thus the SM rho $\rho_\mu$ is totally replaced by the composite operator $
\alpha_{\mu,||}\equiv  i \frac{G}{N} \phi\partial_\mu \phi^t$ even at quantum level. This implies  the dynamically generated $\rho_\mu$ kinetic term is fully replaced by the
Skyrme term. 
 \\
 
 We here return to the original form of the HLS model, Eq.(\ref{N-SMa}) with Eq.(\ref{GrassmannHLS}),  
 where for the present purpose to see the $a-$dependence, we may disregard parts related to the dilaton (SM Higgs) $\varphi(x)$ which is irrelevant to the dynamical generation of $\rho_\mu$ in the 
 large $N$ dynamics and $a$ dependence, 
 in which case the constraint is
$ \hbox{\boldmath$\phi$}\, \hbox{\boldmath$\phi$}^t = N \frac{1}{G} \1 $
 instead of 
$ \hbox{\boldmath$\phi$}\, \hbox{\boldmath$\phi$}^t = N \hbox{\boldmath$\sigma$}^2 \1$ (namely, disregarding $2 p \eta/{\hat \lambda}=6 \eta/{\hat \lambda}$ in the gap equation Eq.(\ref{gapa2})), and thus the scale symmetry is no longer relevant. 

Note also that
{\it the result exactly applies to  the $\rho$ meson in the 2-flavored QCD described by the same $G/H\simeq G_{\rm global} \times H_{\rm local}$ (without the scale symmetry nor dilaton) on the same footing as the SM Higgs Lagrangian}.

We then discuss  the action $S[\phi]= \int d^D  x
  {\cal L}$ with the two terms of the Lagrangian Eq.(\ref{GrassmannHLS}) (we focus on $p=3$ case as mentioned before):\footnote{
 Introducing multiplier into the CCWZ parameterization may be heretical, since   ${\cal L}_A + a {\cal L}_V $ is written already 
 based on the constraint $ \hbox{\boldmath$\phi$}\, \hbox{\boldmath$\phi$}^t = N \frac{1}{G} \1 $. As far as the discussions 
 on the broken phase are concerned, we do not need it at all. Here we included it for the unbroken phase discussions as well.
 The results are the same in either way,  anyway, as we demonstrate equivalence between Eq.(\ref{twopointa2}) and Eq.(\ref{twopointfunc}).  
 } 
  \begin{eqnarray}
  {\cal L}&=& 
  {\cal L}_A + a {\cal L}_V -\frac{1}{2} {\rm tr}_{_{p\times p}} \left[ \eta\left(\hbox{\boldmath$\phi$}\, \hbox{\boldmath$\phi$}^t - N \frac{1}{G} \1 \right)  \right]    \nonumber\\
&=&
\frac{1}{2} {\rm tr}_{_{p\times p}} \left[ \left(\partial_\mu \phi  \partial^\mu \phi^t + \frac{G}{N} \left(\phi \partial_\mu \phi^t\right)^2
 \right)+
  \frac{a}{2}\cdot \frac{N}{G}   \left(\rho_\mu- i \frac{G}{N} \phi \partial_\mu \phi^t 
  \right)^2-\eta\left(\hbox{\boldmath$\phi$}\, \hbox{\boldmath$\phi$}^t - N \frac{1}{G} \1 \right)
   \right]
  \,\nonumber\\
  &=& 
   \frac{1}{2}  {\rm tr}_{_{p\times p}} 
  \left[  
  \left(
  \partial_\mu \phi  \partial^\mu \phi^t 
  \right)  
  + \frac{1}{2} \left(\frac{a}{2}\cdot \frac{2N}{G}\right) \rho_\mu^2
  -\frac{a}{2}\cdot2 i   \rho_\mu \phi \partial^\mu \phi^t  + 
  \left(1-\frac{a}{2}\right) 
 \frac{G}{N} 
  \left(\phi \partial_\mu \phi^t \right)^2   -\eta\left(\hbox{\boldmath$\phi$}\, \hbox{\boldmath$\phi$}^t - N \frac{1}{G} \1 \right)   \right],
  \label{a-depaction}
   \end{eqnarray}
which as mentioned before is reduced to the action for Eq.(\ref{covariantform}) for arbitrary $a$, as
 far as we use  the classical equation of motion: 
 \begin{eqnarray}
 \rho_\mu &=&\left(\rho_\mu\right)_{ij}=\rho^a_\mu S^a_{ij} = \alpha_{\mu,||}
 =i \frac{G}{N} \phi \partial_\mu \phi^t=
 i \frac{G}{N} \phi_{i\alpha}  \partial_\mu \left(\phi^t\right)_{\alpha j}
 \,, \quad \left((i,j)=1,\cdots,p=3\,; \alpha=1,\cdots, N\right)\,, \nonumber\\
&& \frac{1}{2} {\rm tr}_{_{p\times p}} \left(S^a S^b\right) =\delta^{ab}
\,.
\label{classicaleom}
 \end{eqnarray}

\subsection{Problem with one-loop for $a\ne 2$}
 
 Le us first consider the one-loop for  Eq.(\ref{a-depaction}) with arbitrary $a$ (with extra assumption $N\gg p=3$, which coincides with the large $N$
 limit only for $a= 2$).   We would have 
  \begin{eqnarray}
\Gamma^{(\rho)}_{\mu\nu}(p)&=& 2 {\tilde \Gamma}^{(\rho)}_{\mu\nu}(q)
=   
\left[
\left(\frac{a}{2}\right)\left(\frac{2N}{G}\right)g_{\mu\nu} +  \left(\frac{a}{2}\right)^2  2B_{\mu\nu}(q)
 \right]\,, 
 \label{twopointatree}
  \end{eqnarray}
where the bubble function $B_{\mu\nu}(p)$ is given as:
\begin{eqnarray}
 B_{\mu\nu}(q) &=&
  \frac{N}{2} \int \frac{dk^D}{i (2\pi)^D} 
 \frac{
 (2 k+q)_\mu (2k+q)_\nu
 }{
 \left(k^2-\eta\right)\left((k+q)^2-\eta\right)}\nonumber\\
 &&=  N\left[
 \left(
 g_{\mu\nu} - \frac{q_\mu q_\nu}{q^2} \right) q^2 f(q^2,\eta)
 - g_{\mu\nu} \int\frac{d^D k}{i (2\pi)^D} 
 \frac{1}{- k^2+\eta}   
  \right]
 \nonumber\\
   &&
   = N\left[\left(
 g_{\mu\nu} - \frac{q_\mu q_\nu}{q^2} \right) q^2 f(q^2,\eta)  - g_{\mu\nu} 
 \left(
 \int\frac{d^D k}{i (2\pi)^D}
 \frac{1}{- k^2}   
+ \int\frac{d^D k}{i (2\pi)^D} 
\left( \frac{1}{- k^2+\eta}   - \frac{1}{- k^2} 
\right)
 \right)
  \right]   
  \nonumber\\
  && =
  N \left[
 \left(
 g_{\mu\nu} - \frac{q_\mu q_\nu}{q^2} 
 \right) q^2 f(q^2,\eta) 
 -  \frac{1}{{\check G}_{\rm crit}} 
\, g_{\mu\nu}
 \right]
 \,, 
 \label{bubble}
\end{eqnarray}
 with
 \begin{eqnarray}
 \frac{1}{{\check G}_{\rm crit}}\equiv\frac{1}{G_{\rm crit}} -  
 \frac{\Gamma(2-\frac{D}{2})}{D/2-1} \cdot \frac{ \eta^{D/2-1}}{(4\pi)^{D/2}}=\frac{1}{G_{\rm crit}} -v^2_\eta\,.
 \label{checkG}
 \end{eqnarray}
 where $v^2_\eta$ is defined in the gap equation Eq.(\ref{gapa2}) (up to the term, $2 p \eta/{\hat \lambda}=6 \eta/{\hat \lambda}$, as already noted),
 which now reads:
 \begin{eqnarray}
v^2 =\frac{1}{G} -\frac{1}{{\check G}_{\rm crit}}\,.
\label{gap-2}
\end{eqnarray}
 \\
 
Let us first check the case 
$a=2$, where the additional four-$\phi$ vertex in Eq.(\ref{a-depaction}) is absent and the one loop dominance is literally
valid in the large $N$ limit. In fact Eq.(\ref{twopointatree}) with Eq.(\ref{bubble})  yields
\begin{eqnarray}
\Gamma^{(\rho)}_{\mu\nu}(q)
&=& \left[
\frac{2N}{G} g_{\mu\nu} + 2 B_{\mu\nu}(q)
\right]
\nonumber\\
&=&N \left[
2 \left(\frac{1}{G} -\frac{1}{{\check G}_{\rm crit}}\right) 
g_{\mu\nu}
+ 2 f(q^2,\eta)\, q^2 \left(
 g_{\mu\nu} - \frac{q_\mu q_\nu}{q^2} \right) \right]
 \,,\nonumber\\
 {\rm case} \,{\rm (i)} &&= 
 N \left[
 2 v^2\,  g_{\mu\nu}  +  2  f(q^2,0)\,  q^2  \left(
g_{\mu\nu} - \frac{q_\mu q_\nu}{p^2} \right)  \right]
 \,, \quad \left(v\ne 0\,;\, \eta=M_\pi^2=0 \right)
 \,,\nonumber\\
 {\rm case} \,{\rm (ii)} &&= 
 N\cdot 2 f(q^2,\eta)\, q^2\, \left(
 g_{\mu\nu} - \frac{q_\mu q_\nu}{q^2}  \right)\,, \quad \left(v=0\,;\, \eta=M_\pi^2\ne 0
 \right)\,,
 \label{twopointa2}
  \end{eqnarray}
where use has been made of the gap equation Eq.(\ref{gap-2}).

The result Eq.(\ref{twopointa2})  of course coincides with Eq.(\ref{twopointfunc}) based on Eq.(\ref{covariantform}) which is equivalent to Eq.(\ref{GrassmannHLS})  for $a=2$ 
without use of the equation of motion. Note that in Eq.(\ref{twopointfunc}) making full use of the Lagrange multiplier (no tree-level $\rho_\mu$ mass term), the contact term $g_{\mu\nu} $ in the loop integral of
$\hbox{\boldmath$\phi$}$ arising from the $\rho_\mu \rho^\mu \hbox{\boldmath$\phi$} \hbox{\boldmath$\phi$}^t$ coupling 
makes the loop contribution to be  transverse. On the other hand,  such a loop graph contact term does not exist in the 
present CCWZ
parameterization, Eq.(\ref{GrassmannHLS}), while the $\rho_\mu$ tree mass term exists instead of $\rho_\mu \rho^\mu \hbox{\boldmath$\phi$} \hbox{\boldmath$\phi$}^t$ coupling, which  
 is combined with the tree contact term, resulting in the same answer.
  
  Hence the one-loop $\rho_\mu$ propagator at $a=2$ coincides with  Eq.(\ref{propagator}) as its should, where $F_\rho^2=M_\rho^2/g_{_{\rm HLS}}^2=N 2v^2=2 F_\pi^2$. The result indicates that the quantum correction for $F_\rho^2$ and $F_\pi^2$  keeps the relation $F_\rho^2/F_\pi^2=(F_\rho^2/F_\pi^2)_0=2=a$. 
\\

 On the other hand, for 
 $a\ne 2$,  the one-loop result is  depending on $a$, obviously in disagreement with the large $N$ limit result, Eq.(\ref{propagator}):
  \begin{eqnarray}
\Gamma^{(\rho)}_{\mu\nu}(q)&=& 2 {\tilde \Gamma}^{(\rho)}_{\mu\nu}(q) =
\left[
\left(\frac{a}{2}\right)\left(\frac{2N}{G}\right)g_{\mu\nu} + \left(\frac{a}{2}\right)^2  2 B_{\mu\nu}(q)
 \right] \nonumber\\
 &=& N   \left[  
  \frac{a}{2} \cdot 2\left(
\frac{1}{G} - \frac{a}{2} \frac{1}{{\check G}_{\rm crit}} \right)
\cdot g_{\mu\nu} 
 +   
\left(\frac{a}{2}\right)^2   2 f(q^2,\eta)\, q^2\cdot  \left(
 g_{\mu\nu} - \frac{q_\mu q_\nu}{q^2} 
 \right)  \right]\,,
\label{twopointanot2tree}
\end{eqnarray}
which yields 
the $\rho_\mu$ propagator $\langle \rho^a_\mu \rho^b_\nu \rangle(q)= \delta^{ab} (- \Gamma^{(\rho)}_{\mu\nu}(q))^{-1}=\frac{1}{2} \langle \rho_\mu \rho_\nu \rangle (q)$: 
\begin{eqnarray}
\langle \rho^a_\mu \rho^b_\nu \rangle(q)
&=& \frac{1}{2N} \delta^{ab} \left[
\frac{-f^{-1}(q^2,\eta) \left(\frac{2}{a}\right)^2}{q^2- 
\left(\frac{2}{a}\right)^2\cdot \left(- f^{-1}(q^2,\eta) 
(v^\prime)^2\right) 
} 
\left(g_{\mu\nu} -\frac{q_\mu q_\nu}{\left(\frac{2}{a}\right)^2\cdot \left(- f^{-1}(q^2,\eta) 
(v^\prime)^2\right) }\right)
\right]\,. 
\end{eqnarray}
This reads in the broken phase:
\begin{eqnarray}
\langle \rho^a_\mu \rho^b_\nu \rangle(q)
&\approx& 
\frac{1}{2N} \delta^{ab} \left[
\frac{-f^{-1}(M_\rho^2,0) \left(\frac{2}{a}\right)^2}{q^2- 
M_\rho^2
} 
\left(g_{\mu\nu} -\frac{q_\mu q_\nu}{M_\rho^2}\right)
\right]\quad \left({\rm near}\,\, q^2=M_\rho^2\,, \eta=0\right)\,,
\label{twopointan2}
\end{eqnarray}
where  
\begin{eqnarray}
M_\rho^2&=& \left(
\frac{2}{a}
\right)^2
\cdot \frac{1}{-2 f(M_\rho^2,0)} \cdot
2 (v^\prime)^2\,, \quad 
\left(
(v^\prime)^2=\frac{a}{2} \left(
 \frac{1}{G} - \frac{a}{2} \frac{1}{{\check G}_{\rm crit}} 
 \right) 
  \right)\,, \nonumber\\
 g^{-2}_{_{\rm HLS}}&=& N \left(\frac{a}{2}\right)^2\cdot \left(- 2f(M_\rho^2,0)\right)\,.  \label{Frhoa}
 \end{eqnarray}
 
 In the unbroken  phase it takes the form:
 \begin{eqnarray}
 \langle \rho^a_\mu \rho^b_\nu \rangle(q)
&=&\frac{1}{2N} \delta^{ab} 
\frac{-f^{-1}(0,\eta) \left(\frac{2}{a}\right)^2}{q^2} 
\cdot g_{\mu\nu} + {\rm gauge}\,\, {\rm terms}\,,\nonumber\\
g^{-2}_{_{\rm HLS}}&=& N \left(\frac{a}{2}\right)^2\cdot \left(- 2f(0,\eta)\right)\,.
 \end{eqnarray}
Then 
both the pole position $M_\rho^2= g^2_{_{\rm HLS}}\cdot  F_\rho^2
$ and the residue  $g^2_{_{\rm HLS}}$ are  dependent on
$a$. Also the VMD  is violated, since the direct coupling in Eq.(\ref{a-depaction})
 $\left(1-\frac{a}{2}\right) 
 \frac{G}{N} 
  \left(\phi \partial_\mu \phi^t \right)^2$ does exist. 
Note that 
 $(F_\rho/F_\pi)^2=2 (v^\prime/v)^2\ne a $, which differs from the classical relation $(F_\rho/F_\pi)^2 =a$ except for $a=2$. 
  \\
 
 However, all these results for $a\ne 2$ are  {\it artifacts of the one-loop calculations} (can be identified with the large $N$ limit only for $a= 2$), in fact we shall next show  {\it $a-$independence of the on-shell quantities and $F_\rho^2=M_\rho^2/g_{_{\rm HLS}}^2=2 F_\pi^2$ ($a=2$ relation!!) for arbitrary $a$ in the genuine large $N$ limit}.

\subsection{Large $N$ limit calculation} 
\label{fullN}

We now show that {\it the on-shell  quantities are independent of the parameter $a$ in the large $N$ limit, irrespectively of the phases.}

Indeed, in the large $N$ limit the dominant diagrams are not just the one-loop but do include an infinite sum of the bubble diagrams $B_{\mu\nu}$ due to the additional four-$\phi$ vertex
proportional $(1-a/2)$:\footnote{This is based on the observation  by Taichiro Kugo (private communication), who showed the same result as Eq.(\ref{rhopropagatoruniversal}) by more simplified calculation in the broken phase ignoring quantum corrections $\frac{1}{{\check G}_{\rm crit}}$ in $B_{\mu\nu}$,
Eq.(\ref{bubble}), which implies identifying $v^2=1/G$ through the gap equation Eq.(\ref{gap-2}). We thank him for 
his very illuminating discussions.
} 
\begin{eqnarray}
\Gamma^{(\rho)}_{\mu\nu}(q)&=& 2 \cdot {\tilde \Gamma}^{(\rho)}_{\mu\nu}(q) = 
\left(\frac{a}{2}\right)\left(\frac{2N}{G}\right)g_{\mu\nu} + 2 \left(\frac{a}{2}\right)^2\left\{ B_{\mu\nu}(q) +   B_{\mu\lambda}(q)\cdot \left(\frac{a}{2}-1\right)\frac{G}{N} \cdot
 B^{\lambda}_\nu(q) +\cdots \right\}
 \nonumber\\
&=&
2 
\left[
\left(\frac{a}{2}\right)\left(\frac{N}{G}\right) g_{\mu\nu} +  \left(\frac{a}{2}\right)^2 B_{\mu\lambda} (q) \cdot C^\lambda_\nu (q)
\right]\,,\nonumber\\
&&C_{\mu\nu} (q) = g_{\mu\nu}+ \left(\frac{a}{2} -1\right) \frac{G}{N} B_{\mu\lambda}(q)\cdot   C^\lambda_\nu (q)\,,
\label{twopointa}
\end{eqnarray}
where $B_{\mu\nu}(q)$ is given in Eq.(\ref{bubble}).
Now $C_{\mu\nu}(q)$ is given as 
\begin{eqnarray}
C^{-1}_{\mu\nu}(q) &=& 
g_{\mu\nu} + \left(1- \frac{a}{2}\right) \frac{G}{N} B_{\mu\nu} (q)
= \left[
1- \left(1-
\frac{a}{2}
\right) \frac{G}{{\check G}_{\rm crit}}
\right] \frac{q_\mu q_\nu}{q^2}\nonumber\\
&&+ \left[
\left(1- \left(1-
\frac{a}{2}
\right) \frac{G}{{\check G}_{\rm crit}}
\right) +
\left(1-
\frac{a}{2}\right) G f(q^2,\eta)\cdot  q^2\right] \left(
 g_{\mu\nu} - \frac{q_\mu q_\nu}{q^2} 
 \right) 
 \,\,,
\end{eqnarray} 
or
\begin{eqnarray}
C_{\mu\nu} (q) &=&
\frac{1}{
1- \left(1-
\frac{a}{2}\right) \frac{G}{{\check G}_{\rm crit}}
}
\frac{q_\mu q_\nu}{q^2}
+ 
\left[
\frac{1}
{1- 
\left(1-
\frac{a}{2}
\right) 
\left(
\frac{G}{{\check G}_{\rm crit}} 
- G f(q^2,\eta)\cdot  q^2
\right)
}
\left(
 g_{\mu\nu} - \frac{q_\mu q_\nu}{q^2} 
 \right) 
\right]
 \,.
\end{eqnarray} 
Then we get:
\begin{eqnarray}
 \Gamma^{(\rho)}_{\mu\nu}(q)&=&2 N 
 \left[
 \left(\frac{a}{2}\right)\frac{1}{G}
 -
  \frac{\left(\frac{a}{2}\right)^2 \frac{1}{{\check G}_{\rm crit}}
  }
  {
1- \left(1-
\frac{a}{2}\right) \frac{G}{{\check G}_{\rm crit}}
}  
\right]
g_{\mu\nu} \nonumber\\
 &&+
2 N \frac{
\left(\frac{a}{2}\right)^2
 f(q^2,\eta)\cdot  q^2
 }{
 \left(1-
\left(1-
\frac{a}{2} 
\right) 
\left(\frac{G}{{\check G}_{\rm crit}} 
- G f(q^2,\eta)\cdot  q^2
\right)
\right)
\cdot 
\left(
1- \left(1-
\frac{a}{2}\right)
 \frac{G}{{\check G}_{\rm crit}}
\right)
}
\left(
 g_{\mu\nu} - \frac{q_\mu q_\nu}{q^2}
  \right)\nonumber \\
  &=&
\frac{N 
}{
1- \left(1-
\frac{a}{2}\right)
 \frac{G}{{\check G}_{\rm crit}}
}\left[
a\left(
\frac{1}{G}-\frac{1}{{\check G}_{\rm crit}}
\right) g_{\mu\nu}+
\frac{2
\left( \frac{a}{2}\right)^2
f(q^2,\eta)\cdot  q^2
 }{
 \left(
 1- 
\left(1-
\frac{a}{2} 
\right) 
\frac{G}{{\check G}_{\rm crit}} \right)
+\left(1-
\frac{a}{2} 
\right) G f(q^2,\eta)\cdot  q^2
}
\left(
 g_{\mu\nu} - \frac{q_\mu q_\nu}{q^2}
  \right)
\right]
\nonumber\\
&=&  N 
A  \left[ g_{\mu\nu}+\frac{\alpha \cdot q^2}{\beta+ \gamma \cdot q^2} \left(
 g_{\mu\nu} - \frac{q_\mu q_\nu}{q^2}
  \right)  \right]\,,
 \label{Gammarho-a} 
 \end{eqnarray}
 where we defined (using the gap equation Eq.(\ref{gap-2})):
 \begin{eqnarray}
A&=&\beta^{-1} a\left(\frac{1}{G}-\frac{1}{{\check G}_{\rm crit}}\right)= \beta^{-1} a v^2\,,\nonumber\\
\alpha &=&2 A^{-1}\beta\left(\frac{a}{2}\right)^2  f(q^2,\eta)= \frac{a}{2v^2}  f(q^2,\eta)\,,\quad
\beta= 1 -\left(1-\frac{a}{2}\right) \frac{G}{{\check G}_{\rm crit}}\,, \quad \gamma =
\left(1-\frac{a}{2}
\right) G  f(q^2,\eta) \,.
 \end{eqnarray}
Noting that
\begin{eqnarray}
 \alpha+\gamma = \frac{
 \frac{a}{2}+ \left(1-\frac{a}{2}\right) G \left(\frac{1}{G}-\frac{1}{{\check G}_{\rm crit}}\right)}
 {v^2}  f(q^2,\eta)=
  \frac{\beta}{v^2}   f(q^2,\eta)\,,
  \end{eqnarray}
we have the $\rho_\mu$ propagator by inverting $\Gamma^{(\rho)}_{\mu\nu}(q)$ (of course only for $a\ne 0$, since 
 $\rho_\mu$ does not exist and  $\Gamma^{(\rho)}_{\mu\nu}(q)\equiv 0$ for $a= 0$):
\begin{eqnarray}
\langle \rho^a_\mu \rho^b_\nu
\rangle (q)
 &=&- \delta^{ab} \Gamma^{(\rho)}_{\mu\nu}(q)^{-1}=- \frac{1}{2}  \delta^{ab} \tilde {\Gamma}^{(\rho)}_{\mu\nu}(q)^{-1}\nonumber\\
 &=&\delta^{ab}\frac{1}{N} A^{-1}
 \left[
 \left(
 -1+\frac{\alpha}{\alpha+\gamma}
 \right)
 g_{\mu\nu} +
 \frac{-\alpha \beta \left(\alpha+\gamma\right)^{-2}}
 {q^2+ \beta \left(\alpha+\gamma\right)^{-1}}\cdot\left(g_{\mu\nu} -\frac{q_\mu q_\nu}{-\beta \left(\alpha+\gamma\right)^{-1}} \right)
  \right]
  \nonumber\\
  &=&
  \delta^{ab}
  \left[
  \left(-\frac{2}{a} +1\right)\,  \frac{G}{2N} \, g_{\mu\nu}
  +
\frac{1}{N}  \frac{-\frac{1}{2}f^{-1}(q^2,\eta)
  }{q^2-\left(- v^2 f^{-1}(q^2,\eta)\right)}
  \left(
  g_{\mu\nu} - \frac{q_\mu q_\nu}{- v^2 f^{-1}(q^2,\eta)}
  \right)
  \right]\,,\label{rhopropagatoruniversal}\\
  &=&
  \delta^{ab}
  \left[
  \left(-\frac{2}{a} +1\right)\,  \frac{G}{2N} \, g_{\mu\nu}
  +
 \frac{g^2_{_{\rm HLS}}(q^2,\eta)
  }{q^2-M_\rho^2(q^2,\eta)}
  \left(
  g_{\mu\nu} - \frac{q_\mu q_\nu}{- v^2 f^{-1}(q^2,\eta)}
  \right)
  \right]\,,\\ 
  &\approx& 
\delta^{ab}
  \left[
 \left(-\frac{2}{a} +1\right) \frac{G}{2N} g_{\mu\nu}+
\frac{g^2_{_{\rm HLS}}}{q^2-M_\rho^2} 
\left(
g_{\mu\nu} -
\frac{q_\mu q_\nu}{M_\rho^2}
\right)
\right]\,,\quad \left({\rm near}\,\, q^2 =M_\rho^2\right)\,,
\end{eqnarray}
where
\begin{eqnarray}
M_\rho(q^2,\eta)&=&- v^2 \cdot f^{-1}(q^2,\eta) =g^{2}_{_{\rm HLS}}(q^2,\eta)\cdot  \left(N\cdot 2v^2\right)=g^{2}_{_{\rm HLS}}(q^2,\eta)\cdot  F_\rho^2: \quad a\left(\ne 0\right)-{\rm independent}\,,\label{aindipendence}\\
M_\rho^2&=&M_\rho(M_\rho^2,0)=- v^2\cdot  f^{-1}(M_\rho^2,0)=g^{2}_{_{\rm HLS}}\cdot  F_\rho^2 \,, \quad M_\rho^2(q^2, M_\pi^2\ne 0)\equiv 0\,, \nonumber\\
g^{2}_{_{\rm HLS}}&=&g^{2}_{_{\rm HLS}}(M_\rho^2,0)=- \frac{1}{2N} f^{-1} (M_\rho^2,0) \,,\quad \left(M_\rho^2, v^2 \ne 0, M_\pi^2=\eta=0\right)\,, \nonumber\\
&=&g^{2}_{_{\rm HLS}}(0,M_\pi^2)=- \frac{1}{2N} f^{-1} (0,M_\pi^2)\,,\quad \left(M_\rho^2 =v^2=0, M_\pi^2=\eta\ne 0\right)\,, 
\end{eqnarray}
which for $a=2$ indeed agrees with Eq.(\ref{propagator}), with the concrete expression of $f(q^2,\eta)$  given in Eqs.(\ref{f-function4Dbroken}) and (\ref{f-function4Symmetric}).  

In the unbroken phase $v=0$, we have  $N A g_{\mu\nu}=0$  and $ \Gamma^{(\rho)}_{\mu\nu}(q)$ becomes purely transverse and as it stands
cannot be inverted, but {\it thanks to the gauge symmetry, HLS, which exists for $a\ne 0$}, we have a freedom to fix the gauge
to take the inversion as usual, the same situation as Eq.(\ref{propagatorsymmetric}) when inverting Eq.(\ref{twopointfunc}).

Note that in the large $N$ limit we have {\it a universal ratio $F_\rho^2/F_\pi^2=2$ corresponding to ``$a=2$''
independently of the classical parameter $a$}:
\begin{eqnarray}
F_\rho^2\equiv \frac{M_\rho^2}{g_{_{\rm HLS}}^2}=2 \cdot Nv^2=2\cdot  F_\pi^2\simeq  2\cdot \left(246\,\, {\rm GeV}\right)^2
\simeq \left( 350\,\, {\rm GeV}\right)^2\,,
\label{Frhoaindependence}
\end{eqnarray}
which is compared with the simple one-loop result Eq.(\ref{Frhoa}) where $F_\rho^2/F_\pi^2$
is a complicated dependence on $a$ unless $a=2$.
\\

Thus we establish  that the on-shell quantities, such as the pole position 
$M_\rho^2= -  f^{-1}(M_\rho^2,0)\cdot  v^2 $ and the pole residue $-f^{-1}(M_\rho^2,0)= 2 N g^2_{\rm HLS}$ are independent  of the parameter
$a$, and become identical to the values at $a=2$ (the tree-level $a-$dependence of $(F_\rho)_0^2=a (F_\pi)_0^2$ disappears as
 $F_\rho^2\rightarrow 2 F_\pi^2$ at quantum level !!), so is the kinetic term of the dynamical gauge boson of HLS:
We may say that {\it the choice $a=2$ good for the reality in QCD is not a mysterious parameter choice but the dynamical consequence of the quantum theory in the large $N$ limit!!}
\\ 
  
For all these $a-$independence of the on-shell quantities, however, 
the resultant $\rho_\mu$ propagator has an ``unusual''  contact term $-\frac{2}{a}\frac{G}{N}$ which is 
cancelled for $a=2$ and only for $a=2$, in which case we are left with the standard massive vector meson propagator of the unitary gauge in the broken phase and the massless propagator in the unbroken phase in perfect agreement with Eq.(\ref{propagator}) as it should be.
 This is  compared with the one-loop result Eq.(\ref{twopointan2}) having no such a contact term. 
Although this term by itself vanishes in the continuum limit $G \sim G_{\rm crit} ={\cal O}\left(\frac{(4\pi)^{D/2}}{\Lambda^{D-2}}\right) \rightarrow 0$ as $\Lambda \rightarrow \infty$   through the gap equation Eq.(\ref{gap-2}),  discussing  its origin 
might be useful to understand  the theory at quantum level.
 \\

To see the origin of the $a-$dependent contact term, we now look into the relation:\footnote{This is due to Taichiro Kugo,
 private communications.} 
 \begin{eqnarray}
\langle \rho_\mu \rho_\nu\rangle (q) =
\langle  \alpha_{\mu,||}\, \alpha_{\nu,||} \rangle (q)
 -  \frac{2G}{a N} g_{\mu\nu}\,, \quad 
  \alpha_{\mu,||}= 
  i\frac{G}{N}\phi\partial_\mu\phi^t \,,
  \label{adependence}
\end{eqnarray}
which follows from the combined use of the Ward-Takahashi identities (for $a\ne 0$):
\begin{eqnarray}
0&=& \int {\cal D} \phi \frac{\delta}{\delta \rho_\nu(y)} \left(\rho_\mu(x) e^{i S[\phi]}\right)=\int {\cal D} \phi \left[\delta^{(4)}(x-y) \cdot g_{\mu\nu} +\rho_\mu(x)\cdot \left(\frac{aN}{2G} \right) \left(\rho_\nu(y)-i\frac{G}{N} \phi\partial_\nu \phi^t(y)\right)
\right]\cdot e^{i S[\phi]}
 \,,\nonumber\\
0&=&\int {\cal D} \phi \frac{\delta}{\delta \rho_\mu(x)} \left(\phi\partial_\nu \phi^t(y) e^{i S[\phi]}\right)
=\int {\cal D} \phi \left(\frac{aN}{2G} \right) \left(\rho_\mu(x)-i\frac{G}{N} \phi\partial_\mu \phi^t(x)\right) \phi\partial_\nu \phi^t(y)\cdot e^{i S[\phi]}
\,,
\end{eqnarray}
where $S[\phi]$ is given in Eq.(\ref{a-depaction}).
The $a-$dependent contact term corresponds to the ``tree $\rho_\mu$ propagator'' (not propagating)  with tree mass $a/(2 G)$: 
 $\langle \rho_\mu \rho_\nu\rangle^{(\rm tree)}=\frac{1}{N}\cdot (q^2- (\frac{a}{2G}))^{-1}|_{q^2=0}\cdot  g_{\mu\nu}=- \frac{2G}{aN}\cdot g_{\mu\nu}$.
Thus the $\rho_\mu$ propagator at quantum level should depend on $a$ through  the 
``unusual'' contact term. 
\\

\subsection{$\alpha_{\mu,||}$ as an $a-$independent genuine vector bound state}

Here $ \langle \alpha_{\mu,||} \, \alpha_{\nu,||} \rangle (q) $ should be $a-$independent, since it is independent of the presence of the auxiliary field $\rho_\mu$ 
in a manner similar to Eq.(\ref{twopointa}) taking an infinite sum of the bubble diagram. Now we show that it is indeed the case
(in the genuine $a=0$ case, where no HLS exists, we have a problem in the unbroken phase as noted before and to be repeated in the below). 
\\

Let us sum up  the bubble diagrams in the large $N$ limit:
\begin{eqnarray}
 \langle \alpha_{\mu,||}\, \alpha_{\nu,||}\rangle(q)&=&
 \left(i\frac{G}{N}\right)^2 \langle \phi\partial_\mu\phi^t 
  \,\, \phi\partial_\nu\phi^t
   \rangle (q)
=  \left(i\frac{G}{N}\right)^2 \left[B_{\mu\nu}(q) + B_{\mu\lambda}(q)\cdot \left(-\frac{G}{N}\right) B^\lambda_\nu(q) +\cdots\right]\nonumber\\
&=&\left(i\frac{G}{N}\right)^2 B_{\mu\lambda}(q)\cdot C^\lambda_\nu(q)\,, \nonumber\\
C_{\mu\nu}(q)&=&g_{\mu\nu} -\frac{G}{N} B_\mu^\lambda(q)\cdot  C_{\lambda\nu}(q)\,, \nonumber\\
C^{-1}_{\mu\nu}(q)&=& g_{\mu\nu}+\frac{G}{N} B_{\mu\nu}(q)=
\left(1-\frac{G}{{\check G}_{\rm crit}}\right) g_{\mu\nu} + G q^2 f(q^2,\eta) \left(g_{\mu\nu} - \frac{q_\mu q_\nu}{q^2}\right)
\nonumber\\
&=&G \left[v^2 g_{\mu\nu} +q^2 f(p^2,\eta) \left(g_{\mu\nu} - \frac{q_\mu q_\nu}{q^2}\right)\right]
 \,.
\end{eqnarray}
Here we note  that the ``direct $4-\phi$ vertex'' at the each end of the bubble graph, there are two relevant contributions, one is the 
genuine direct
$4-\phi$ coupling $(\frac{a}{2}-1)\frac{G}{N}$ and the other from the ``tree $\rho_\mu$ propagator'' $(-i\frac{a}{2})\cdot \frac{1}{\frac{a}{2}\frac{N}{G}}\cdot (-i\frac{a}{2})=-\frac{a}{2}\frac{G}{N}$ such that 
\begin{eqnarray}
\left[\left(\frac{a}{2}-1\right)\frac{G}{N}\right] +  \left[-\frac{a}{2}\frac{G}{N}\right]=- \frac{G}{N}\,,
\label{4phivertex}
\end{eqnarray}
as in the above calculations, which makes the each vertex attached to the bubble $B_{\mu\nu}(q)$ in the sum to be independent of $a$.

Then we have 
\begin{eqnarray}
C_{\mu\nu}(q)&=& \frac{1}{1-\frac{G}{{\check G}_{\rm crit}}} g_{\mu\nu} - 
\frac{G p^2 f(q^2,\eta)}{ \left(1-\frac{G}{{\check G}_{\rm crit}} \right)+ G q^2  f(q^2,\eta)}
\cdot \left(g_{\mu\nu} - \frac{q_\mu q_\nu}{q^2}\right)\nonumber\\
&=&\frac{1}{G v^2}g_{\mu\nu} - \frac{q^2 f(q^2,\eta)}{v^2+q^2 f(p^2,\eta)}
\end{eqnarray}
and hence 
\begin{eqnarray}
 \langle \alpha_{\mu,||} \, \alpha_{\nu,||}\rangle(q) 
&=&\left(i\frac{G}{N}\right)^2 B_{\mu\lambda}(q)\cdot C^\lambda_\nu(q)
=\frac{G}{N} 
\frac{1}{v^2} \left[
-\frac{1}{{\check G}_{\rm crit}}g_{\mu\nu}  +\frac{1}{G} \frac{p^2 f(q^2,\eta)}{ v^2 + q^2 f(p^2,\eta)} \cdot \left(g_{\mu\nu} - \frac{q_\mu q_\nu}{q^2}\right)\right]\nonumber\\
&=& 
\frac{1}{N} \left[
 G\cdot g_{\mu\nu}
+
\frac{- f^{-1}(q^2,\eta)}{q^2 - (- v^2 f^{-1}(q^2,\eta))} \left(g_{\mu\nu} -\frac{q_\mu q_\nu}{- v^2 f^{-1}(q^2,\eta)}
\right)\right]\,.
\label{compositepropagator}
\end{eqnarray}
The result is independent of $a$, as it should be. We thus have a massive composite vector bound state $\alpha_{\mu,||}$ in the broken phase, even without use
of the auxiliary field $\rho_\mu$, i.e., $a=0$, or the concept of HLS at all.
\\

Now, Eq.(\ref{compositepropagator}) is consistent  with Eq.(\ref{adependence}) and $\langle \rho_\mu \rho_\nu\rangle(q)$ in Eq.(\ref{rhopropagatoruniversal}) (in the
basis change $\langle \rho_\mu\rho_\nu\rangle(q)= 2 \cdot \langle \rho^a_\mu\rho^a_\nu\rangle(q)$ (no sum) as noted in Eq.(\ref{twopointfunc})):
\begin{eqnarray}
\langle \rho_\mu\rho_\nu\rangle (q)&=&
\frac{1}{N} \left[
\left( -\frac{2}{a}+1\right) G\cdot g_{\mu\nu}
+
\frac{- f^{-1}(q^2,\eta)}{q^2 - (- v^2 f^{-1}(q^2,\eta))} \left(g_{\mu\nu} -\frac{q_\mu q_\nu}{- v^2 f^{-1}(q^2,\eta)}
\right)\right]\,.
\label{rhopropagator-a}
\end{eqnarray}
Note again that $a=2$ is special in the sense that the $\rho_\mu$ propagator takes the standard unitary gauge form without
extra contact term as in Eq.(\ref{propagator}).
\\

  Thus we have established that both $M_\rho^2$ and $g_{_{\rm HLS}}^2$ are independent of the parameter $a$, while the $\rho_\mu$
  propagator does have an $a-$dependent contact term, which implies the violation of the classical equation of motion. The physical meaning of this contact term will be discussed in the next subsection.
 \\

Further comments:

One might be puzzled by the contact term in Eq.(\ref{compositepropagator}), though. It is actually nothing peculiar. 
To see the bound states in the vector channel in the large $N$ limit,  we need the full amplitude containing  the tree-level
direct $4-\phi$ vertex in Eq.(\ref{4phivertex}), as is done in the NJL model (see e.g., \cite{Bardeen:1989ds}):
\begin{eqnarray}
T_{\mu\nu}(q) &=&  -\frac{G}{N} g_{\mu\nu}+\left(i\frac{G}{N}\right)^2 \left[B_{\mu\nu}(q) + B_{\mu\lambda}(q)\cdot \left(-\frac{G}{N}\right) B^\lambda_\nu(q) +\cdots\right]= -\frac{G}{N} g_{\mu\nu} +  \langle \alpha_{\mu,||}\, \alpha_{\nu,||}\rangle(q)  \nonumber\\
 &=& 
\frac{- f^{-1}(q^2,0)}{p^2 - (- v^2 f^{-1}(q^2,0))} \left(g_{\mu\nu} -\frac{q_\mu q_\nu}{- v^2 f^{-1}(q^2,0)}\right)\,,  \quad \left(
{\rm broken} \,\, {\rm phase}\right)\,.
\label{massivevector}
\end{eqnarray}
which is in fact 
the standard  massive vector bound state
propagator of the unitary gauge form in the broken phase. Here the tree contact term is precisely cancelled by the
contact term arising from the infinite sum of the bubbles. So the existence of the contact term in Eq.(\ref{compositepropagator}) is of no peculiarity, or rather welcome.
\\

This result in fact implies the ``vector meson dominance (VMD)'' for the $\pi\pi$ scattering in the broken phase even without the
auxiliary field $\rho_\mu$, in sharp contrast to the conventional HLS formalism which is realized at $a=4/3$  for the $\pi\pi$ scattering~\cite{Harada:2003jx}. 

We shall further show later that the same kind of cancellation takes place for $\pi$ form factor (analogue of the pion electromagnetic form factor), with the tree contact term precisely cancelled by that arising from an infinite sum of the bubble diagrams, realizing  the VMD not by the parameter choice but by the nonperturbative dynamics  in the large $N$ limit. Moreover, similar cancellation of the contact terms always take place not just for $\alpha_{\mu,||}$ but
also for $\rho_\mu$ even including the
``peculiar'' $a-$ dependent contact term, a phenomenon like a ``miracle'' of the large $N$ dynamics. 
\\

In the unbroken phase $v=0$,  on the other hand, $C^{-1}_{\mu\nu}(q)$ has no $g_{\mu\nu}$ term and hence cannot be inverted into $C_{\mu\nu}(q)$. This is no problem, however, since {\it we have a gauge symmetry, HLS, for $a\ne 0$.
It only means that we need to fix the gauge to take inversion for getting the massless propagator}, the same situation as 
$\langle \rho_\mu\rho_\nu\rangle (q)$ in Eq.(\ref{propagatorsymmetric}) as repeatedly mentioned: 
\begin{eqnarray}
T_{\mu\nu}(q) &=&\frac{- f^{-1}(0,\eta)}{q^2} g_{\mu\nu} + {\rm gauge}\,\, {\rm  terms}\quad \left(
{\rm symmetric} \,\, {\rm phase}\right)\,.
\label{amplitudesymmetric}
\end{eqnarray}
\\

The independence of the on-shell quantities from the auxiliary field parameter $a$ is also true in the NJL model (see Appendix \ref{NJL}) where the auxiliary fields $\hat \pi^a$  
and $\hat  \sigma$ can be 
introduced with arbitrary weight, say $\alpha$, as 
$ \frac{\alpha}{2 G} [(\pi^a + G \bar \psi i \gamma_5\tau^a \psi/\sqrt{2})^2+ (\hat  \sigma + G\bar \psi \psi/\sqrt{2})^2]$, which may or may not cancel completely the effects of the four-fermion operators, $\frac{G}{4}[ 
(\bar \psi i \gamma_5 \tau^a \psi)^2  + (\bar \psi \psi)^2]$ except for $\alpha=-1$ but the on-shell physics are  
the same for  arbitrary  $\alpha$ in the large $N$ limit. 

There is a caveat, however:  A distinctive  difference between the present model and the NJL model is the unbroken phase, where {\it our model is a consistent 
quantum theory only when the auxiliary field is introduced (i.e., $a\ne 0$) to make the gauge symmetry HLS explicit} as we noted repeatedly, while in the NJL model no such a
problem exists in the unbroken phase even without the auxiliary field.
 \\

 Incidentally, we may note  that once the vector bound state $\alpha_{\mu,||}$  in the broken phase is generated in the form of the standard  unitary gauge propagator, Eq.(\ref{massivevector}), it is well-known to give the Skyrme term in the region $p^2\ll M_\rho^2$~\cite{Ecker:1989yg}:
\begin{eqnarray}
{\cal L}^{(\rho)}_{(p^4)}=\frac{F_\rho^2}{32 M_\rho^2} {\rm tr} \left([L_\mu,L_\nu]\right)^2=
\frac{1}{32 g_{_{\rm HLS}}^2} \left([L_\mu,L_\nu]\right)^2\,,\quad \left(
q^2\ll M_\rho^2\right)\,, \quad
\left(L_\mu\equiv \partial_\mu U\cdot U^\dagger,  \quad U(x)=e^{2i \pi(x)/F_\pi}\right)
\label{Skyrme1}
\end{eqnarray}
with identification $e^2=g_{\rm HLS}^2$, and no non-Skyrme term $ {\rm tr} \left(\{L_\mu,L_\nu\}\right)^2$. Here use has been made of Eq.(\ref{Frhoaindependence}) and the notation and the normalization is the one for the standard  $SU(2)_L \times SU(2)_R/SU(2)_V$ (see Appendix \ref{sHLS}).
\\

\subsection{Physical implications of $a-$(in)dependence  in the large $N$ limit}
\label{implicationsaindependence}

We have shown that in the large $N$ limit of the SM Higgs Lagrangian the on-shell quantities of the dynamically generated HLS gauge boson $\rho_\mu$ are $a-$independence as in Eq.(\ref{aindipendence}), while $a-$dependence does exist in the off-shell quantities in the $\rho_\mu$ propagator through the contact term as in Eq.(\ref{rhopropagatoruniversal}). 

We here discuss what are  independent of $a$ and what else  are not, and their physical implications, by focussing on the broken phase in the {\it SM Higgs Lagrangian and the 2-flavor QCD on the same footing}, since the {\it dilatonic (SM Higgs boson) contributions are irrelevant to the  large $N$ limit physics of the HLS gauge bosons} as we have already mentioned, although they are described by the scale-invariant (with dilaton)  and the non-scale-invariant (without dilaton) version, respectively of {\it the same Grassmannian nonlinear sigma model}. 
  
It turns out that all the ``successful results of $a=2$'' in QCD, namely $\rho-$universality and KSRF I, II, 
and the VMD  
of the form factor of the NG boson $\pi$, and their analogues in the SM are realized independently of $a$.
We have already noted that the 
 ``VMD of $\pi\pi$ scattering'' is realized even without
auxiliary field $\rho_\mu$ (in the broken phase). On the other hand, the off-shell physics depend on $a$ in principle: the skyrmion physics are dependent on $a$ in such a way that at the $a\rightarrow \infty $ where the classical equation of motion for $\rho_\mu$ is recovered also at quantum level, whence the dynamically generated $\rho_\mu$ kinetic term is totally replaced by the Skyrme term (not just low energy limit but also the high energy limit crucial to the stabiliztion of the 
skyrmion).

\subsubsection{$a-$independent results}

 Let us first discuss the $a-$independent properties, which are relevant to the possible signatures of $\rho_\mu$ resonances as the SM rho   to be detectable at the collider experiments as well as the $\rho$ meson properties in QCD. 
\\

 Eq.(\ref{aindipendence}) and Eq.(\ref{Frhoaindependence}) read:
 \begin{eqnarray}
 M_\rho(q^2)\equiv M_\rho^2(q^2,0)= g^2_{_{\rm HLS}}(q^2) F_\rho^2=2 g^2_{_{\rm HLS}}(q^2)  F_\pi^2\,,\quad 
 \left(g^2_{_{\rm HLS}}(q^2)\equiv g^2_{_{\rm HLS}}(q^2, \eta=0) \right)\,,
 \label{Mrho-a}
 \end{eqnarray}
 which is  independent of $a$.  This  implies the  KSRF II (Eq.(\ref{KSRFII})),   if  the $\rho-$ universality  $g_{\rho\pi\pi}= g_{_{\rm HLS}}$ (Eqs.(\ref{rhouniversality})) is satisfied also independently of $a$. Now we show that this is indeed the case.

 Note that $g_\rho$
 is defined as the matrix element of the current ${\cal V}^a_\mu$ of the gauged $H_{\rm global} (\subset G_{\rm global})$ for the $\rho_\mu$ state as 

$ \langle 0| {\cal V}^a_\mu|\rho^b(q)\rangle \equiv \delta^{ab}\cdot g_\rho(q^2)\cdot \epsilon_\mu(q) 
=\delta^{ab}\cdot M_\rho(q^2) \cdot F_\rho\cdot \epsilon_\mu(q) $ and hence we have
\begin{eqnarray}
g_\rho(q^2) = M_\rho(q^2) \cdot F_\rho\cdot 
= g_{_{\rm HLS}}(q^2) \cdot F_\rho^2= 2g_{_{\rm HLS}}(q^2)  \cdot F_\pi^2\,,
 \label{grhoa}
 \end{eqnarray}
which is also $a-$ independent, where use has been made of Eq.(\ref{Mrho-a}).
        
      On the other hand, the KSRF I, Eq.(\ref{KSRFI}) is of course  independent of $a$ even in the conventional HLS approach,  and actually a  low energy theorem of HLS~\cite{Bando:1984pw,Bando:1987br} proved to all orders of loop expansion~\cite{Harada:1993jk}:
\begin{eqnarray}
      g_\rho(q^2)&=& 
2 g_{\rho\pi\pi}(q^2)\cdot F_\pi^2 \,,
 \quad \left( 
 {\rm KSRF}\, {\rm I}  
 \right) \,.
 \label{KSRFIq2}
  \end{eqnarray}     
   
 Comparing Eq.(\ref{KSRFIq2}) with Eq.(\ref{grhoa}), we have the $\rho-$universality independently of $a$:    
      \begin{eqnarray}
     g^2_{\rho\pi\pi}(q^2) &=& g^2_{_{\rm HLS}}(q^2) =-\frac{1}{2 N} f^{-1}(q^2,0) =  \frac{3 \left(4\pi\right)^2}
{N \,\ln \left(\frac{\Lambda^2}{q^2}\right) }
\,,\quad \left(\rho-{\rm universality}
\right)\,,
      \label{universality-a} 
               \end{eqnarray}
 where $N\rightarrow 4$ is understood for the SM  as before (see Eq.(\ref{SMkinetic}) through Eq.(\ref{SMFrho})) and for the 2-flavor QCD as well. 
 Then Eq.(\ref{Mrho-a}) reads the (generalized) KSRF II: 
 \begin{eqnarray}
 M_\rho^2(q^2)= g^2_{\rho\pi\pi}(q^2) \cdot F_\rho^2 = 2 g^2_{\rho\pi\pi}(q^2) \cdot F_\pi^2\,, \quad \left({\rm KSRF}\, {\rm II}\right)\,.
 \label{KSRFIIq2}
  \end{eqnarray}
  
   The result  is further  
confirmed  by a direct computation without recourse to the KSRF I, which is  given in Appendix \ref{universality}.

{\it Thus the dynamically generated $\rho_\mu$ in the large $N$ limit reproduces the celebrated ``$a=2$ relations'' Eqs.(\ref{rhouniversality}) and (\ref{KSRFII})
 in QCD even for arbitrary value of $a$. In other words, Nature's mysterious choice $a=2$ in QCD is nothing but a dynamical consequence of the nonperturbative dynamics in the large $N$ limit !!}
In the case of the SM Higgs Lagrangian the same result should also be checked at collider experiments if the $\rho_\mu$ 
has a mass in the detectable region, as will be discussed later.
\\

 Next we discuss the VMD which is realized in the reality of  the 2-flavored QCD, and  should also be in the SM Higgs Lagrangian if the $\rho_\mu$ mass is within the detection range of the collider experiments (see later discussions). 
 Since it is off-shell physics, there is no a priori reason to believe it be realized 
 $a-$ independently. 
 Nevertheless, it turns out to be the case.  In contrast,  
 it is well-known that the VMD for the $\pi$ form factor is  valid  only for $a=2$ in the conventional HLS formalism,  at tree level with the kinetic term of $\rho_\mu$ (simply assumed to be dynamically generated)~\cite{Bando:1984ej,Bando:1985rf,Bando:1984pw,
 Bando:1987br,Harada:2003jx}. Furthermore,  at the one-loop order ${\cal O}(p^4)$ in the chiral perturbation theory,  
 it is even violated badly in general,
 particularly near the phase transition point~\cite{Harada:2001rf,Harada:2003jx}

The NG boson form factor with the external gauge boson such as the $\gamma\pi\pi$ in the hadron physics is given
as usual  by  gauging $H_{\rm global} (\subset G_{\rm global})$ in the Lagrangian Eq.(\ref{a-depaction}) as  $D_\mu \phi =\left(\partial_\mu -i \rho_\mu\right) \phi \Rightarrow {\hat D}_\mu \phi= (D_\mu \phi + i  \phi {\cal B}_\mu)$ as in Eq.(\ref{fullcovariant}).

For $a=2$, there is no ${\cal B}_\mu \pi\pi$, a direct coupling to the NG boson $\pi$ (contained in $\phi$, recall our parameterization Eq.(\ref{ONpNp})),  as easily read from Eq.(\ref{covariantform}) by gauging $H_{\rm global}
\subset G_{\rm global}$ corresponding to the photon $\gamma_\mu$ in hadron physics). Therefore the vector meson dominance is
trivially realized with the unitary gauge propagator in Eq.(\ref{rhopropagator-a}), which has no 
contact term for $a=2$ but does have a nontrivial log $q^2$ dependence through $g_{_{\rm HLS}}^2= (-2 N f(q^2,0))^{-1}$.
\\

For arbitrary $a$ 
we use the Ward-Takahashi identity Eq.(\ref{rhopipiGreenfn})
for the Green function $\langle \rho^{(R)}_\mu\, \pi\,\pi\rangle =\langle \alpha_{\mu,||}^{(R),a}\, \pi\,\pi\rangle$ and the explicit computation  Eq.(\ref{universalitycheck}) in the Appendix \ref{universality}
(with base change $\rho_\mu^{ij}\rightarrow \rho_\mu^a$):
\begin{eqnarray}
\langle \rho^{(R),a}_\mu(q) \, \pi(k)\,\pi(q+k)\rangle\big|^{k^2=(k+q)^2=0}_{\phi-{\rm amputated}}\ &=&\langle \alpha_{\mu,||}^{(R),a}(q)\,  \pi(k)\, \pi(q+k)\rangle\big|^{k^2=(k+q)^2=0}_{\phi-{\rm amputated}}\nonumber\\
&=&
\frac{ g_{\rho\pi\pi}(q^2)}{q^2- g^2_{_{\rm HLS}} (q^2)\cdot F_\rho^2}\cdot \left(g_{\mu\nu} -\frac{q_\mu q_\nu}{g^2_{_{\rm HLS}} (q^2)\cdot F_\rho^2}
\right)\cdot (q+2k)^\nu\,,
\label{identical}
\end{eqnarray}
both of which have no $a-$dependence and no contact term.
Here the $\alpha_{\mu,||}$ is ``renormalized'' (rescaled to the canonical kinetic term) as $\alpha_{\mu,||}^{(R)}(q)= g_{_{\rm HLS}} (q^2) \cdot \alpha_{\mu,||}$ 
and similarly for $\rho_\mu$ (see Appendix \ref{universality}). 
In the generic case the form factor $F_{_{{\cal B}\pi\pi}}(q^2)$ for  the gauged $H_{\rm global}$ current ${\cal V}_\mu^a$ is just a linear combination of these two of the identical form :
\begin{eqnarray}
F_{_{{\cal B}\pi\pi}}(q^2) \left(q+2k\right)_\mu
&=&
\langle  {\cal V}^a_\mu(q)\, \pi(k)\,\pi(k+q)\rangle\big|^{k^2=(k+q)^2=0}_{\phi-{\rm amputated}}\ \nonumber\\
&=&
- g_\rho(q^2) \cdot \left[
\frac{a}{2} \langle \rho^{(R),a}_\mu(q) \, \pi(k)\, \pi(q+k)\rangle+\left(1-\frac{a}{2}\right) \langle  \alpha_{\mu,||}^{(R),a}(q)\, \pi(k) \,\pi(q+k)\rangle
\right]\Big|^{k^2=(k+q)^2=0}_{\phi-{\rm amputated}}\cdot \left(q+2k\right)^\nu\nonumber\\
&=&\frac{
g_\rho(q^2) \cdot  g_{\rho\pi\pi}(q^2)}{
g^2_{_{\rm HLS}}(q^2)\cdot F_\rho^2-q^2}
\cdot \left(g_{\mu\nu} -\frac{q_\mu q_\nu}{g^2_{_{\rm HLS}} (q^2)\cdot F_\rho^2}
\right)\cdot (q+2k)^\nu\nonumber\\
&=&\frac{
g^2_{_{\rm HLS}}(q^2) \cdot F_\rho^2}{
g^2_{_{\rm HLS}}(q^2)\cdot F_\rho^2-q^2}
\cdot \left(g_{\mu\nu} -\frac{q_\mu q_\nu}{g^2_{_{\rm HLS}} (q^2)\cdot F_\rho^2}
\right)\cdot (q+2k)^\nu\,,
\end{eqnarray}
namely, the VMD is realized independently of $a$. The contact terms are cancelled within each of
$\langle \rho^{(R)}_\mu\, \pi\,\pi\rangle$ and $\langle \alpha_{\mu,||}^{(R)}\, \pi\, \pi\rangle$, separately, and hence the VDM follows independently of any combination of those.

 The result has the log $q^2$ dependence through  $g_{_{\rm HLS}}^2= (-2 N f(q^2,0))^{-1}$ as in Eq.(\ref{universality-a}), 
 \begin{eqnarray}
 F_{_{{\cal B}\pi\pi}}(q^2)=\frac{ g_\rho(q^2)\cdot g_{\rho\pi\pi}(q^2)}{g^2_{_{\rm HLS}} (q^2)\cdot F_\rho^2 - q^2}=\frac{ g^2_{_{\rm HLS}} (q^2)\cdot F_\rho^2}{g^2_{_{\rm HLS}} (q^2)\cdot F_\rho^2 - q^2}
 =\frac{ g^2_{\rho\pi\pi} (q^2)\cdot F_\rho^2}{g^2_{\rho\pi\pi} (q^2)\cdot F_\rho^2 - q^2}\,, \quad \left({\rm VMD}\right)\,,
 \label{VMDq2}
  \end{eqnarray}
  with a correct normalization $F_{_{{\cal B}\pi\pi}}(0)=1$ even including the $\log q^2$ dependence.  
 This 
differs from   the naive VMD without such a  log $q^2$ dependence, although it takes the same form near   the on-shell $q^2\simeq M_\rho^2=g^2_{_{\rm HLS}}(q^2=M^2_\rho)\cdot F_\rho^2=g^2_{_{\rm HLS}}\cdot F_\rho^2$: 
   \begin{eqnarray}
 F_{_{{\cal B}\pi\pi}}(q^2) \approx 
 \frac{M_\rho^2}{M_\rho^2-q^2}\,, \quad \left( q^2\approx M_\rho^2\right)\,.
  \end{eqnarray}
However, such a $q^2$ dependence is actually necessary for the modern version of the  
VMD 
to describe the correct $q^2$ behavior  of the
$\pi$ form factor and the related quantities  in low energy QCD,  in both the space-like and the time-like momentum regions, see e.g., ~\cite{Gounaris:1968mw,Harada:1995sj,Benayoun:2011mm}.
Then the VMD in a modern version is naturally realized in large $N$ limit of the present theory even without the auxiliary field $\rho_\mu$

 It is compared with the standard HLS formulation which satisfies VMD only for $a=2$, while  $ F_{_{{\cal B}\pi\pi}}(0) =1$ for any $a$ in a different way: \cite{Harada:2003jx}
 \begin{eqnarray}
 F_{_{{\cal B}\pi\pi}}(q^2)&=& 
 \left(\left(1\right)_{{\cal L}_A} -\left(\frac{a}{2}\right)_{{\cal L}_V}\right)\bigg|_{\rm direct} + \frac{g_\rho g_{\rho\pi\pi}}{M_\rho^2- q^2}\nonumber\\
 &=&\left(1 -\frac{a}{2}\right)\bigg|_{\rm direct} + \frac{a}{2}\frac{M_\rho^2}{M_\rho^2- q^2}\,,
  \end{eqnarray}
  where $g_\rho g_{\rho\pi\pi} =\frac{a}{2} M_\rho^2$ is $a-$dependent in contrast to the above our corresponding  relation  which is $a-$independent.
\\

 \subsubsection{$a-$dependent results}
 
Another interesting off-shell physics is the skyrmion.
First note that $\langle \rho_\mu \rho_\nu\rangle(q)$ in Eq.(\ref{rhopropagator-a}) for $a=2$ and the pole in $T_{\mu\nu}(q)$
 in Eq.(\ref{massivevector}) is identical. Then the $a=2$ case gives the same Skyrme term for
 $q^2\ll M_\rho^2$:
 \begin{eqnarray}
{\cal L}_{(p^4)}=\frac{F_\rho^2}{32 M_\rho^2} {\rm tr} \left([L_\mu,L_\nu]\right)^2=
\frac{1}{32 g_{_{\rm HLS}}^2} \left([L_\mu,L_\nu]\right)^2\,,\quad \left(a=2;\,\, q^2\ll M_\rho^2\right)\,.
\label{Skyrme2}
\end{eqnarray}

For $a\ne 2$,  the above discrepancy between the two propagators, Eq.(\ref{rhopropagator-a}) and Eq.(\ref{compositepropagator}),
 disappears in the limit $a\rightarrow \infty$ and hence the {\it equation of motion at classical level is recovered
even at the quantum level} as can be seen from
Eq.(\ref{adependence}). It  then  implies that  the 
  field strength of the dynamically generated HLS gauge boson $\rho_\mu$ 
 reads \cite{Igarashi:1985et}:
 \begin{eqnarray}
 \rho_{\mu\nu}\Big|_{\rho_\mu=\alpha_{\mu,||}} 
 =\partial_\mu \alpha^a_{\nu, ||} -\partial_\nu \alpha_{\mu, ||} -i\left[\alpha_{\mu, ||},\alpha_{\nu, ||}\right]
 =i \left[\alpha_{\mu, \perp},\alpha_{\nu, \perp} \right]\,,
\end{eqnarray}
which  for $N=4$ and $p=3$ yields the  kinetic term into the form:
\begin{eqnarray}
{\cal L}^{(\rho)}_{\rm kinetic} =-\frac{1}{2 g^2_{_{\rm HLS}}} \, {\rm tr} \,
\rho^2_{\mu\nu} \longrightarrow\frac{1}{2 g^2_{_{\rm HLS}}} \, {\rm tr} \,
\left([\alpha_{\mu,\perp},\alpha_{\nu,\perp}]\right)^2=
\frac{1}{32 e^2 }
{\rm tr} ([L_\mu, L_\nu])^2\,,\quad \left(a\rightarrow\infty\right)\,,
 \quad \left(e^2= g^2_{_{\rm HLS}}\right)\,, 
 \label{Skyrme}
 \end{eqnarray}
 where for comparison with the standard Skyrme term expression, we have used notation/normalization of the equivalent $G/H=[SU(2)_L\times SU(2)_R]/SU(2)_V$, with
$L_\mu=\partial_\mu U\cdot U^\dagger=2 i \cdot \xi_L^\dagger\cdot  \alpha_{\mu,\perp} \cdot \xi_L$ in Eq.(\ref{chiralalpha}).
Hence 
{\it the $a\rightarrow \infty$ limit in the large $N$ dynamics of the SM does generate the
Skyrme term.} This is compared with Eq.(\ref{Skyrme1}) and also with Eq.(\ref{Skyrme2}) at $a=2$,  where the dynamically generated $\rho_\mu$ effects are replaced 
by the Skyrme term only  in the low energy limit $q^2\ll M_\rho^2$.

In the case at hand, $a\rightarrow \infty$ limit,  the $\rho_\mu$ kinetic term effects are completely replaced by the
Skyrme term for entire energy region not just in the low energy limit. This is crucial for the {\it SM skyrmion which is stabilized by the short distance off-shell dynamics}. Thus in contrast to the on-shell physics independent of the parameter $a$, the off-shell effects such as the skyrmion do depend on the parameter $a$. 
\\

One might suspect that a Higgs-like scalar bound state is also generated in the large $N$ limit, which would produce the non-Skyrme term
$ {\rm tr} \left(\{L_\mu,L_\nu\}\right)^2$ destabilizing the skyrmion  in the low energy limit. Were it not for the
elementary scalar, pseudo dilaton $\varphi$ as in the present SM case, then the  model would be  simply a usual Grassmmannian nonlinear sigma model without
dilaton field, which  in fact would generate an $O(N)$-singlet dynamical Higgs-like particle in the large $N$ limit,
in the same as the $CP^{N-1}$ model acting like a pseudo dilaton (see Appendix \ref{CPNApp}). In the case at hand,
having the elementary Higgs already, however, such a nonperturbative dynamics would not generate new particle but only gives quantum corrections of the already existing particle. So once such an elementary scalar is included from the onset in the soliton equation, as done in  \cite{Matsuzaki:2016iyq,MOY2018}, the skyrmion physics based on the dynamical gauge boson of the
HLS 
would not be affected further by the large $N$ dynamics acting on the dilatonic scalar sector.
\\
 
Thus if the SM is close to the ``Skyrme limit'' $a\rightarrow \infty$, then the skyrmion in the SM having the elementary scalar is well described by $1/a$ expansion near the Skyrme 
limit. 
Actually, the $a-$ dependence of Skyrmion as the dark matter in the SM will be rather weak all the way down to $a={\cal O}(1)$, as will
be shown in the forthcoming paper~\cite{MOY2018}.
\\

\subsubsection{Phenomenological implications of SM rho}

Phenomenological implications of our result  for the SM rho would be divided into two different scenarios depending on the possible value of a single extra free parameter existing in the 
nonperturbative theory,  $M_\rho=g_{\rho\pi\pi}\cdot F_\rho$ (or $g_{\rho\pi\pi}=g_{_{\rm HLS}}\equiv g_{_{\rm HLS}}(M_\rho^2)
=M_\rho/F_\rho=M_\rho/(350 {\rm GeV})$
or the cutoff $\Lambda$ (or the Landau pole $\tilde \Lambda$) in view of  Eqs.(\ref{SMkinetic}) -(\ref{SMFrho}): 
\begin{eqnarray}
\Lambda= e^{-4/3}\cdot  \tilde \Lambda=
e^{-4/3}\cdot  M_\rho  \cdot \exp{\left[\frac{\frac{3}{8}(4\pi  F_\rho)^2}{M_\rho^2}\right]}\,,
\label{LambdavsMrho}
\end{eqnarray}
which implies that  $\Lambda < M_\rho\, (g_{_{\rm HLS}}>6.7,\,  M_\rho> 2.3\,  {\rm TeV})$ and  $\Lambda > M_\rho\, ( g_{_{\rm HLS}}< 6.7, \, M_\rho< 2.3\,  {\rm TeV})$.  

1) ``Low $M_\rho$ scenario'' ($M_\rho<2.3 \, {\rm TeV},\,  \Lambda> M_\rho$):

Signatures of the SM rho $\rho_\mu$ at the collider experiments are qualitatively  similar to the technirho in the walking technicolor based on the same type of the s-HLS effective theory \cite{Kurachi:2014qma,Fukano:2015hga,Fukano:2015uga,Fukano:2015zua},
but more {\it definite prediction  due to a single parameter in the present case}, 
arising from the $a-$independence of the resonance parameters:
Typical LHC signatures would be in the diboson channel  through {\it the Drell-Yang process} with the $W/Z/\gamma-\rho_\mu$ mass mixing: $q \bar q \rightarrow W/Z/\gamma \rightarrow \rho_\mu\rightarrow  W_L W_L/W_L Z_L$, characterized by  the VMD,  with the   coupling,
 $\sim \alpha_{\rm em} g_\rho/M_\rho^2=\alpha_{\rm em} F_\rho/M_\rho=
\alpha_{\rm em}/g_{_{\rm HLS}}$. The production cross section depends on the parameter as  $\sim 1/M_\rho^2\sim 1/g^2_{_{\rm HLS}}$.
Decays (branching ratios) are dominated by the diboson processes $\rho_\mu\rightarrow W_L W_L/W_L Z_L$, with the coupling $g_{\rho\pi\pi}=g_{_{\rm HLS}}$, 
 characterized by the {\it absence of  the 
 processes  of $\rho_\mu\rightarrow W_L/Z_L+ \varphi $} ($\varphi$=SM\, Higgs=(pseudo-)dilaton), in contrast to the ``equivalence theorem results'' in many other models,  because of the ``conformal barrier'' arising from  the scale symmetry of the $\rho_\mu$ mass term in the s-HLS parameterization, similarly to the walking technirho~\cite{Fukano:2015uga,Fukano:2015zua} (see also Appendix \ref{DGHLS}).

Given a reference value $M_\rho=2$ TeV for instance, we would have  $g_{\rho\pi\pi}\simeq 5.7$ and $\Lambda\simeq 3.3\, {\rm TeV}\simeq 4\pi F_\pi$ (simple scale-up of  the QCD $\rho$ meson), perfectly natural scale with respect to the weak scale.
This  yields the width $\Gamma_\rho \simeq \Gamma_{\rho\rightarrow WW}\simeq g_{\rho\pi\pi}^2 M_\rho/(48\pi) = M^3_\rho/(48\pi F_\rho^2) \simeq  433 $ GeV \footnote{This is  in contrast to  the walking technirho \cite{Fukano:2015hga,Fukano:2015zua} of the typical one-family model ($N_D=4$ weak-doublets), where the decay width has suppression of $1/N_D=1/4$ and hence of order $O(100)$ GeV.
 }, so broad as barely detectable at
 LHC.   For larger (smaller) $M_\rho$ the width gets larger (smaller) as $\sim M_\rho^3$, and the  production cross section gets smaller (larger) as $\sim 1/M_\rho^2$, thus more difficult for $M_\rho>2$ TeV to be seen at LHC. 
  The SM rho with narrow resonance $\Gamma_\rho\lesssim100$ GeV could be detected at LHC for  $M_\rho\lesssim1.2$ TeV, which corresponds to $g_{_{\rm HLS}}\lesssim 3.5$ and $\Lambda \gtrsim 50\, {\rm TeV}$.

  2)  ``High $M_\rho$  scenario''  ($M_\rho\gg 2.3 \, {\rm TeV},\,  \Lambda < M_\rho$,  as a stabilizer of the skyrmion dark matter $X_s$)\cite{Matsuzaki:2016iyq}:

 Even if no direct evidence were seen at the collider experiments, 
 physical effects of the dynamical $\rho_\mu$ are still observable through the skyrmion dark matter $X_s$ in the SM.  In fact the SM skyrmion is stabilized by the
  {\it off-shell} $\rho_\mu$ in the {\it short distance} physics as shown in Ref.\cite{Matsuzaki:2016iyq}, the result of which corresponds to $a\rightarrow \infty$ calculation. 
It was shown~\cite{Matsuzaki:2016iyq,MOY2018} that 
the (complex scalar) skyrmion $X_s$ whose coupling to the SM Higgs as a pseudo-dilaton is given by the low energy theorem of the scale symmetry  Eq.(\ref{LET}). Then the
current direct detection experiments,  XENON1T and PandaX-II \cite{Aprile:2017iyp,Cui:2017nnn}, give a constraint on the skyrmion mass (and  simultaneously the skyrmion coupling to the SM Higgs which is directly related to the mass):
\begin{eqnarray} 
 M_{X_s} \lesssim 11 \,{\rm GeV}\,,\quad
 {\rm or}\,\, {\rm equivalently}\,,\,\, \lambda_{\varphi X_s X_s}\equiv \frac{g_{\varphi X_s X_s}}{2F_\varphi}  =\frac{M_{X_s}^2}{F_\pi^2} \lesssim 
 0.002  \,,\quad \left(F_{\varphi}=F_\pi=\sqrt{N} v=246\,\, {\rm GeV}\right)\,,
\end{eqnarray}
independently of the details of the skyrmion profile, solely due to the dilatonic nature of the SM Higgs (Here, $v^2$ in Eq.(\ref{LET}) has been rescaled by $N$ in the context of large $N$ limit arguments).
In the limit $a\rightarrow \infty$ the entire physics of the $\rho_\mu$ is traded for the Skyrme term with
the coefficient $e^2=g^2_{_{\rm HLS}}$, see Eq.(\ref{Skyrme}). In this limit  the complex  scalar skyrmion 
mass $M_{X_s}$ (and hence the coupling $g_{\varphi X_s X_s}$) and its square of the mean radius $\langle r_{_{X_s}}^2 \rangle_{X_s}$  have been calculated as~\cite{Matsuzaki:2016iyq,MOY2018}:
\begin{eqnarray}
M_{X_s} \simeq 35 \frac{F_\pi}{g_{_{\rm HLS}}} &\simeq& 11 \, {\rm GeV}\times 
\left( \frac{780}{g_{_{\rm HLS}}}  \right)
\,, 
\quad 
\lambda_{\varphi X_s X_s} =
 \left(\frac{35}{g_{_{\rm HLS}}}\right)^2=
0.002 \times \left(\frac{780}{g_{_{\rm HLS}}}\right)^2\,,\end{eqnarray}
which would imply  
  \begin{eqnarray}
g_{_{\rm HLS}} \simeq 780\,, 
\label{HLScouplingDM}
\end{eqnarray}
and 
\begin{eqnarray}
\langle r_{_{X_s}}^2 \rangle_{X_s}&\simeq &\left(\frac{2.2}{g_{_{\rm HLS}} F_\pi}\right)^2 \simeq 1.3 \times 10^{-10} \, ({\rm GeV})^{-2}
 \times \left(\frac{780}{g_{_{\rm HLS}}}\right)^2\,.
\end{eqnarray}
This leads to the annihilation cross section of the skyrmion dark matter and  the relic abundance $\Omega_{X_s} h^2$~\cite{Matsuzaki:2016iyq,MOY2018}:
\begin{eqnarray}
\langle \sigma_{\rm ann} v_{\rm rel} \rangle_{\rm radius}
&\simeq& 4\pi\cdot \langle r_{_{X_s}}^2 \rangle_{X_s} 
\simeq 
1.7 \times 10^{-9} \,\, {\rm GeV}^{-2}\,,\nonumber\\
\Omega_{X_s} h^2 &\simeq& {\cal O}(0.1)\,,
\end{eqnarray}
which is roughly consistent with the observed cold dark matter relic abundance $\Omega_{X_s} h^2\simeq 0.12$.
Then the popular belief that ``the dark matter is the physics beyond the SM'' will be no more than a  folk lore. For $a<\infty$, by $1/a$ expansion  we can explicitly show that 
the results are rather stable against changing $a$ all the way down to $a \sim 2
$~\cite{MOY2018}. 
 Note the cutoff is  $\Lambda =e^{-4/3} \tilde \Lambda \simeq e^{-4/3}\cdot M_\rho ={\cal O} (10^2\, {\rm TeV})$, where
$M_\rho = g_{_{\rm HLS}} \cdot F_\rho 
$ is  a typical mass scale (no longer  the ``on-shell'' mass, since the SM rho  is deeply off-shell).  

In either scenario, the phenomenologically interesting nonperturbative SM physics has  typical strong SM rho gauge coupling
$ g_{_{\rm HLS}}\simeq  1/3 - 10^3$, which corresponds to  the cutoff $\Lambda ={\cal O} (10^0 -10^2)\, {\rm TeV}$, or the quadratic divergence 
corrections to the weak scale  (see Eq.(\ref{quadraticdiv})): 
\begin{eqnarray}
\delta F_\pi^2 \sim 4 \times \frac{\Lambda^2}{(4\pi)^2} \sim 
(0.1 \,{\rm TeV})^2 - (10^2 \, {\rm TeV})^2\,, 
\end{eqnarray}
thus resolving the naturalness problem without recourse to the BSM, in sharp contrast to the pSM.
(As already mentioned in the Introduction, the HLS as a dynamically generated gauge symmetry in the SM is trivially anomaly-free,
since the SM fermions have no HLS charges.)
In this sense the SM with nonperturbative dynamics may be ``dual'' to some underlying BSM theory on that  scale,  similarly to the hadron-quark duality (nonlinear sigma model/chiral Lagrangian vs QCD). Immediate 
candidate for such a BSM would be the walking technicolor also having the approximate scale symmetry and pseudo-dilaton (technidilaton) \cite{Yamawaki:1985zg,Bando:1986bg}  (See Summary and Discussions)
\\

 \section{Summary and Discussions}

In this paper, we found that the ``SM rho'' as the gauge bosons $\rho_\mu(x)$ of the $O(3)\simeq SU(2)_V$ Hidden Local Symmetry (HLS)~\cite{Bando:1984ej,Bando:1985rf,Bando:1984pw,Bando:1987br,Harada:2003jx} within the Standard Model (SM) Higgs Lagrangian, though
 auxiliary field
at classical level, acquire the kinetic term at quantum level by the nonperturbative dynamics in the large $N$ limit, becoming fully propagating dynamical gauge bosons.
\\

We first recapitulated the previous observation \cite{Fukano:2015zua}
that the SM Higgs Lagrangian is rewritten straightforwardly into
a nonlinear realization based on the manifold $G/H=O(4)/O(3) \simeq \left[SU(2)_L\times SU(2)_R\right]/SU(2)_V$ and also
a nonlinear realization of  the (approximate) scale symmetry, with the SM Higgs  $\varphi$ being nothing but a (pseudo-) dilaton near the BPS limit.
The $G/H$ part is further  gauge-equivalent to another 
Lagrangian having a larger symmetry $G_{\rm global}\times H_{\rm local}=O(4)_{\rm global} \times O(3)_{\rm local}\simeq 
\left[SU(2)_L\times SU(2)_R\right]_{\rm global} \times \left[SU(2)_V\right]_{\rm local}$, with $H_{\rm local}$ being the HLS~\cite{Bando:1984ej,Bando:1985rf,Bando:1984pw,Bando:1987br,Harada:2003jx},
a spontaneously broken gauge symmetry existing in any nonlinear sigma model~\cite{Bando:1984pw,Bando:1987br}. 
\\

We  then studied nonperturbative dynamics of the SM Higgs Lagrangian in this HLS form in the large $N$ limit, by extending it to a scale-invariant version of the Grassmannian model $G/H=O(N)/\left[O(N-3)\times O(3)\right]$, which is gauge equivalent to $O(N)_{\rm global} \times \left[O(N-3)\times O(3)\right]_{\rm local}$. Our staring point is the most general HLS Lagrangian of such a scale-invariant version of the  Grassmannian model presented  as an extension of the SM, see Eq.(\ref{Model}) with Eq.(\ref{gaugefixedLA}), Eq.(\ref{LVp}) and  Eq.(\ref{LVNp}).
The model is simply reduced to the SM Higgs Lagrangian when we take $N=4\,,p=3$, Eq.(\ref{SMHLSO4}), where ${\cal L}^{(N-p)}_V$ term is missing.
Also the dynamical generation of the $O(N-3)_{\rm local}$ HLS gauge bosons is not possible in the large $N$ arguments where all the planar diagrams 
come into play and hence are uncontrollable.  If the kinetic term is not generated, then the term Eq.(\ref{LVNp}) is simply solved away
by the equation of motion of the $O(N-3)_{\rm local}$ HLS gauge bosons as  the mere auxiliary field.  So we focused on the dynamical generation of the HLS gauge boson of the $O(3)_{\rm local}$, the SM rho, with  the simplified Lagrangian omitting  Eq.(\ref{LVNp}), consisting
of the standard form of the HLS Lagrangian, ${\cal L}_A +a {\cal L}_V$, besides the dilatonic factor of the dilaton (SM Higgs) $\varphi$. The ${\cal L}_A$ is the original nonlinear sigma model
without HLS gauge bosons, $a {\cal L}_V$ an additional term associated with the HLS.
\\

For the convenience in  taking the large $N$ limit, we followed the standard way as in the $CP^{N-1}$ model, to rewrite the classical Lagrangian ${\cal L}_A +a {\cal L}_V$ into the form of the covariant derivative, Eq.(\ref{covariantform}) and used Lagrange multiplier $\eta(x)$ for the constraint
for the  nonlinear realization (in the scale-invariant form). This we showed  directly corresponds to the $a=2$ choice in the HLS model, although
it is equivalent to any value of $a$ through the use of the equation of motion of the auxiliary field
$\rho_\mu(x)$ (or adding ${\cal L}_V(=0)$ with arbitrary weight to change $a$), as far as the classical level without 
kinetic term of $\rho_\mu$ is concerned. The theory at classical level does not depend on $a$ as a matter of course.

Based on Eq.(\ref{covariantform}), we obtained  the effective action in large $N$ limit  for the $D$ dimensions with $2\leq D\leq 4$, which yields the gap equation, Eq.(\ref{gapa2}), 
in a form similar to that of the NJL model and to 
other Grassmannian models
in $D$ dimensions. Namely, the inverse coupling $1/G=F_\pi^2/N$ receives quantum correction (power divergence) 
denoted by $1/G_{\rm crit}=
\Lambda^{D-2}
/g_{\rm crit}
$ as $v^2=1/G-1/G_{\rm crit}$ ($G<G_{\rm crit}$), and  $1/G-1/G_{\rm crit}=-v_\eta^2$ ($G>G_{\rm crit}$). 

As such, it changes the phase continuously from the broken phase ($v\ne 0, v_\eta^2=0=\eta=\langle \eta(x)\rangle$,  the same as the bare theory) in the weak coupling region, Eq.(\ref{broken}), into the unbroken phase ($v=0, \eta\ne 0$, genuine quantum theory) in the strong coupling one, Eq.(\ref{symmetric}).  
The phase transition is the second order as is the case for
other Grassmannian models including the $CP^{N-1}$ model and the NJL model.
Hence the critical point $G_{\rm crit}$ is a nontrivial ultraviolet
fixed point for the dimensionless coupling $g=G\Lambda^{D-2}$, 
see Eq.(\ref{beta})
: 
\begin{eqnarray}
\beta (g)=
- (D-2)\, \frac{g}{g_{\rm crit}} \, \left( g - g_{\rm crit}\right)\,,\quad g_{\rm crit}=(4\pi)^{\frac{D}{2}} \left(\frac{D}{2} - 1 \right) \Gamma\left(\frac{D}{2}\right) \,. 
  \end{eqnarray}
and the same form for the renormalized coupling $g^{(R)}$ defined in Eq.(\ref{renormalized}) for $D\ne 4$. 

However,   for $D=4$, the ``renormalized coupling'' $g^{(R)}$, after tuning $v^2$ through the quadratic divergence,
actually still has a log divergence, Eq.(\ref{triviality}), which is regularized here by the cutoff $\Lambda$ (Such a cutoff is needed to define the dynamically generated 
HLS gauge kinetic term any way, which is absent in the tree-level SM Higgs Lagrangian as a counter term.):
\begin{eqnarray}
D=4:  \quad\quad g^{(R)}(\mu) &=& \frac{(4\pi)^2}{
 \ln (\Lambda^2/\mu^2)}\rightarrow 0 \quad \left(\frac{\Lambda}{\mu} \rightarrow \infty\right)\,;
\quad \rightarrow \infty \quad \left(\mu \rightarrow \Lambda\right)  \,,
\nonumber\\
\beta(g^{(R)}(\mu)) &=& \mu \frac{\partial g^{(R)}(\mu)}{\partial \mu}= \frac{2}{(4\pi)^2} \left(g^{(R)}(\mu)\right)^2
\,. 
\label{gRsummary}
 \end{eqnarray} 
This implies  a trivial infrared fixed point, although the bare coupling $g(\Lambda)$ has a nontrivial UV fixed point $g(\Lambda) \rightarrow g_{\rm crit}$ (now understood as a Gaussian fixed point, trivial theory). The cutoff $\Lambda$  is nothing but a Landau pole, corresponding to the {\it extra free parameter} to define the nonperturbative quantum theory for the dynamically induced HLS gauge boson kinetic term
absent in the tree SM Higgs Lagrangian.
 \\

We then found that similarly to the $CP^{N-1}$ model, the HLS gauge boson is dynamically generated in the large $N$ limit for the $D$ dimensions with $2\leq D\leq 4$ as in 
Eq.(\ref{propagator}),  which takes the form of Eq.(\ref{propagatorbroken})  in the broken phase, and that of Eq.(\ref{propagatorsymmetric}) in the unbroken phase, respectively. For $2\leq D<4$ the theory is renormalizable and no extra free parameters
are induced. 

On the other hand, in $D=4$  the log divergence in the nonperturbatively generated kinetic term as  regularized by the cutoff $\Lambda$ cannot be renormalized in a usual sense due to absence of the  counter term of the kinetic term in the SM Lagrangian, thus giving rise to   an extra free parameter, the induced HLS gauge coupling $g_{_{\rm HLS}}$ related to $\Lambda$, in sharp contrast to the perturbative SM which never generates such a kinetic term of the HLS gauge boson. 

In the broken phase, the induced gauge coupling $g_{_{\rm HLS}}$ for the kinetic term Eq.(\ref{kineticbroken}) is given in Eq.(\ref{gaugecouplingbroken}) and $\rho_\mu$ has a mass $M_\rho^2= 2 g_{_{\rm HLS}}^2 F_\pi^2$, Eq.(\ref{rhomassa2}), in a typical form of the Higgs mechanism. 

In the unbroken phase, on the other hand, kinetic term with $g_{_{\rm HLS}}$ is given in Eq.(\ref{kineticsymmetric}) and 
the unbroken gauge symmetry  is realized by the massless $\rho_\mu$, namely, the unbroken gauge symmetry is no longer ``hidden'' but explicit, as is the case in the
well-known phenomenon of $CP^{N-1}$ model. The NG boson $\pi$ in the classical theory is no longer the NG boson but has a mass $M_\pi^2=\eta$, Eq.(\ref{pimass}).
\\

Then our main results for the SM Higgs case were obtained as the  $D=4$ and $N\rightarrow 4$ case of the above generic results.
The dynamically generated kinetic term and the mass of the SM rho $\rho_\mu$
read as Eq.(\ref{SMkinetic}) and Eq.(\ref{SMrhomass}):
\begin{eqnarray}
{\rm SM}:\quad  \frac{1}{\lambda_{\rm HLS}(\mu^2)}
 &=&\frac{1}{N g_{_{\rm HLS}}^2(\mu^2)}
 =\frac{1}{3} \frac{1}{(4 \pi)^2} \ln \left(\frac{{\tilde \Lambda}^2}{\mu^2}\right)\,,
 \nonumber
\\
M_\rho^2 (\mu^2)&=& 
g_{_{\rm HLS}}^2(\mu^2) \cdot F_\rho^2 \,,\nonumber\\
F_\rho^2
&=&2 \cdot N v^2 =2\cdot F_\pi^2\simeq 2\cdot  \left(246\,{\rm GeV}\right)^2 \simeq \left(350\, {\rm GeV}\right)^2\,, 
\label{kineticsummary}
\end{eqnarray}
where $\mu^2= M^2_\pi
=\eta\ne 0$ is understood in the unbroken phase with $M_\rho^2(\mu^2)
\equiv 0$, and the ``on-shell'' $M_\rho^2$ in the broken phase with $\eta \equiv 0$ is defined by the solution of 
$M_\rho^2=M^2_\rho(\mu^2=M_\rho^2) = g_{_{\rm HLS}}^2(M_\rho^2) \cdot F_\rho^2$.   
 Then the induced HLS gauge coupling $\alpha_{_{\rm HLS}}(\mu^2)=g^2_{_{\rm HLS}}(\mu^2)/(4\pi)$ has a Landau pole $\alpha_{_{\rm HLS}}(\mu^2)\rightarrow \infty \,\,(\mu^2 \rightarrow {\tilde \Lambda}^2)$ as in Eq.(\ref{Landaupole}), and  is asymptotically non-free, i.e.,  has an  infrared zero:
\begin{eqnarray}
{\rm SM}: \quad \quad \beta(\alpha_{_{\rm HLS}}(\mu^2))= \mu^2 \frac{\partial \alpha_{_{\rm HLS}}(\mu^2)}{\partial \mu^2}= \frac{N}{12\pi} \alpha^2_{_{\rm HLS}}(\mu^2) 
 >0\,,
\label{IRfree}
\end{eqnarray}
similarly to the $g_R(\mu)$ in Eq.(\ref{gRsummary}), which is in accord with the fact that the UV fixed point for the original bare coupling $g=g_{\rm crit} =(4\pi)^2$
is a Gaussian fixed point (free theory). In fact both $\pi$ and $\rho_\mu$ become massless free particles just
on the fixed point with vanishing coupling $\alpha_{_{\rm HLS}}(\mu^2)\rightarrow 0$ ($g\rightarrow g_{\rm cr}$) as ${\tilde \Lambda}^2/\mu^2 \rightarrow \infty$.
\\

We further studied possible $a-$dependence of our results.  The classical theory without the kinetic term is obviously independent of $a$, since the auxiliary field $\rho_\mu(x)$ can simply be solved away via equation of motion. On the other hand, the theory at quantum level sometimes crucially depend on $a$. The different parameterization of the same classical theory may lead to different 
quantum theory, particularly in the  nonperturbative dynamics. 
\\

In fact, as we repeatedly emphasized (see discussions related to Eqs.(\ref{propagatorsymmetric}), (\ref{aindipendence}) and  (\ref{amplitudesymmetric}) ),  in the case $a=0$, i.e., the CCWZ
nonlinear realization without gauge symmetry, the HLS, the quantum theory becomes ill-defined in the unbroken phase, where
the transversality of the two-point function clearly indicates the existence of the massless vector meson pole, while
were it not for the gauge symmetry, the HLS, it cannot be inverted to the well-defined propagator. This is indeed
consistent with the Weinberg-Witten theorem which forbids the massless particle with spin $J\geq 1$ within the
positive definite Hilbert space (i.e., without gauge  symmetry).

Even in the broken phase,
the quantum theory which acquired the kinetic term could in principle depend on $a$: the classical equation of motion is generally violated as dictated by the Ward-Takahashi identities as in Eq.(\ref{adependence}).  More explicitly we showed the $\rho_\mu$ propagator acquires an unusual $a-$dependent contact term in Eq.(\ref{rhopropagatoruniversal}), in conformity with Eq.(\ref{adependence}).

An outstanding off-shell physics of such in the broken phase is the skyrmion physics, which does depend on $a$. In the limit $a\rightarrow \infty$ we recover the classical equation of motion $\rho_\mu=\alpha_{\mu,||}=i\frac{G}{N} \phi\partial_\mu \phi^t$ even at the quantum level, see Eq.(\ref{adependence}), or explicit calculations of each corresponding propagator, Eq.(\ref{compositepropagator}) and Eq.(\ref{rhopropagator-a}) (or Eq.(\ref{rhopropagatoruniversal})). Then the dynamically generated kinetic term of the SM rho  $\rho_\mu$ 
is entirely replaced by the Skyrme term, Eq.(\ref{Skyrme}).
\\

For all those $a-$dependences, however, we also showed that 
as far as the on-shell quantities are concerned, all the $a-$ dependences are (apparently) miraculously cancelled out  in the large $N$ limit, 
as in Eq.(\ref{rhopropagatoruniversal}), to leave them completely independent of $a$, Eq.(\ref{aindipendence}), in sharp contrast to the simple one-loop result in Eq.(\ref{twopointan2}).  

We further showed  notable $a-$independent  relations, Eq.(\ref{KSRFIq2}), Eq.(\ref{universality-a}) and  Eq.(\ref{KSRFIIq2}) as generalized form of the  KSRF I relation, 
the universality of the $\rho_\mu$ coupling, and the KSRF II relation, respectively. 
To our surprise, the outstanding off-shell physics, vector meson dominance (VMD), is also realized  independently of $a$, see Eq.(\ref{VMDq2}).
\\

Now to the possible phenomenological implications of the SM rho which were discussed in the subsection \ref{implicationsaindependence}. 
First of all, as we mentioned in the Introduction,  we  emphasize that the success of the conventional perturbative SM (pSM) results is intact in 
our unconventional  parameterization based 
on the nonlinear realization Eq.(\ref{Higgs2}), which is equivalent to the standard one Eq.(\ref{Higgs1}) as far as the perturbation is concerned. See e.g., Ref.\cite{Alonso:2016oah} for more generic parameterization for the perturbative calculations. 
The same is true also for  its gauge equivalent HLS Lagrangian, since the perturbation does not generate the kinetic term of the HLS gauge boson with
the $a {\cal L}_V$ term identically zero without any physical effects. 
Also note that the pSM in the infrared region
are intact even by the 
nonpertubative dynamics in the ultraviolet region, somewhat analogously to the QCD where success of the perturbative QCD in the  ultraviolet does not contradict the nonperturbative QCD in the infrared (reversed infrared-ultraviolet is due to the asymptotically free  QCD versus asymptotically non-free (infrared free) SM Higgs, or infrared Landau pole vs. ultraviolet Landau pole).

Then what about the physical   effects of the nonperturbative dynamics, namely the dynamical gauge boson SM rho $\rho_\mu$?
As given in the subsection \ref{implicationsaindependence},  we may expect $\rho_\mu$ signature at either collider physics or dark matter physics,
depending on the possible value of a single free parameter $M_\rho$, or equivalently $g_{_{\rm HLS}}$ or the cutoff $\Lambda=e^{-4/3} \cdot \tilde \Lambda$ ;
 either $\Lambda < M_\rho\, (g_{_{\rm HLS}}>6.7,\,  M_\rho> 2.3\,  {\rm TeV})$ or  $\Lambda > M_\rho\, ( g_{_{\rm HLS}}< 6.7, \, M_\rho< 2.3\,  {\rm TeV})$ (see Eq.(\ref{LambdavsMrho})).

1) ``Low $M_\rho$ scenario'' ($M_\rho<2.3 \, {\rm TeV},\,  \Lambda> M_\rho$, collider detection):

A typical example is  
$M_\rho=2$ TeV  
($g_{\rho\pi\pi}\simeq 5.7$), which is a simple scale-up of  the QCD $\rho$ meson, thus is perfectly natural with  $\Lambda\simeq 3.3\, {\rm TeV}\simeq 4\pi F_\pi$.
This  yields the ``broad width'' $\Gamma_\rho \simeq \Gamma_{\rho\rightarrow WW}\simeq g_{\rho\pi\pi}^2 M_\rho/(48\pi)  \simeq  433 $ GeV, which, although a scale-up of the $\rho$ meson width, may be barely detectable at
 LHC.   For larger (smaller) $M_\rho$ the width gets larger (smaller) as $\sim M_\rho^3$, and the  production cross section gets smaller (larger) as $\sim 1/M_\rho^2$, thus more difficult for $M_\rho>2$ TeV to be seen at LHC. 
 The SM rho with narrow resonance $\Gamma_\rho\lesssim100$ GeV if any could be detected at LHC for  $M_\rho\lesssim1.2$ TeV, which corresponds to $g_{_{\rm HLS}}\lesssim 3.5$ and $\Lambda \gtrsim 50\, {\rm TeV}$.

  2) ``High $M_\rho$  scenario''  ($M_\rho\gg 2.3 \, {\rm TeV},\,  \Lambda < M_\rho$,  as a stabilizer of the skyrmion dark matter $X_s$)\cite{Matsuzaki:2016iyq}:
  
 Even if no direct evidence were seen at the collider experiments,  physical effects of the dynamical $\rho_\mu$ are still observable through the skyrmion dark matter $X_s$ in the SM.  In fact the SM skyrmion is stabilized by the
  {\it off-shell} $\rho_\mu$ in the {\it short distance} physics as shown in Ref.\cite{Matsuzaki:2016iyq}, the result of which corresponds to $a\rightarrow \infty$ calculation, while the results are numerically similar even for $a\sim 2$ \cite{MOY2018}. The HLS coupling is extremely large $g_{_{\rm HLS}}={\cal O}(10^3) $, which yields 
  $M_{X_s}\lesssim {\cal O} (10)$ GeV  consistent with  the direct detection of the dark matter, and  in rough agreement with  
 the relic abundance of the dark matter: $\Omega_{X_s} h^2 \simeq 0.1$ \cite{Matsuzaki:2016iyq,MOY2018}. Note the cutoff is  $\Lambda =e^{-4/3} \tilde \Lambda \simeq e^{-4/3}\cdot M_\rho ={\cal O} (10^2\, {\rm TeV})$, where
$M_\rho = g_{_{\rm HLS}} \cdot F_\rho 
$ is  a typical mass scale (no longer  the ``on-shell'' mass, since the SM rho  is deeply off-shell).  

In either scenario, the phenomenologically interesting nonperturbative SM physics has  typical strong SM rho gauge coupling
$ g_{_{\rm HLS}}\simeq  1/3 - 10^3$, which will have the cutoff $\Lambda ={\cal O} (10^0 -10^2)\, {\rm TeV}$ 
close to the weak scale, thus resolving the naturalness problem. 
Note that $g_{_{\rm HLS}}$ diverges at the near  the low scale Landau pole $\tilde \Lambda=e^{4/3} \Lambda \simeq 3.8 \Lambda$, even if the Higgs self coupling is still perturbative and pSM perfectly makes sense.

This indicates that the quadratic divergence  corrections  to the  weak scale $\delta F_\pi^2\sim 4\cdot \Lambda^2/(4\pi)^2\sim (0.1\, {\rm TeV} - 10\,  {\rm TeV})^2$ (see the gap equation Eq.(\ref{quadraticdiv})).   This also suggests a possibility that the SM in the full nonperturbative
formulation  eventually reveals itself as  
 a  ``dual'' to a possible BSM underlying theory with  such a scale,
    similarly to the hadron-quark duality (nonlinear sigma model/chiral Lagrangian vs QCD), or as an analogue of the Seiberg duality to be discussed below. 
  \\

Another possible phenomenological impact would be the phase transition from the broken phase to the unbroken phase in the early Universe; the unbroken phase consists of the 3 massless HLS gauge bosons $\rho_\mu$'s and the 3 massive $\pi$'s (no longer the NG bosons),  plus 3 massive $\hat \rho$'s (no longer the would-be NG bosons to be absorbed into $\rho_\mu$ in the broken phase), plus 6 other spinless massive modes (corresponding to the 6 constraints in the broken phase), one of which is a pseudo-dilaton in the unbroken phase,  see Eq.(\ref{components}).
This is quite different from the conventional linear sigma model picture of the unbroken phase
having degenerate massive 3 $\hat \pi$'s and 3 $\hat \sigma$'s in Eq.(\ref{Higgs1}). 
Although the zero temperature phase transition is the second order, the finite temperature phase transition could be
different, in which case 
electroweak phase transition in the early Universe would be quite different from the conventional one.
 \\

As is easily seen from the calculations in the present paper, 
it is straightforward to do the same analyses as done here for 
other Grassmannian manifolds 
$U(N)/[U(N-p)\times U(p)]$ including $CP^{N-1}$ based on $G/H=U(N)/[U(N-1)\times U(1)]$, the result
being precisely the same for the above relations, i.e., $\rho$-universality, KSRF I,II, and VMD.

One notable example is the SUSY QCD with $N_c, N_f$ for $N_c<N_f<3N_c/2$, near the conformal window, 
whose effective theory is described by the Grassmannian manifold  $G/H=SU(N_f)/[SU(N_c)\times SU(N_f-N_c)]$ and 
further by the ``magnetic gauge theory'' in the Seiberg duality~\cite{Seiberg:1994bz}.
It was already pointed out \cite{Harada:1999zj,Harada:2003jx} that the HLS is a concrete realization of the magnetic gauge theory of the Seiberg duality in such a way that  $G/H=SU(N_f)/[SU(N_c)\times SU(N_f-N_c)]\simeq G_{\rm global}\times H_{\rm local}=
[SU(N_f)]_{\rm global} \times [SU(N_c)\times SU(N_f-N_c)]_{\rm local} $ for $N_c<N_f<3N_c/2$. It in fact corresponds to $N=N_f \rightarrow \infty$ with 
$p=N_f-N_c=$ fixed in the present case. We have mentioned that such a limit realizes the dynamical generation of the HLS only for $SU(N_f-N_c)_{\rm local}=SU(p)_{\rm local}$ but not $SU(N_c)_{\rm local}=SU(N-p)_{\rm local}$.  

Incidentally, this limit is nothing but the ``anti-Veneziano limit'' ~\cite{Matsuzaki:2015sya} of the ``large $N_f$ QCD'', $N_c,N_f \rightarrow \infty 
$ with $N_f/N_c=$fixed $(>1)$, near conformal window,  a concrete realization of the walking technicolor which predicted a technidilaton~\cite{Yamawaki:1985zg},
a composite Higgs behaving in the same way as the SM Higgs as a pseudo-dilaton described in the present paper.
 So this limit  is relevant to the SQCD near the conformal window as well. 
In fact the magnetic gauge theory $SU(N_f-N_c)$ in the Seiberg duality is an IR free theory,  similarly to the present case Eq.(\ref{IRfree}).
It was further argued~\cite{Komargodski:2010mc}
that Seiberg duality implies the VMD in the SQCD. Here we showed that the same result is indeed realized dynamically in the large $N$ limit, independently of the parameter $a$, 
even in the non-SUSY QCD.
\\

So, whatever underlying theory beyond the SM Higgs might be, 
the SM Higgs equivalent to the Grassmannian model as it stands should be regarded in its own right as a
consistent quantum theory on the basis of the nonperturbative formulation. Hence all the nonperturbative results given in the present paper 
are the dynamical consequences of the SM itself and   must also be  realized in a possible underlying theory such as the walking technicolor,
as far as  such a theory has  the same symmetry realization $G/H$.
\\

Furthermore,
on the basis of the duality between macroscopic theory and its microscopic one, if there exists an underlying theory of the Standard Model, it must also have  a spontaneously broken approximate scale symmetry to realize 
the 125 GeV Higgs as a pseudo-dilaton, besides its internal symmetry $G/H=[SU(2)_L\times SU(2)_R]/SU(2)_V\simeq O(4)/O(3)$,  as given in the form of Eq.(\ref{Higgs2}). 
An immediate candidate for such a UV completion is the walking technicolor~\cite{Yamawaki:1985zg,Bando:1986bg} 
 \footnote{Similar studies for suppressing the FCNC are also made without notion of the anomalous dimension and the scale symmetry/technidilaton~\cite{Holdom:1984sk}. For a recent review of the walking technicolor and technidilaton in view of the LHC experiments on the Higgs boson see Ref.\cite{Matsuzaki:2015sya}.
 },
which  indeed has a spontaneously broken approximate scale symmetry and its pseudo-dilaton, ``technidilaton'', as a composite Higgs, and  at the same time has a large anomalous dimension 
$\gamma_m\simeq 1$ to suppress the problematic Flavor-Changing Neutral Currents (FCNC) when extended to include the mass of the 
SM quarks and leptons. 

In fact, possible candidate gauge theories for the walking technicolor have been searched for on the lattice, particularly in the ``large $N_f$ QCD'' with $N_f \gg N_c=3$,
which is expected to be close to the conformal window in accord with the anti-Veneziano limit mentioned above, $N_c, N_f
\rightarrow \infty$ with $N_f/N_c=$ fixed ($>1$). In particular,
it was discovered~\cite{Aoki:2014oha,Appelquist:2016viq} that $N_f=8$ QCD has a light flavor-singlet scalar on the lattice as a candidate for the
technidilaton  (Such a light scalar was also found in $N_f=12$ QCD~\cite{Aoki:2013zsa,Kuti:2014epa}, possibly as a remnant of the conformal window). Further studies on this line  will be decisive 
for revealing a possible underlying theory beyond the SM.
\\

On the same token,
it would be even more important to check  whether or not the dynamical generation of the HLS gauge boson $\rho_\mu$ of the SM presented here is the case on the lattice, not just in the large $N$ limit dynamics. 
So far only triviality studies were made based on the conventional linear sigma model parameterization Eq.(Higgs1}). 

However,   different parameterization of  the same classical theory could lead to 
different quantum theory, as we have seen in the present mode which would become ill-defined in the unbroken phase, unless 
the gauge symmetry, the HLS, is explicitly introduced. It is well known that 
latticizing gauge theories is equivalent to nonlinear 
realization or vice versa~\cite{Bardeen:1979xx}. In other words, the nonlinear realization is nothing but the gauge theories on the lattice, the same as the HLS. Thus the fully nonperturbative lattice simulations of the SM Higgs Lagrangian in terms of the parameterization based on the
nonlinear realization Eq.(\ref{Higgs2}) (and its HLS version Eq.(\ref{SM-HLS})) could be different from the conventional
parameterization Eq.(\ref{Higgs1}).
\\

Incidentally, one might think that the SM, when combined with the Yukawa coupling and the electroweak gauge coupling, has no Landau pole below the Planck scale in the perturbative calculations, so that there would be
no urgent motivation for studying the nonperturbative dynamics. This arguments would make sense, if
these logically independent  different parts theories in the SM  were inter-correlated at deeper level such as in ``the final theory'', which is unfortunately no more than a dream theory at this moment.
Otherwise, such a result is just a phenomenon of the accidental cancellation in the running coupling coefficients among separate theories having a Landau pole of their own,  namely, the result itself would need logical explanation.
\\

We now conclude with a generic remark:
although useful, the concept of HLS itself in the broken phase is not literally of a rigorous physical sense. This  is actually the case for any spontaneously broken gauge symmetry including the SM electroweak gauge symmetry, namely,  the concept of the spontaneously broken ``gauge symmetry'' does not make real sense,
unless the coupling is very small !! It is really the convenience of the description, while the dynamical formation of the massive vector bound is a real fact, independently of the ``gauge symmetry'' as we showed in Eq.(\ref{massivevector}). As in the case of the SM electroweak theory, however, there is a notable exception where  the HLS is  very useful even in the broken phase, though not mandatory, that is, near the phase transition point, Eq.(\ref{HLStrivial}), where
the induced HLS coupling is (conceptually) small:  $N g^2_{_{\rm HLS}}=3 (4\pi)^2/\ln(\Lambda^2/M_\rho^2)\ll 1$ $(\Lambda^2/M_\rho^2 \gg 1)$, or $g\rightarrow g_{\rm crit}$ ($\Lambda^2/v^2 \rightarrow \infty$).
Indeed, this is the case for the $\rho$ meson in the QCD, where the HLS is rather useful even though the $\rho$ coupling is not  very small numerically (similarly to the QCD itself where the large $N_c$ limit works well for $1/N_c=1/3$, not very small numerically). Also some off-shell dynamics like skyrmion physics, the HLS
is a useful concept even for the massive (unstable) $\rho_\mu$ which is  far away from the phase transition point.
 
On the other hand,  in the unbroken phase, the gauge symmetry is mandatory, not just useful, namely the HLS comes into a rigorous  reality. {\it Were it not for the HLS,
the quantum theory  in the unbroken phase is ill-defined as in the case of $CP^{N-1}$, consistently with the Weinberg-Witten Theorem}~\cite{Weinberg:1980kq} which forbids massless spin $J\geq 1$ particles in quantum theory on the positive definite Hilbert space (i.e., without gauge symmetry), see Appendix \ref{CPNApp}. 

So, if a theory ought to be a {\it well-defined quantum theory independently of all possible different phases}  in the nonperturbative sense, {\it we necessarily have to introduce the HLS}, notwithstanding the fact that  
its presence  in the broken phase is no more than the convenience of description (redundancy of the description), except for the near phase transition point and some
off-shell physics like the skyrmion physics. 
\\

Finally, our results are not restricted to the SM Higgs Lagrangian but to the generic nonlinear sigma model of the same $G/H=O(4)/O(3) \simeq \left[SU(2)_L\times SU(2)_R\right]/SU(2)_V$, with/without nonlinearly realized (approximate) scale symmetry, since we showed that the  dynamical results obtained in the large $N$ limit are not sensitive to the presence of the pseudo-dilaton $\varphi$. Then it is readily applied to the 
{\it two-flavored QCD} in the chiral limit. \footnote{
Incidentally, our results also imply that some ``composite Higgs model'' based on $G/H=SO(N)/[SO(N-p)\times SO(p)]$ (e.g., 
$SO(5)/SO(4)$ \cite{Agashe:2004rs}) should have the dynamical  gauge boson of $SO(p)_{\rm local}$ HLS by the nonperturbative dynamics of the large $N$ in a way independent of $a$, without recourse to the UV completion.}

In particular,
 the so-called {\it successful $a=2$ results of the $\rho$ meson, i.e., $\rho$-universality, KSRF I and II, and vector meson dominance (VMD), 
 are now proved to be realized for any $a$ for the dynamical gauge boson of the HLS,  and thus  are  simply nonperturbative  dynamical consequences
in  the large $N$   limit but not a mysterious parameter choice $a=2$.}  
The dynamically generated kinetic term has a new free parameter, the $\rho$ coupling (related to the cutoff or Landau pole, Eq.(\ref{kinetic1})), which is adjusted to the reality as 
$g_{\rho\pi\pi}=g_{_{\rm HLS}}  \simeq 5.9$ corresponding to
 $m_\rho=g_{\rho\pi\pi} f_\rho=\sqrt{2} g_{_{\rm HLS}} f_\pi\simeq 770\, {\rm MeV}$ ($f_\pi\simeq 92 \, {\rm MeV}$), Eq.(\ref{rhomassresult1}). This  implies the  cutoff 
 (related to the Landau pole) $\Lambda =\tilde \Lambda \cdot e^{-4/3}= m_\rho \cdot e^{3 (4\pi)^2/(8 g^2_{_{\rm HLS}})} \cdot e^{-4/3} 
\simeq 1.1\, {\rm GeV}$ which coincides with the breakdown scale of the chiral perturbation theory $\Lambda_\chi \simeq 4 \pi f_\pi$.

The fact  is  a most remarkable triumph of the nonlinear sigma model as an effective field theory including full nonperturbative dynamics. It in fact becomes
a direct evidence of the  dynamically generation of the
HLS gauge boson in QCD !! 
Phrased differently, QCD knows the Grassmannian manifold! Or, Nature chooses Grassmannian manifold as the effective theory of QCD-like theories.

If the phase transition at zero temperature shown in this paper is also applied to the finite temperature or finite density 
phase transition, the unbroken phase is quite different from the conventional view of the relevant QCD phase transition.
The unbroken phase in the chiral limit would be accompanied by massless $\rho$ mesons of the unbroken HLS gauge symmetry (``magnetic gauge symmetry'' of the Seiberg dual, as pointed out in \cite{Harada:1999zj,Harada:2003jx}), while  $\pi$'s are no longer the NG bosons and hence are all massive, degenerate with the $\check \rho$'s which are no longer the would-be NG bosons absorbed 
into the $\rho_\mu$. The $\check \rho$'s, if existed, have exotic quantum number and hence are not  simple $\bar q q$ bound states, maybe $qq$ (color-flavor locking?), or multi-quarks or even 
glueballs, or mixtures of them?  The quark-gluon plasma discovered at RHIC is seemingly still a strongly coupled system (non-Abelian Coulomb phase?),
and may retain some bound states such as those in the unbroken phase found in the present work.
We shall come back to this point in the future.

The result obtained here in the large $N$ limit is based on the equivalence  $G/H=SU(2)_L\times SU(2)_R/SU(2)_V
\simeq O(4)/O(3)$ and its extension to $G/H=O(N)/[O(N-3)\times O(3)]$ in the large $N$ limit. 
As it stands, such a large $N$ limit is not available for the $N_f\geq 3$ QCD, just the same situation as the skyrmion model whose $N_f\geq 3$ extension is highly involved. Extension of the present result to the $N_f\geq 3$ QCD
is certainly a challenging project for the future.

\acknowledgments 
We would like to thank Taichiro Kugo for  very helpful numerous discussions including crucial suggestions and some concrete calculations, without which the paper would not have been materialized in its present form.
We also thank Shinya Matsuzaki and Hiroshi Ohki for collaborations on Ref.\cite{Matsuzaki:2016iyq,MOY2018}, which motivated the present work, and also for fruitful discussions. 
Stimulating discussions with Masayasu Harada and Masaharu Tanabashi are also appreciated.

\appendix
\section{Generic CCWZ Parameterization and its HLS model for the SM}
 \label{CCWZ}
 
  Here we review  the generic CCWZ nonlinear realization \cite{Coleman:1969sm,Callan:1969sn} and its generic HLS version
  \cite{Bando:1984pw,Bando:1987br} and its scale-invariant version equivalent to the SM Higgs Lagrangian.

 \subsection{CCWZ Nonlinear realization for the SM}
  \label{CCWZ2} 
  
The system having the symmetry $G$ spontaneously broken into $H (\subset G)$ may be described in terms of the
Nambu-Goldstone variables $\pi$ in the CCWZ nonlinear realization:
   \begin{eqnarray}
  \xi(\pi)=e^{i \pi^a X_a/F_\pi} (\in G/H)\,,\quad T_A=\{S_a \in {\cal G}\,, X_a \in {\cal G}-{\cal H} \} \,,\quad {\rm tr} (T_A T_B)= \frac{1}{2} \delta_{AB}\,,
  \quad \pi=\pi_a X_a\,,
       \end{eqnarray}
  which transforms under $G$ as
 \begin{eqnarray}
 \xi(\pi) \rightarrow h(\pi,g)\, \xi(\pi)\, g^\dagger\,, \quad (h\in H \,\,{\rm and}\,\,  g\in G)\,.
 \end{eqnarray}
  We define Maurer-Cartan one-form taking Lie-algebra value $\alpha_\mu(\pi)=\alpha_\mu^\dagger(\pi)$: 
\begin{eqnarray}
 \alpha_\mu(\pi)&\equiv&\partial_\mu \xi(\pi) \cdot \xi^\dagger(\pi)/i= 
 \frac{1}{i} \left[\left(\frac{i}{F_\pi}\right)\partial_\mu \pi + \frac{1}{2!}\left(\frac{i}{F_\pi}\right)^2  \left[\pi,\partial_\mu \pi\right]  +\frac{1}{3!}
\left(\frac{i}{F_\pi}\right)^3\left[\pi,\left[\pi,\partial_\mu \pi \right] \right]+ \cdots\right] \nonumber \\
&=& 
  \alpha_{\mu,||}(\pi)+\alpha_{\mu,\perp}(\pi)\,, \nonumber \\
 \alpha_{\mu,\perp}(\pi)&\equiv& 2 {\rm tr} (\alpha_\mu(\pi) X_a) \cdot X_a = \frac{1}{F_\pi}\partial_\mu \pi  +\cdots\,,  \nonumber\\       \alpha_{\mu,||}(\pi)&\equiv& 2 {\rm tr} (\alpha_\mu(\pi) S_a) \cdot S_a= 
 \frac{i}{2 F_\pi^2}  \left[\pi,\partial_\mu \pi\right]  +\cdots\,,
   \end{eqnarray}
   where we confined ourselves to the  symmetric coset space $[ {\cal G}-{\cal H},  {\cal G}-{\cal H}] \subset {\cal H}$ such that $\alpha_{\mu,\perp}(\pi)$ contains only odd number of $\pi$
  and $\alpha_{\mu,||}
  (\pi)$ does even number of $\pi$, respectively.  
 They 
transform as
\begin{eqnarray}  
 \alpha_\mu(\pi)  &\rightarrow& h(\pi,g) \,  \alpha_\mu(\pi) \,  h^\dagger(\pi,g)  - i \partial_\mu  h(\pi,g)  \cdot    h^\dagger (\pi,g) \,,\\
 \alpha_{\mu,||}(\pi) &\rightarrow& h(\pi,g) \,  \alpha_{\mu,||}(\pi)\,  h^\dagger(\pi,g)  - i \partial_\mu  h(\pi,g)  \cdot    h^\dagger (\pi,g) \,,\\ \alpha_{\mu,\perp}(\pi) \  &\rightarrow&     h(\pi,g)  \, \alpha_{\mu,\perp}(\pi) \, h^\dagger(\pi,g) \,,
 \end{eqnarray}
and hence the invariant as the Lagrangian takes the form
\begin{eqnarray}
{\cal L}_{_{\rm CCWZ}} = F_\pi^2 \cdot {\rm tr}\left( \alpha_{\mu,\perp}(\pi)\right)^2={\rm tr}\left((\partial_\mu\pi)^2+\cdots\right)= \frac{1}{2} \left(\partial_\mu \pi_a\right)^2+\cdots\,.
\end{eqnarray} 
\\

For $G/H=SU(2)_L\times SU(2)_R/SU(2)_V$, with $S^a/X^a=(T_R^a\pm T_L^a)/2$, we may write 
 \begin{eqnarray}
   U(x)&=&e^{ i\frac{2\pi}{F_\pi}}=\xi(\pi) \cdot \xi(\pi)=[\xi^\dagger(\pi)]^\dagger \cdot\xi(\pi)\,,   \quad \left(\xi(\pi)=e^{ i\frac{\pi}{F_\pi}} \right)\,,
   \nonumber\\
   &\rightarrow& g_L U(x) g_R^\dagger
   \,,  \\
 \left( \xi(\pi),\xi^\dagger(\pi)\right)
  &\rightarrow&  
  h(\pi,g)\, \left( \xi(\pi),\xi^\dagger(\pi)\right)\, g_{R,L}^\dagger \,.
  \end{eqnarray}
 The Maurer-Cartan one-form reads
  \begin{eqnarray}
   \alpha_{\mu}(\pi)_{(R,L)}  &=& \left(\partial_\mu \xi(\pi)\cdot \xi^\dagger (\pi)\,, \partial_\mu \xi^\dagger (\pi)\cdot \xi (\pi)  \right)/i    \,,\nonumber \\
    \alpha_{\mu,||}(\pi)&=&\left(\partial_\mu \xi(\pi)\cdot \xi^\dagger (\pi) + \partial_\mu \xi^\dagger(\pi)\cdot \xi(\pi) \right)/(2i)\,,\nonumber  \\
   \alpha_{\mu,\perp}(\pi)&=&\left(\partial_\mu \xi(\pi)\cdot \xi^\dagger (\pi) - \partial_\mu \xi^\dagger(\pi)\cdot \xi(\pi) \right)/(2i)\,.     \end{eqnarray}
   and then an invariant yields the CCWZ Lagrangian:
   \begin{eqnarray}
 {\cal L}_{_{\rm CCWZ}} = F_\pi^2  {\rm tr}  \left(\alpha^2_{\mu,\perp}(\pi)\right)&=& \frac{F_\pi^2}{4}{\rm tr} \left( \partial_\mu U \partial^\mu U^\dagger \right) \,,
 \label{Uform}
    \end{eqnarray} 
where use has been made of 
$ \alpha_{\mu,\perp}(\pi)=\left(\partial_\mu \xi(\pi)\cdot \xi^\dagger (\pi) - \partial_\mu \xi^\dagger(\pi)\cdot \xi(\pi) \right)/(2i)=\xi^\dagger(\pi) \partial_\mu U(x) \, \xi(\pi)/(2i)= \xi(\pi) \partial_\mu U^\dagger(x) \, \xi^\dagger(\pi)/(2i)$.  
Then the SM Lagrangian with $F_\pi=v$, Eq.(\ref{Higgs2}), is further rewritten as
 \begin{eqnarray}
 {\cal L}_{\rm SM} 
&=&\chi^2(\varphi) \cdot 
 \left[ 
\frac{1}{2} 
\left(
\partial_\mu \varphi 
\right)^2  
+ v^2 {\rm tr}\left(\alpha_{\mu,\perp}^2(\pi)\right)
\right] 
-V(\varphi)\,.
\label{Higgs4}
 \end{eqnarray}

\subsection{SM Higgs Lagrangian as a Scale-invariant HLS model \cite{Fukano:2015zua}}
\label{sHLS}

The SM Higgs Lagrangian was further shown \cite{Fukano:2015zua} to be {\it gauge equivalent} to the {\it scale-invariant} 
version \cite{Kurachi:2014qma} of the Hidden Local Symmetry (HLS) Lagrangian \cite{Bando:1984ej,Bando:1987br,
Harada:2003jx}, which contains {\it possible new vector boson} $\rho_\mu$, ``SM rho'', {\it hidden behind the SM Higgs Lagrangian},
as an analogue of the QCD $\rho$ meson.
\\

In the generic case, nonlinear sigma model based on the manifold $G/H$, the HLS model having a symmetry  $G_{global} \times H_{\rm local}$ can be written 
in terms of the base $\xi(x)$ transforming as~\cite{Bando:1984pw,Bando:1987br}
\begin{eqnarray}
\xi(x) \rightarrow h(x)\cdot \xi(x) \cdot g^\dagger\,, \quad  h(x)\in H_{\rm local}\,,\,  g\in G_{\rm global}, 
\end{eqnarray}
which   may be parameterized as
\begin{eqnarray}
\xi (x)&=&\xi(\check \rho) \cdot \xi(\pi)\,,
\quad
\xi(\check \rho)
=\exp \left( i\frac{\check \rho}{F_\rho}   \right),  \,
\xi(\pi)=\exp \left( i\frac{\pi}{F_\pi}   \right)\,,  \nonumber\\
 \check \rho&=&\check \rho_a S_a\,, \pi=\pi_a X_a\,,\quad \left(S_a\in {\cal H}\,, X_a \in {\cal G}-{\cal H}\right)\,,
 \label{HLSparameterization}
  \end{eqnarray}
with   $\check{\rho}$ and $F_\rho$ being the would-be NG boson  to be absorbed into the HLS gauge boson  $\rho_\mu$ by the Higgs mechanism  and its  decay constant, respectively (See the discussions below), where   $\xi(\check \rho) = e^{i \check{\rho}/F_\rho}$ transforms as $\xi(\check \rho) \rightarrow h(x)\cdot  \xi(\check \rho)\cdot h^\dagger(\pi,g)$ and the CCWZ base $\xi(\pi)$ does as $\xi(\pi) \rightarrow h(\pi,g)\cdot \xi(\pi)\cdot g^\dagger$.
The Maurer-Cartan one-form reads:
\begin{eqnarray}
 \alpha_\mu(x)&=&\frac{1}{i}\partial_\mu \xi(x) \cdot \xi^\dagger(x)= \frac{1}{i} \partial_\mu \xi(\check \rho)\cdot \xi^\dagger(\check \rho)
 + \xi(\check \rho)\cdot \alpha_\mu(\pi) \cdot \xi^\dagger(\check \rho)\,, \nonumber\\
  \alpha_{\mu,\perp}(x) &=&\frac{2}{i} {\rm tr} \left(
\partial_\mu \xi(x) \cdot \xi^\dagger (x) X_a
\right)
 X_a   
  =
  \xi(\check \rho)\alpha_{\mu,\perp}(\pi) \xi^\dagger(\check \rho)\,,
  \nonumber\\
   \alpha_{\mu,||}(x) &=&\frac{2}{i} {\rm tr} \left(
\partial_\mu \xi(x) \cdot \xi^\dagger (x) S_a
\right)
 S_a   =
\frac{1}{i}\partial_\mu \xi(\check \rho)\cdot \xi^\dagger(\check \rho)   +\xi(\check \rho)\alpha_{\mu,||}(\pi) \xi^\dagger(\check \rho)\nonumber\\
   &=&
  \frac{1}{i} \left[\frac{1}{F_\rho} \partial_\mu\check \rho + \frac{1}{2!}\left(\frac{i}{F_\rho}\right)^2  \left[\check \rho,\partial_\mu \check \rho\right]  +\frac{1}{3!}
\left(\frac{i}{F_\rho}\right)^3\left[\check \rho,\left[\check \rho,\partial_\mu \check \rho\right] \right]+ \cdots\right]
  + \xi(\check \rho)\alpha_{\mu,||}(\pi) \xi^\dagger (\check \rho) \,,     
  \label{MaurerCartangeneric}
   \end{eqnarray}
where use has been made of ${\rm tr}
\left(\partial_\mu \xi(\check \rho) \cdot \xi^\dagger (\check \rho) X_a\right) =0$.

When we fix the gauge of HLS as $\xi(x)=\xi(\pi) $ (unitary gauge $\xi(\check \rho)=1,\, \check{\rho}=0$). 
 $H_{\rm local}$ and $H_{\rm global} (\subset G_{\rm global})$ 
get simultaneously broken spontaneously (Higgs mechanism), leaving the 
diagonal subgroup $H=H_{\rm local}\oplus H_{\rm global}$, which is nothing but the subgroup $H$ of the original $G$ of $G/H$: $H\subset G$. Then the
extended symmetry $G_{\rm global} \times H_{\rm local}$ is simply reduced back to the original nonlinear realization of $G$ on the manifold $G/H$,
both are gauge equivalent to each other. 
\\

The HLS gauge boson,  $\rho_\mu(x)$, is introduced as usual  by a covariant derivative as 
\begin{equation}
D_\mu \xi (x) = \partial_\mu \xi (x)-i \rho_\mu(x) \xi(x) \,,\quad \rho_\mu=\rho_\mu^a S_a \,,  
\label{HLScovariant}
\end{equation}
which transforms in the same way as $\xi(x)$.
The covariantized Maurer-Cartan one-form reads:
\begin{eqnarray}
{\hat \alpha}(x) &\equiv&
\frac{1}{i} D_\mu \xi(x) \cdot \xi^\dagger(x) =\alpha_\mu(x)-\rho_\mu(x)= {\hat \alpha}_{\mu,\perp}(x) + {\hat \alpha}_{\mu,||}(x) \,,\nonumber\\
{\hat \alpha}_{\mu,\perp} (x)
&=&\frac{2}{i} {\rm tr} 
\left(
D_\mu \xi(x) \cdot \xi^\dagger (x) X_a
\right)
 X_a 
=\alpha_{\mu,\perp}(x)= \xi(\check \rho) \alpha_{\mu,\perp}(\pi) \xi^\dagger(\check \rho)\,,\nonumber\\
 {\hat \alpha}_{\mu,||} (x)&=& \frac{2}{i} {\rm tr} 
 \left(
 D_\mu \xi(x) \cdot \xi^\dagger (x) S_a
 \right) S_a 
   =\alpha_{\mu,||}(x) - \rho_\mu(x)\, .
 \end{eqnarray} 
They both transform as
 \begin{eqnarray}
 {\hat \alpha}_{\mu,\perp,||}  \rightarrow h(x) \cdot  {\hat \alpha}_{\mu,\perp,||} \cdot h^\dagger (x)\,.
 \end{eqnarray}
 
 Thus the HLS Lagrangian consists of two invariants:
 \begin{eqnarray}
 {\cal L}_{\rm HLS} &=& {\cal L}_A + a \, {\cal L}_V\,, \nonumber\\
 {\cal L}_A&=&F_\pi^2\, \,  {\rm tr}  \left({\hat \alpha}_{\mu,\perp}^2(x) \right)=F_\pi^2 \, {\rm tr}  \left({\alpha}_{\mu,\perp}^2(\pi) \right)
 ={\rm tr} \left((\partial_\mu \pi)^2 +\cdots \right)=\frac{1}{2} \left(\partial_\mu \pi_a\right)^2+\cdots\,,  
 \nonumber\\ 
  {\cal L}_V
    &=&F_\pi^2\,  {\rm tr}  \left({\hat \alpha}_{\mu,||}^2(x) \right)
 = F_\pi^2 \, {\rm tr} \,  \left( \alpha_{\mu,||}(x) - \rho_\mu(x)\right)^2  \nonumber \\      
  &=& F_\pi^2 \,  {\rm tr}
  \left[
  \left(
  \frac{1}{F_\rho} \partial_\mu {\check \rho} -
 \frac{i}{2 F_\rho^2}
 \left[
 \partial_\mu {\check \rho}, \check \rho 
 \right]- 
 \frac{i}{F_\pi^2}
 \left[
 \partial_\mu \pi,\pi
 \right]  \right)
  -  \rho_\mu(x)     +\cdots  \right]\,.
 \label{HLSgeneral}
     \end{eqnarray} 
Note that  ${\cal L}_A$ is identical to the original nonlinear sigma model based on $G/H$, since $\xi(\check \rho)$ carrying the gauge transformation 
in the representation Eq.({\ref{HLSparameterization}) has been traced out in  ${\cal L}_A$, which is equivalent to taking the unitary gauge $\xi(\check \rho)=1 (\check \rho=0)$.
On the other hand, a new term $a {\cal L}_V$ contains  the kinetic term of the $\check \rho$, which  is normalized as  the canonical one by the requirement:
\begin{eqnarray}
  a=  \frac{F_\rho^2}{F_\pi^2}\,.
\end{eqnarray}
The $\rho_\mu$ field at classical level is merely an auxiliary field without kinetic term and hence can be solved away by the equation of motion
\begin{eqnarray}
\rho_\mu=\alpha_{\mu,||}(x)\,, \,{\rm i.e.}, \quad   a {\cal L}_V\equiv 0\,.
\label{EM}
\end{eqnarray}
Thus the HLS Lagrangian Eq.(\ref{HLSgeneral}) at classical level is gauge equivalent to the original nonlinear sigma model based on $G/H$:
${\cal L}_{\rm HLS}={\cal L}_A$. 
\\

In the case at hand, the SM in the form of Eq.(\ref{Higgs2}), the gauge-equivalent  HLS Lagrangian having the SM rho,
$\rho_\mu$, 
reads:
\begin{eqnarray}
{\cal L}_{\rm SM-HLS}= 
&=&\chi^2(\varphi) \cdot \left[ \frac{1}{2} \left(\partial_\mu \varphi\right)^2  +{\cal L}_A+ a {\cal L}_V)\right] -V(\varphi)\,, \nonumber\\
{\cal L}_A
&=& F_\pi^2\, {\rm tr} \left(\hat \alpha_{\mu,\perp} (x)\right)^2 
=   F_\pi^2\, {\rm tr} \left(\alpha_{\mu,\perp} (\pi)\right)^2 = 
{\rm tr} \left((\partial_\mu \pi)^2 +\cdots \right)\,,  
\nonumber\\
a {\cal L}_V&=&F_\rho^2\, {\rm tr} \left(\hat \alpha_{\mu,||} (x)\right)^2 = F_\rho^2 \,  {\rm tr}
  \left[
  \left(\rho_\mu-
  \frac{1}{F_\rho} \partial_\mu {\check \rho}
  \right) -  \frac{i}{2 F_\rho^2}
 \left[
 \partial_\mu {\check \rho}, \check \rho 
 \right]- 
 \frac{i}{2F_\pi^2}
\left[
 \partial_\mu \pi,\pi\right]
 +\cdots
 \right]^2\,, \nonumber \\
 F_\rho^2&=& a F_\pi^2= a v^2\,,
  \label{SM-HLS} 
  \end{eqnarray}
  which is obviously reduced back to  Eq.(\ref{Higgs2}) (and hence the original SM Lagrangian Eq.(\ref{Higgs1})) by 
  solving away the auxiliary field $\rho_\mu$ as  Eq.(\ref{EM}) with fixing the gauge $\xi(\check \rho)=1 (\check \rho=0)$ and/or 
  using the parameterization Eq.(\ref{HLSparameterization}).
 \\
 
 i) 
 Parameterization for $SU(2)_L\times SU(2)_R/SU(2)_V\simeq [SU(2)_L\times SU(2)_R]_{\rm global}\times [SU(2)_V]_{\rm local}$:
 
 A more familiar notation in this case  is 
 by dividing $U(x)$ into two parts \cite{Bando:1984ej,Bando:1987br,
Harada:2003jx}:
\begin{equation}
 U(x)=e^{2 i  \frac{\pi(x)}{F_\pi}}= \xi_L^\dagger(x) \cdot \xi_R(x)\,, 
   \label{U:decomp}
\end{equation}  
where $\xi_{R,L}(x)$ 
transform under $G_{\rm global} \times H_{\rm local}$ as
\begin{eqnarray}
\xi_{R,L}(x) \rightarrow h(x) \cdot \xi_{R,L}(x) \cdot g_{R,L}^\dagger\,,\quad 
U(x) \rightarrow g_L U(x)  g_R^\dagger\,,  \nonumber \\ 
 \left(h(x)\in H_{\rm local},\, g_{R,L}\in G_{\rm global} \right)\,.
\end{eqnarray}
The $H_{\rm local} $ is  regarded as a gauge symmetry of group $H$ arising from the redundancy (gauge symmetry) how to divide
$U$ into two parts. $\xi_{R,L}$, which  can be parameterized as
\begin{equation}
 \left(\xi_R(x),\xi_L\right(x))=\xi(\check \rho)\cdot \left(\xi(\pi)\,, \xi^\dagger (\pi)\right)\,.
 \label{chiralHLSrepresentation}
\end{equation}

The covariant derivative reads
\begin{equation}
D_\mu \xi_{R,L}(x) = \partial_\mu \xi_{R,L} (x)-i \rho_\mu(x) \xi_{R,L}(x) \,,\,\, \rho_\mu=\rho_\mu^a\frac{\tau^a}{2}\,,
\label{HLScovariant}
\end{equation}
and the Maurer-Cartan one-forms are 
  \begin{eqnarray}
   \{  {\hat \alpha}_{\mu,R,L}(x), 
   {\hat \alpha}_{\mu, ||, \perp}(x)
   \}
  &\rightarrow& h(x) \cdot \{ {\hat \alpha}_{\mu, R,L} (x),
  {\hat \alpha}_{\mu, ||,\perp}(x)
  \}\cdot h^\dagger(x)\,,\nonumber \\ 
  {\hat \alpha}_{\mu,R,L}(x)
  &\equiv& \frac{1}{i}D_\mu \xi_{R,L}(x)
  \cdot \xi_{R,L}^\dagger(x)
  =\frac{1}{i}\partial_\mu \xi_{R,L}(x)
   \cdot \xi_{R,L}^\dagger (x)
 - \rho_\mu(x)
 \,, \nonumber\\ 
 {\hat \alpha}_{\mu, ||,\perp}(x)
 &\equiv & \frac{1}{2}\left({\hat \alpha}_{\mu,R}(x)
 \pm  {\hat \alpha}_{\mu, L}(x)
 \right) 
= 
\Bigg\{ \begin{array}{c}
  {\alpha}_{\mu ||}(x)
   - \rho_\mu(x)
  \nonumber \\  
 {\alpha}_{\mu \perp}  (x)
 \end{array}\,,\\
      \end{eqnarray}
      where
      \begin{eqnarray}
   {\hat \alpha}_{\mu ||}(x)
    &=&  \frac{1}{2i} \left(D_\mu \xi_{R}(x)
    \cdot \xi_{R}^\dagger(x)
    + D_\mu \xi_{L}(x)
     \cdot \xi_{L}^\dagger(x)
      \right) =\frac{1}{F_\rho} \partial_\mu \check{\rho}-\frac{i}{2 F_\rho^2}[\partial_\mu {\check \rho},{\check \rho}]   - \frac{i}{2 F_\pi^2}[\partial_\mu \pi,\pi] +\cdots\,,  \nonumber\\
    {\hat \alpha}_{\mu \perp} (x)
    &=&
    \frac{1}{2i} \left(D_\mu \xi_{R} (x)
    \cdot \xi_{R}^\dagger(x)
    - D_\mu \xi_{L}(x)
     \cdot \xi_{L}^\dagger(x)
       \right)=  \alpha_{\mu \perp} (x)       =\xi(\check \rho) \cdot  {\alpha}_{\mu \perp} (\pi)\cdot \xi^\dagger (\check \rho)  \nonumber\\
      &=&
       \frac{1}{2i}\xi_L \left( \partial_\mu U\cdot U^\dagger\right) \xi_L^\dagger=\frac{1}{2i}\xi_R\left(  \partial_\mu U^\dagger\cdot U\right) \xi_R^\dagger\,.
         \label{chiralalpha}
                 \end{eqnarray}
Then  ${\cal L}_A$ takes a familiar form:
 \begin{eqnarray}
 {\cal L}_A=F_\pi^2 \,{\rm tr} \left({\alpha}_{\mu \perp}^2 (x) \right)=F_\pi^2 \,{\rm tr} \left({\alpha}_{\mu \perp}^2 (\pi) \right) =\frac{F_\pi^2}{4}{\rm tr} \left(\partial_\mu U \partial^\mu U^\dagger\right) \,,
  \end{eqnarray}
 where again the gauge-variant field $\xi(\check \rho)$ in the parameterization Eq.(\ref{chiralHLSrepresentation}) is traced out in  ${\cal L}_A$, which is equivalent to taking the unitary gauge.
   While $a {\cal L}_V$ takes the same form as Eq.(\ref{SM-HLS}). 
 \\
 
   ii) Parameterization for  $O(4)/O(3) \simeq   O(4)_{\rm global}\times O(3)_{\rm local}$:
   
   The CCWZ base $\xi^{(4)}(\pi)$ for the $O(4)/O(3)$ given in Eq.(\ref{O4}) is extended to the HLS base $\xi^{(4)}(x)=\xi^{(4)}(\check \rho) \cdot \xi^{(4)}(\pi)$ as in the generic case Eq.(\ref{HLSparameterization}), with a different normalization ${\rm tr} (T_A T_B)= 2 \delta_{AB},\, T^t_A=-T_A$ . Everything is the same as the generic case, Eqs.(\ref{MaurerCartangeneric}) - (\ref{HLSgeneral}),    except for the normalization and we have the HLS version of the SM in terms of this parameterization corresponding to Eq.(\ref{SM-HLS}):
\begin{eqnarray}
  {\cal L}_{\rm SM-HLS}= 
&=&\chi^2(\varphi) \cdot \left[ \frac{1}{2} \left(\partial_\mu \varphi\right)^2  +{\cal L}_A+ a {\cal L}_V)\right] -V(\varphi)\,, \nonumber\\
{\cal L}_A
&=& \frac{F_\pi^2}{4}\, {\rm tr} \left(\hat \alpha^{(4)}_{\mu,\perp} (x)\right)^2 
=   \frac{F_\pi^2}{4}\, {\rm tr} \left(\alpha^{(4)}_{\mu,\perp} (\pi)\right)^2 = 
\frac{1}{4}{\rm tr} \left((\partial_\mu \pi)^2 +\cdots \right)\,,  
\nonumber\\
a {\cal L}_V&=&\frac{F_\rho^2}{4}\, {\rm tr} \left(\hat \alpha_{\mu,||} (x)\right)^2 = \frac{F_\rho^2}{4} \,  {\rm tr}
  \left[
  \left(\rho_\mu-
  \frac{1}{F_\rho} \partial_\mu {\check \rho}
  \right) -  \frac{i}{2 F_\rho^2}
 \left[
 \partial_\mu {\check \rho}, \check \rho 
 \right]- 
 \frac{i}{2F_\pi^2}
\left[
 \partial_\mu \pi,\pi\right]
 +\cdots
 \right]^2\,, \nonumber \\
 F_\rho^2&=& a F_\pi^2= a v^2\,.
 \label{SM-HLSO4}
 \end{eqnarray} 
This form is the basis for the Grassmannian N-extension $O(N)/[O(N-p)\times O(p)] \simeq O(N)_{\rm global}\times [O(N-p)\times O(p)]_{\rm local}$ as the main target of the present paper.

 \subsection{Physical Implications of the 
 Dynamical Generation of the HLS Gauge Boson}
  \label{DGHLS}
  
If the SM rho as an auxiliary field $\rho_\mu$ acquires the kinetic term 
\begin{equation}
{\cal L}^{(\rho)}_{\rm kinetic} =- \frac{1}{2 g_{_{\rm HLS}}^2}\,  {\rm tr}\,  \rho_{\mu\nu}^2
\end{equation}
 by the quantum corrections, with $g_{_{\rm HLS}}$
being the induced gauge coupling of the HLS,   
then the quantum theory for the SM Higgs would take the form:
\begin{eqnarray}
{\cal L}^{\rm quantum}_{\rm SM-HLS}&=& \chi^2
\cdot \left(
\frac{1}{2} \left(\partial_\mu \varphi\right)^2 +
 \frac{v^2}{4}\cdot {\rm tr} \left(\partial_\mu U \partial^\mu U^\dagger \right) + F_\rho^2 
 \cdot  {\rm tr} \, \left(\rho_\mu
  - {\alpha}_{\mu ||}
  \right)^2 \right) -V(\varphi) \nonumber \\
 &-& \frac{1}{2 g^2_{_{\rm HLS}}}\, {\rm tr}\, \rho_{\mu\nu}^2 +\cdots
 \,,
 \label{qHiggsHLS}
 \end{eqnarray}
where ``$\cdots$'' stands for other induced terms at quantum level.

When it happens, after rescaling the kinetic term of $\rho_\mu$, $\rho_\mu(x) \rightarrow g_{_{\rm HLS}}\, \rho_\mu(x)$ to the canonical one $-\frac{1}{2 }\, {\rm tr} \,\rho_{\mu\nu}^2$, 
the $\chi^2 \, a {\cal L}_V$ term yields  the scale-invariant mass term of $\rho_\mu$, which reads, when fixing the gauge $(\check \rho=0)$:
\begin{eqnarray}
\chi^2\, a  {\cal L}_V&=& \chi^2
F_\rho^2 \cdot   {\rm tr} \, \left(g_{_{\rm HLS}} \rho_\mu
 - {\alpha}_{\mu,||}
 \right)^2=
 M_\rho^2 \,{\rm tr} (\rho_\mu
 )^2+ g_{\rho\pi\pi} \cdot 2i {\rm tr} \left(\rho^\mu\left[\partial_\mu \pi,\pi\right]\right) +\cdots\,,\nonumber \\
 M_\rho^2&=& g_{_{\rm HLS}}^2 F_\rho^2 = g_{_{\rm HLS}}^2 (a\, v^2)= a g_{_{\rm HLS}}^2  F_\pi^2 \,,\quad g_{\rho\pi\pi}= \frac{F_\rho^2}{2 v^2}\, g_{_{\rm HLS}}=\frac{a}{2} g_{_{\rm HLS}}\,, 
 \label{rhomass}
 \end{eqnarray}
 with the mass acquired by the Higgs mechanism mentioned above. 
  \\

 For a specific choice $a=2$ (which turns out to be the case in the Grassmannian models in the large $N$ limit as shown in the text), we have  \cite{Bando:1984ej,Bando:1987br} the ``universality'' of the $\rho_\mu$ coupling:
 \begin{eqnarray}
 g_{\rho\pi\pi}=\frac{a}{2} g_{_{\rm HLS}}=  g_{_{\rm HLS}}\,, \quad \left(a=2\right)\,,
 \quad\left({\rm Universality}\right)
 \label{rhouniversality}
 \end{eqnarray}
and 
the standard KSRF II relation 
:
\begin{equation}
 M_\rho^2=\left(\frac{2}{a} g_{\rho\pi\pi}\right)^2 F_\rho^2 
 =\frac{4}{a} g^2_{\rho\pi\pi} F_\pi^2  = 2g^2_{\rho\pi\pi} F_\pi^2\,, \quad \left(a=2\right)\,,\quad
 \left({\rm KSRF}\,\,{\rm II}\right)\,.
 \label{KSRFII}
 \end{equation} 
 and 
 \begin{eqnarray}
 F_\rho^2= a F_\pi^2 \quad =2 F_\pi^2 \,,\quad \left( a=2\right) \,.
 \label{Frhoa2}
 \end{eqnarray}
\\

Note that {\it the HLS gauge boson acquires the
scale-invariant mass term thanks to the dilaton factor $\chi^2$}, the nonlinear realization of the scale symmetry, in sharp contrast to the {\it Higgs (pseudo-dilaton)  which acquires mass only from the explicit breaking of the scale symmetry}. 
\\

 The electroweak gauge bosons ($\in {\cal R}_\mu ({\cal L}_\mu)$) are introduced by extending the covariant derivative of Eq.(\ref{HLScovariant}) 
this time by gauging $G_{\rm global}$, which is {\it independent of $H_{\rm local}$} in  the HLS extension:
\begin{equation}
D_\mu \xi_{R,L}(x)\Rightarrow {\hat D}_\mu \xi_{R,L}(x)\equiv  \partial_\mu \xi_{R,L} (x)-i \rho_\mu(x) \, \xi_{R,L}(x)  +i \xi_{R,L}(x)\, {\cal R}_\mu ( {\cal L}_\mu)\,.
\label{fullcovariant}
\end{equation}
We then finally have a gauged s-HLS version of the Higgs Lagrangian (gauged-s-HLS):\footnote{
This form of the Lagrangian  is the same as that of the effective theory of the one-family ($N_F=8$) walking technicolor \cite{Kurachi:2014qma}, except for the shape of the scale-violating 
potential $V(\varphi)$  which has a scale dimension 4 (trace anomaly) in the case of the walking technicolor instead of  2 of the SM Higgs case (Lagrangian mass term). 
}
 \begin{equation}
 {\cal L}^{\rm gauged}_{\rm Higgs-HLS}=  \chi^2(x)\cdot 
 \left[
\frac{1}{2} \left(\partial_\mu \varphi\right)^2  + {\hat {\cal L}}_A+ a {\hat {\cal L}}_V
  \right] - V(\varphi) + {\cal L}_{\rm kinetic}^{(\rho, {\cal L},\cal{R})} +\cdots \,, 
  \label{gaugedsHLS}
  \end{equation}
 with 
 \begin{equation}
  {\hat {\cal L}}_{A,V}= {\cal L}_{A,V} \left( D_\mu \xi_{R,L}(x) \Rightarrow {\hat D}_\mu \xi_{R,L}(x)\right) \,.
     \end{equation}
When $({\cal R}_\mu, {\cal L}_\mu)$ are treated as external source fields (thus NG bosons $\pi$ are not absorbed into $W/Z$ as in the QCD case), we would have
 for $a=2$ the celebrated ``vector meson dominance (VMD)'' where direct coupling of electroweak gauge fields to $\pi \pi$ are cancelled between 
 those from ${\cal L}_A$ and ${\cal L}_V$ terms, so that the couplings go only through mixing with the SM rho  $\rho_\mu$: \cite{Bando:1984ej,Bando:1987br}
  \begin{equation}
 g_{({\cal R}_\mu/{\cal L}_\mu)\pi\pi} = 1-\frac{a}{2}\quad =0\,, \quad (a=2)\,, \quad \left({\rm Vector}\,\, {\rm Meson}\,\,{\rm Dominance}\right)
 \label{VMD2}
 \end{equation}
\\ 
 
 In the present SM case, the standard Higgs mechanism in  ${\cal L}_A$ yields the conventional $W/Z/\gamma$ mass mixings in the SM, while
 ${\cal L}_V$ term yields additional  mass mixing  $\rho-W/Z/\gamma$ 
with the $\gamma$ mass staying exactly zero after diagonalization as usual. The VMD in Eq.(\ref{VMD2}) implies that $q \bar q \rightarrow W/Z/\gamma \rightarrow W_LW_L/W_LZ_L \,(\pi \pi)$, can only go 
through  {\it the Drell-Yang process} as a production  of $\rho_\mu$ due to the $W/Z/\gamma-\rho_\mu$ mixing: $W/Z/\gamma \rightarrow \rho_\mu\rightarrow  W_LW_L/W_LZ_L$,  with the coupling $\sim \alpha_{\rm em} g_\rho/M_\rho^2= \alpha_{\rm em} F_\rho/M_\rho=\alpha_{\rm em}/g_{_{\rm HLS}}$. This is similar to the walking technirho
described also by the s-HLS effective theory  \cite{Kurachi:2014qma,Fukano:2015hga,Fukano:2015uga,Fukano:2015zua}. 
 \\
 
Eq.(\ref{gaugedsHLS}) also yields a {\it notable $a-$independent relation (KSRF I)} between the $\rho-\gamma$ mixing strength $g_\rho$ and $g_{\rho\pi\pi}$ from the mass term $\chi^2 a {\cal L}_V$ \cite{Bando:1984ej,Bando:1987br} (low energy theorem of HLS~\cite{Bando:1984pw,Bando:1987br}: Proof in Ref. \cite{Harada:1993jk}) :
\begin{equation}
g_{\rho}=M_\rho F_\rho=g_{_{\rm HLS}} F_\rho^2 =\left(\frac{2}{a} g_{\rho\pi\pi}\right)\cdot  a F_\pi^2 =2 g_{\rho\pi\pi} F_\pi^2 \quad\left(a-{\rm independent}\right)\,,
\quad \left({\rm KSRF}\,\,{\rm I}\right)\,,
 \label{KSRFI}
 \end{equation}
in accord with  
the $a$-independence to be shown later in the case of the dynamically generated HLS gauge boson.
\\

     Also note that the extension to $W/Z/\gamma-\rho$ mixing strength should be intact in the scale-invariant mass term
which simply carries the extra dilaton factor $\chi^2$.      
In fact {\it the mass terms including the couplings of all the SM particles, except for the Higgs  mass term $V(\varphi)$, 
are dimension 4 operators due to $\chi^2$   and thus are scale-invariant, yet giving the same mass as in the
case without the $\chi^2$ factor.    
}\\

A salient feature of the scale-invariance of the $\rho_\mu$ mass term  in $\chi^2\cdot  a {\hat {\cal L}}_V$ is the 
{\it absence of the coupling of $\rho_\mu-W/Z-\varphi$ (``conformal barrier'')}, also similarly to the walking technirho \cite{Fukano:2015uga,Fukano:2015zua}. The SM Higgs $\varphi$ resides in the overall factor $\chi^2$ and hence can be coupled to the each gauge boson $\rho_\mu, W/Z$ only in the form of the simultaneous  mass diagonalization, hence has no off-diagonal couplings $\rho_\mu-W/Z$. This is in sharp contrast to many other vector meson models having the scale-violating
mass term of $\rho_\mu^2$ (dimension 2), dominance of which would lead to the so-called ``equivalence theorem result'' 
$\Gamma(\rho_\mu\rightarrow WW/WZ)\simeq \Gamma(\rho_\mu \rightarrow W/Z +\varphi)$.   Thus the $\rho_\mu$  
in the present case with only scale-invariant mass term  predominantly decays to diboson channels $\rho_\mu \rightarrow W_LW_L/W_LZ_L$ with the coupling 
$g_{\rho\pi\pi}$ ($=g_{_{\rm HLS}}$ for universality). 
\\

\section{Dynamical Generation of the HLS Gauge Boson in $CP^{N-1}$ Model}
\label{CPNApp}

In this appendix we elaborate on Refs.\cite{Bando:1987br,Harada:2003jx} for the dynamical generation of the HLS gauge boson in $CP^{N-1}$ model.
The $CP^{N-1}$  model is {\it renormalizable in $D$ dimensions for $2\leq  D < 4$, and thus is a well-defined laboratory to test the nonperturbative quantum effects}. It is in fact well established~\cite{Eichenherr:1978qa,
Golo:1978de,DAdda:1978vbw,DAdda:1978dle,
Witten:1978bc,Arefeva:1980ms,Haber:1980uy,Bando:1987br,Weinberg:1997rv,Harada:2003jx} that 
 the $U(1)$ hidden local symmetry (HLS) gauge boson 
introduced as an auxiliary field at classical level of the  $CP^{N-1}$  model does in fact generate
 the kinetic term at quantum level by the nonperturbative dynamics in $1/N$ expansion.  See e.g.,  Ref.\cite{Witten:1978bc} for an excellent description of this phenomenon in $D=2$. 
 For $D=4$ the theory is a cutoff theory, nevertheless the dynamical generation of the kinetic term of the HLS gauge boson is operative in exactly the same manner as for
 $2<D<4$, see \cite{Bando:1987br,Harada:2003jx}. The cutoff formulation  we adopt here in $D=4$ is close to the Wilsonian renormalizaton group formulation of the dynamical generation of the auxiliary fields such as those in the NJL model (Appendix \ref{NJL}).   Essentially the same result for $D=4$ is also established in a slightly different formulation (``effective theory'' made finite by all possible counter terms, i.e., with extra free parameters,  as in the Chiral Perturbation Theory)~\cite{Weinberg:1997rv}. 
\\

The $CP^{N-1}$ model is a nonlinear sigma
model based on the complex Grassmannian coset space $G/H=U(N)/[U(N-p)\times U(p)]\big|_{p=1}\simeq SU(N)/[SU(N-1)\times U(1)]$ 
written in terms of the massless NG bosons living in the manifold $G/H$ at classical level.
As we emphasized in the text, any nonlinear sigma model has HLS~\cite{Bando:1987br}, and as such the classical $CP^{N-1}$ Lagrangian is most commonly given in the form invariant under
$G_{\rm global} \times 
H_{\rm local}=SU(N)_{\rm global}\times\mbox{U}(1)_{\rm local}$, with
$
 H_{\rm local}= \mbox{U}(1)_{\rm local}$ being the Hidden Local Symmetry (HLS) ($H\subset G$): 
\begin{equation}
{\cal L} 
^{{\rm HLS}} = D_\mu \hbox{\boldmath$\phi$}^\dag
D^\mu \hbox{\boldmath$\phi$} - \eta(x)
\left( \hbox{\boldmath$\phi$}^\dag \hbox{\boldmath$\phi$} - N/G 
\right) \ ,
\label{CPN:Lag}
\end{equation}
where $\mbox{\boldmath$\phi$}$ is an $N$-component complex scalar field $\mbox{\boldmath$\phi$}$  with a constraint:
\begin{equation}
^t\hbox{\boldmath$\phi$} \equiv
\left(\phi_1, \phi_2, \ldots , \varphi_N\right) \ ,
\quad \phi_a \in \mbox{\bf C} \ , \quad
\hbox{\boldmath$\phi$}^\dag \hbox{\boldmath$\phi$} = N/G 
\quad \left( G \, : \, \mbox{coupling constant} \right) \, ,
\label{CPN:cond}
\end{equation}
where 
the field $\eta(x)$ is a Lagrange multiplier, and 
$D_\mu\hbox{\boldmath$\phi$}$ is the $U(1)_{\rm hidden}$ covariant derivative  given by
$
D_\mu\hbox{\boldmath$\phi$} = \left( \partial_\mu - i  
A_\mu \right)
\hbox{\boldmath$\phi$}.
$ 
The reason why $SU(N)_{\rm loca}$ is not usually discussed is that its gauge boson carries the index running $1,\dots, N-1$, so that the all planar 
graphs not just one-loop are involved in the large $N$ limit, which makes the analyses impossible in contrast to the $U(1)_{\rm local}$.
This is the same situation as for the $O(N-3)_{\rm local}$ in the Grassmannian model $G/H= O(N)/[O(N-3)\times O(3)]$ discussed in the main text of the present paper.
  Note that the theory also has a scale symmetry at classical level.\footnote{
The dimensions of the quantities ${\cal O}$ are as follows: $d_{\hbox{\boldmath$\phi$}} =D/2-1=d_v$, $d_\eta=2$,
$d_{A_\mu}= 
1$, $d_G=2-D$,  $d_{F_{\mu\nu}^2}=4 
$.
}

Since 
the HLS gauge field $A_\mu$ 
is an auxiliary field having no kinetic
term, 
it can be
eliminated by using the equation of motion: 
\begin{equation}
A_\mu = - \frac{i \,G}{2N} \hbox{\boldmath$\phi$}^\dag
\mathop{\partial_\mu}^{\leftrightarrow}
\hbox{\boldmath$\phi$}
\quad
\left( f \mathop{\partial_\mu}^{\leftrightarrow} g =
f \partial_\mu g - f \mathop{\partial_\mu}^{\leftarrow} g \right)
\ .
\end{equation}
Then the classical Lagrangian (\ref{CPN:Lag}) is equivalent to
\begin{equation}
{\cal L}^{{\rm HLS}} = \partial_\mu \hbox{\boldmath$\phi$}^\dag
\partial^\mu \hbox{\boldmath$\phi$} 
+ \frac{G}{4N}
\left( 
  \hbox{\boldmath$\phi$}^\dag
  \mathop{\partial_\mu}^{\leftrightarrow} \hbox{\boldmath$\phi$}
\right)^2
- \eta
\left( \hbox{\boldmath$\phi$}^\dag \hbox{\boldmath$\phi$} - N/G
\right) \ ,
\label{CPN:Lag2}
\end{equation}
which still retains the $\mbox{U}(1)_{\rm local}$ invariance: 
$\hbox{\boldmath$\phi$}^{\prime}(x) = e^{i\theta(x)}
\hbox{\boldmath$\phi$}(x)$.
Since $\hbox{\boldmath$\phi$}$ has $2N$ real components and is
constrained by one real condition in Eq.~(\ref{CPN:cond}), 
then reduced to $2N-1$ degrees of freedom, and 
by the $\mbox{U}(1)_{\rm local}$ gauge
invariance 
we can gauge away one further component of 
$\hbox{\boldmath$\phi$}$, leaving $2N-2$ degrees of freedom which are
exactly the dimension of the manifold
$CP^{N-1} = SU(N)/[SU(N-1)\times U(1)]$. Thus by gauge fixing of the HLS $U(1)_{\rm local}$
symmetry, Eq.(\ref{CPN:Lag2}) is further reduced to the genuine $CP^{N-1}$ nonlinear sigma model based on the  manifold $G/H = SU(N)/[SU(N-1)\times U(1)]$
:   
\begin{equation}
{\cal L}^{{\rm NL}\sigma}=
\partial_\mu \hbox{\boldmath$u$}^\dagger \partial^\mu \hbox{\boldmath$u$}
\left[1+\frac{G}{4N} \hbox{\boldmath$u$}^\dagger \hbox{\boldmath$u$}\right]^{-2}
+ \frac{G}{4N}
\left( 
  \hbox{\boldmath$u$}^\dag
  \mathop{\partial_\mu}^{\leftrightarrow} \hbox{\boldmath$u$}
\right)^2
\left[1+\frac{G}{4N} \hbox{\boldmath$u$}^\dagger \hbox{\boldmath$u$}\right]^{-4}\,,
\label{CPN:NLS}
 \end{equation}
where $^t\hbox{\boldmath$u$} \equiv \left(u^1, u^2, \ldots , u^{N-1}\right) 
$ 
is an unconstrained $N-1$ component complex variables standing for $2N-2$ independent degrees of freedom. 

Eq.(\ref{CPN:NLS}) is certainly  {\it gauge equivalent  to the HLS model Eq.(\ref{CPN:Lag}) at classical level}, in exactly the same sense  as the gauge equivalence 
between Eq.(\ref{Higgs1}) (and hence Eq.(\ref{Higgs2}) ) and the HLS Lagrangian Eq.(\ref{SM-HLS}) in the case of the SM. Note that the classical theory  takes the form of the spontaneously broken phase, $G$ broken down to $H$, written in terms of the NG boson fields living on the coset  $G/H$.
 {\it At quantum level}, however, the theory gets in  unbroken phase in the strong coupling region (for all coupling region in $D=2$ dimensions)  where
both become different in such a way that {\it  Eq.(\ref{CPN:NLS})   becomes  ill-defined, while Eq.(\ref{CPN:Lag}) is well-defined,  due to the very existence of the dynamical generation of the massless HLS gauge boson} acquiring the kinetic term through the nonperturbative dynamics, as we see in the below.
\\

Let us consider the effective action for the Lagrangian Eq.(\ref{CPN:Lag}) with a symmetry $G_{\rm global}\times 
H_{\rm local}=SU(N)_{\rm global} \times U(1)_{\rm local}$:
In the leading order of the $1/N$ expansion it is evaluated as
\begin{equation}
\Gamma\left[ \hbox{\boldmath$\phi$} , \lambda \right]
= \int d^D x \left[
  D_\mu \hbox{\boldmath$\phi$}^\dag D^\mu \hbox{\boldmath$\phi$} 
  - \eta
  \left( 
    \hbox{\boldmath$\phi$}^\dag \hbox{\boldmath$\phi$} - N/G
  \right) 
\right]
+ i N \,\mbox{Tr} \mbox{Ln} \left(
  - D_\mu D^\mu - \eta
\right)
\ .
\label{CPN:efac}
\end{equation}
Because of the $\mbox{SU}(N)$ symmetry, the VEV of 
$\hbox{\boldmath$\varphi$}$ can be written in the form
\begin{equation}
\left\langle
  ^t \hbox{\boldmath$\phi$}(x)
\right\rangle
= \left( 0, 0, \ldots , \sqrt{N} v \right) \ .
\label{VEVv}
\end{equation}
Then the effective action (\ref{CPN:efac}) gives
the effective potential for $v$ and $\eta\equiv \langle\eta(x)\rangle$ ($\langle A_\mu \rangle=0$) as
\begin{equation}
\frac{1}{N} V \left( v, \eta \right) = 
\eta \left( v^2 - 1/G \right) +
\int \frac{d^Dk}{i(2\pi)^D} \, \ln \left( k^2 - \eta \right) \ .
\end{equation}
The stationary conditions of this effective potential are given by
\begin{eqnarray}
&& \frac{1}{N} \frac{\partial V}{\partial v} 
= 2 \lambda v = 0 \ ,
\label{CPN:sta1}
\\
&& \frac{1}{N} \frac{\partial V}{\partial \eta} 
=
v^2 - \frac{1}{G} + 
\int \frac{d^Dk}{i(2\pi)^D} \, \frac{1}{\eta- k^2}
= 0 \ .
\label{CPN:sta2}
\end{eqnarray}

The first condition (\ref{CPN:sta1}) is realized in either of the
cases
\begin{equation}
\left\{ \begin{array}{ll}
  \eta=0 \, (v\neq0) \ , & \mbox{case (i)} \ , \\
  v=0 \, (\eta\neq0) \ , & \mbox{case (ii)} \ .
\end{array}\right.
\label{CPN:two phases}
\end{equation}
The case (i) corresponds to the broken phase
of the $SU(N)_{\rm global}\times U(1)_{\rm local}$ symmetry and scale symmetry,
and case (ii) does to the unbroken phase of $SU(N)_{\rm global}\times U(1)_{\rm local}$ symmetry but the broken phase of the
scale symmetry.

The second stationary condition (\ref{CPN:sta2}) gives relation
between $\eta$ and $v$.
By putting $\eta=v=0$ in Eq.~(\ref{CPN:sta2}), the critical point
$G(\equiv G(\Lambda)) =G_{\rm crit}(\equiv G_{\rm crit}(\Lambda))$ separating the two phases in Eq.~(\ref{CPN:two phases})
is determined as
\begin{equation}
\frac{1}{G_{\rm crit}} = 
\int \frac{d^Dk}{i(2\pi)^D} \, \frac{1}{- k^2}
= \frac{1}{\left(\frac{D}{2} - 1 \right) \Gamma(\frac{D}{2}) }
  \frac{\Lambda^{D-2}}{(4\pi)^{\frac{D}{2}}}
\ .
\label{CPN:gcr}
\end{equation}
which implies that $G_{\rm crit}\rightarrow 0$ for $D\rightarrow 2$.

Substituting Eq.~(\ref{CPN:gcr}) into the second stationary condition
(\ref{CPN:sta2}), we obtain
\begin{equation}
v^2 - 
\int \frac{d^Dk}{i(2\pi)^D} \, 
\left( \frac{1}{- k^2} - \frac{1}{\eta- k^2} \right)
= 
\frac{1}{G} - \frac{1}{G_{\rm crit}}= \frac{1}{G^{(R)}} - \frac{1}{G_{\rm crit}^{(R)}}\ ,
\label{CPN:sta3}
\end{equation}
where we have defined the
renormalized coupling  at renormalization point $\mu^2$ as 
\begin{eqnarray} 
\frac{1}{G^{(R)}} 
&\equiv& 
\frac{1}{G^{(R)}(\mu)}= 
\frac{1}{G} 
-\int \frac{d^Dk}{i(2\pi)^D} \, \frac{1}{\mu^2 - k^2}
\,, \notag \\ 
 \frac{1}{G^{(R)}_{\rm crit}} 
 &\equiv& 
 \frac{1}{G^{(R)}_{\rm crit}(\mu)}
 = 
 \int \frac{d^Dk}{i(2\pi)^D} 
 \left(\frac{1}{- k^2}-\frac{1}{\mu^2 - k^2} \right)
\notag \\ 
&=& 
\frac{\Gamma(2- D/2)}{\left(D/2 - 1 \right)}
\cdot \frac{\mu^{D-2}}{(4\pi)^{D/2}}
\,,  
\end{eqnarray}
 so that the equation reads a finite relation 
 for $2 \le D< 4$  
 as it should since it is a renormalizable theory.

The stationary condition in Eq.~(\ref{CPN:sta3}), combined with Eq.~(\ref{CPN:sta1}), leads to the cases 
(i) (broken phase of $ SU(N)_{\rm global}\times U(1)_{\rm local}$) 
and (ii) 
(unbroken phase of $ SU(N)_{\rm global}\times U(1)_{\rm local}$) 
in Eq.~(\ref{CPN:two phases}), 
respectively;
\begin{eqnarray}
&&\mbox{(i)} \ G < G_{\rm cr} 
\Rightarrow
 \langle \mbox{\boldmath$\phi$}_N \rangle =\sqrt{N} v \neq0 \ , \ 
 \langle \eta(x) \rangle = \eta=0\,  
\nonumber\\
&&
\frac{1}{G(\Lambda)} - \frac{1}{G_{\rm crit}(\Lambda)} 
=
 \frac{1}{G^{(R)}(\mu)} - \frac{1}{G_{\rm crit}^{(R)}(\mu)}
=v^2 \,
>0 \,\, ,
\label{CPN:gap ep}\\
&&
\mbox{(ii)} \ G > G_{\rm cr} 
\Rightarrow 
  \langle \mbox{\boldmath$\phi$}_N \rangle =\sqrt{N} v = 0 \ , \ 
  \langle \eta(x)\rangle =\eta \neq 0 
\nonumber\\
&& 
\frac{1}{G(\Lambda)} - \frac{1}{G_{\rm crit}(\Lambda)} 
=
 \frac{1}{G^{(R)}(\mu)} - \frac{1}{G_{\rm crit}^{(R)}(\mu)} 
\notag \\ 
&& 
\hspace{78pt}
= 
- 
\frac{\Gamma(2- D/2)}{\left(D/2 - 1 \right)}
 \cdot 
\frac{\eta^{D/2-1}}{(4\pi)^{D/2}} 
\equiv -v_\eta^2\, <0\,.
\label{CPN:gap ep2}
\end{eqnarray}
The gap equations Eq.(\ref{CPN:gap ep}) and Eq.(\ref{CPN:gap ep2}) take the same form as  that of the $D$-dimensional NJL model
which is also renormalizable for $2\leq D<4$~\cite{Kikukawa:1989fw,Kondo:1992sq},  with opposite sign and the same sign, respectively.
(See also Eq.~(\ref{NJLgap4D}) for $D=4$ NJL model). Both  (i) and (ii) are in broken phase of the scale symmetry by $v\ne 0$ and $\eta\ne 0$, respectively.~\footnote{ Similarly to the $D$-dimensional NJL model~\cite{Kikukawa:1989fw,Kondo:1992sq}, we may define dimensionless coupling $g$ as $G\equiv  g/\Lambda^{D-2}$
and $g_{\rm crit}= \left(D/2 - 1 \right) \Gamma(D/2) (4\pi)^{\frac{D}{2}}$, where the beta function  in both phases takes the form similar to that in the $D$-dimensional NJL model~\cite{Kikukawa:1989fw,Kondo:1992sq}: $\beta (g)=\Lambda \frac{\partial g}{\partial \Lambda}|_{v\, {\rm or}\, \eta={\rm fixed}} =
 - (D-2)\,  g\, (g - g_{\rm crit})/g_{\rm crit}$ (the same form for $g^{(R)}(\mu)=G^{(R)}\mu^{D-2}$ at $D< 4$), where $g_{\rm crit}$  is a UV  fixed point.
}

The case (i) is the {\it perturbative phase} where the classical theory structure remains. 
  Eq.(\ref{CPN:gap ep})  is the gap equation for the spontaneous breaking of the symmetry
 $SU(N)_{\rm global} \times U(1)_{\rm local}$, 
 or equivalently (after gauge fixing) the coset space $SU(N)/[SU(N-1)\times U(1)]$, with the
 Higgs mechanism of $U(1)_{\rm local}$ yielding the ``mass''  of $A_\mu$: $(M^{(0)}_A)^2 = 2 N  v^2$ (with mass dimension $D-2$), 
 as read from Eq.(\ref{CPN:Lag}), with Eq.(\ref{CPN:cond}).  
 The scale symmetry is also spontaneously broken by the same $v\ne 0$, with the
 pseudo-dilaton identified with the real part of the $\mbox{\boldmath$\phi$}_N$ (its mass from the trace anomaly due to the regularization with $\Lambda$, or the renormalization with $\mu$).

The case (ii) is a genuine {\it nonperturbative phase} in strong coupling $G>G_{\rm crit}$. It implies that
{\it the quantum theory  is actually in the unbroken phase} of $SU(N)_{\rm global} \times U(1)_{\rm local}$,
although the theory at classical level 
is written in terms of the NG boson variables $\mbox{\boldmath$\phi$}$ in the coset $SU(N)/[SU(N-1)\times U(1)]$ as if it were in the broken phase.
The HLS gauge symmetry $U(1)_{\rm local}$ thus is never spontaneously broken and the gauge boson if exists as 
a particle should 
be massless.  
In fact,  the would-be NG bosons $\hbox{\boldmath$\phi$}$ 
at classical level are {\it no longer
 the NG bosons at quantum level}  by the nonperturbative dynamics (at large $N$) and acquire dynamically the  mass
  \begin{equation}
 M^2_{\hbox{\boldmath$\phi$}}=\eta =\langle \eta(x)\rangle\ne 0\,, \quad \left(G>G_{\rm crit}\,, \quad v=0 \right)\,,
 \end{equation}
as readily seen from Eq.(\ref{CPN:Lag}).  
 Note that
$\langle \eta(x) \rangle$ in Eq.(\ref{CPN:gap ep2})  {\it breaks no internal symmetry
 but the scale symmetry.} Writing $\eta(x)=\eta e^{\varphi(x)/\eta}$, we may regard $\varphi(x)$ as a pseudo-dilaton in this phase (its mass from the trace anomaly due to the regularization with $\Lambda$, or the renormalization with $\mu$).

Special attention should be paid to {\it $D=2$ dimensions, where $G_{\rm crit}=0$ and hence  the {\it case (i) (the classical/perturbative phase, broken phase with $v\ne 0$) does not exist at all}, in accord with the Mermin-Wagner-Coleman theorem on 
absence of the spontaneous symmetry breaking in $D=2$ dimensions}.  
On the other hand, the gap equation Eq.(\ref{CPN:sta2}) with $D=2$  
takes the form $\frac{1}{G}=- \frac{1}{4\pi} \ln \frac{\eta}{\Lambda^2}$, or:
\begin{equation}
\langle \eta(x)\rangle =\Lambda^2 \cdot \exp\left(-\frac{4\pi}{G(\Lambda)}\right) = \mu^2 \cdot \exp\left(-\frac{4\pi}{G^{(R)}(\mu)}\right)\,,
\end{equation}
where the scale symmetry appears to be spontaneously broken by $\langle \eta(x)\rangle\ne 0$ in the same sense as $D>2$ (up to explicit breaking due to the trace anomaly), but actually undergoes  the BKT phase transition similarly to the $D=2$ NJL model (Gross-Neveu model), see  footnote \ref{MWCtheorem}. 
\\
 
 We now discuss the dynamical generation of the kinetic term of the $U(1)$ HLS gauge boson. 
 \\
 
 First we take a look at the case  (ii): 
  ($G>G_{\rm crit}\,, v=0, \langle \eta(x)\rangle\ne 0$),
 where the {\it massless} gauge boson of unbroken $U(1)_{\rm local}$ HLS gauge symmetry does appear dynamically. Particularly for $D=2$ this is a  whole story, 
 since case (i) does not exists there.

The  (amputated) two-point vertex function of the HLS gauge field $A_\mu$  is an auxiliary field at classical level: $\Gamma_{\mu\nu}(x)=\langle  A_\mu (x) A_\nu (0)\rangle_{\rm amp}$. At quantum level at the $1/N$ leading order it has  one-loop contributions  of the fundamental particles  $\hbox{\boldmath$\phi$}$. Since the Lagrangian Eq.(\ref{CPN:Lag}) has the  $U(1)_{\rm local}$ symmetry,
$\Gamma_{\mu\nu}(x)$ must have the form invariant under the gauge symmetry such that:
   \begin{equation}
   \Gamma_{\mu\nu}(p) =\left( p^2 g_{\mu\nu} - p_\mu p_\nu \right)\cdot f(p^2)\,.
   \label{2-pointCPN}
   \end{equation}
  Since  
  $\hbox{\boldmath$\phi$}$ are now massive, $M^2_{\hbox{\boldmath$\phi$}}= \langle \eta(x)\rangle \ne 0$, in the unbroken phase,  the only singularity of $f(p^2)$ arises
  from the two-$\hbox{\boldmath$\phi$}$ threshold $p^2=(2M_{\hbox{\boldmath$\phi$}})^2>0$ and beyond,
  and hence has no singularity at $p^2=0$, namely $f(0) \ne 0$. Then we see that  the two-point Green function develops a genuine massless pole:
  \begin{equation}
  {\cal F.T.} \langle  T(A_\mu (x) A_\nu (0))\rangle=- \Gamma_{\mu\nu}(p)^{-1}= 
  g_{\mu\nu} 
  \frac{-f^{-1}(0)}{p^2} +\mbox{gauge terms} 
  \,,
  \label{twopointA}
    \end{equation}
  where the ``gauge terms'' depend on the gauge fixing, and the residue $-f^{-1}(0) \,(>0)$ is characterized by  
  $M^2_{\hbox{\boldmath$\phi$}}= \langle \eta(x)\rangle=\eta \ne 0$. 
  Thus  
   the HLS gauge boson kinetic term reads
  \begin{equation}
 {\cal L}^{({\rm kin})}_{\rm HLS}= -\frac{1}{4\, g_{_{\rm HLS}}^2} F^2_{\mu\nu}, \,, \quad 
 \frac{1}{g_{_{\rm HLS}}^2}=-f(0) =
 \frac{N}{3} 
\frac{
\Gamma(2-\frac{D}{2})
}{
(4\pi)^{\frac{D}{2}}\, \Gamma(2)
} 
 M_{\hbox{\boldmath$\phi$}}^{D-4} \,. 
 \label{kineticD}
 \end{equation}
 {\it Hence the kinetic term of the HLS gauge boson 
 indeed has been dynamically generated by the
  nonperturbative dynamics at $1/N$ leading order!! } 
  Note that the scale symmetry existing at classical level has been broken by the
  kinetic term, 
  with the scale dimension 4 (not $D$)  operator, which is traced back to the spontaneous scale-symmetry breaking due to $\langle \eta(x) \rangle\ne 0$.
\footnote{
We may write the kinetic term in a scale-invariant form through the dilaton field $\phi_{_{\lambda}}(x)$:
  $-\frac{1}{4 g_{_{\rm HLS}}^2} \cdot \chi_{_{\eta}}^{D-4} \cdot F^2_{\mu\nu}=-\frac{1}{4  g_{_{\rm HLS}}^2} \cdot F^2_{\mu\nu}+\cdots$, 
  where $\sqrt{\eta(x)}= \sqrt{\eta} \cdot \chi_{_{\eta}}(x)$ with $\chi_{_{\eta}}(x)\equiv  e^{\phi_{_{\eta}}(x)/\sqrt{\eta}}$. 
  The dimensionless field $\chi_{_{\eta}}(x)$ has a scale-dimension  1, 
  similarly to $\chi(x)$ in Eq.(\ref{NLscale}) in the SM case.  
\label{kineticscale} }

Next we see that  in the same case (ii) ($G>G_{\rm crit}\,, v=0, \langle \eta(x) \rangle \ne 0$),  {\it the same dynamical generation of HLS gauge boson does not take place} in the genuine nonlinear sigma model Lagrangian Eq.(\ref{CPN:Lag2}) {\it without explicit gauge symmetry from onset at classical level.} Although Eq.(\ref{CPN:Lag2}) is gauge equivalent to the HLS Lagrangian Eq.(\ref{CPN:Lag}) at classical level, it yields only an ill-defined quantum theory \cite{Haber:1980uy}, in sharp contrasts to the HLS Lagrangian Eq.(\ref{CPN:Lag}) which 
has a well-defined quantum theory with massless gauge boson dynamically generated.

 Corresponding to Eq.(\ref{twopointA}), let us look at the two-point Green function for the composite vector operator $\hbox{\boldmath$u$}^\dagger \partial_\mu \hbox{\boldmath$u$}$ of the Lagrangian Eq.(\ref{CPN:Lag2}):
 \begin{equation}
 T_{\mu\nu}={\cal F.T.}  \langle T(\hbox{\boldmath$u$}^\dagger(x) \partial_\mu \hbox{\boldmath$u$}(x) \,
 \hbox{\boldmath$u$}^\dagger(0) \partial_\nu \hbox{\boldmath$u$}(0)  )
 \rangle
 \end{equation}
  in $D=2$ case where only the unbroken phase $g>g_{\rm crit}=0$ exists. 
At leading order of $1/N$, $T_{\mu\nu}$ is given by the infinite geometric series of  the one-loop bubble diagram contribution $B_{\mu\nu}$ as a solution of the equation:
\begin{eqnarray}
T_{\mu\nu} &=& B_{\mu\nu} + B_\mu^\rho \, T_{\rho\nu}\,,\nonumber\\
B_{\mu\nu}&=& g_{\mu\nu} + (g_{\mu\nu} p^2 - p_\mu p_\nu) f(p^2)\,,
\end{eqnarray}
where $B_{\mu\nu}$ has a gauge non-invariant term $g_{\mu\nu}$ due to the lack of gauge symmetry in Eq.(\ref{CPN:Lag2}).
Then the solution reads
\begin{equation}
T_{\mu\rho}=\left( g_{\mu\nu} - B_{\mu\nu}\right)^{-1} \cdot B^\nu_\rho =
- (g_{\mu\nu} p^2 - p_\mu p_\nu)^{-1} f^{-1}(p^2) \cdot B^\nu_\rho \,,
 \end{equation}
 which is divergent because of the zero eigenvalue of $g_{\mu\nu} p^2 - p_\mu p_\nu$, namely non-invertible, since there is
 no gauge-fixing freedom due to the absence of the gauge symmetry. Thus the quantum theory of the Lagrangian Eq.(\ref{CPN:Lag2}) 
 is simply {\it ill-defined due to the lack of the gauge symmetry}.

 The result is  in perfect conformity with the Weinberg-Witten theorem~\cite{Weinberg:1980kq} which forbids the
 dynamical generation of the massless  particles with spin $J\geq 1$. The theorem is proved  in the Hilbert space with positive definite metric and hence without gauge symmetry. This is in  
 sharp contrast to the HLS Lagrangian Eq.(\ref{CPN:Lag}) which does have
 a gauge symmetry thus is quantized with indefinite metric Hilbert space, and hence generates a massless gauge boson without conflict to the Weinberg-Witten theorem.
   
    Thus the lesson is : when the theory is parameterized differently and still equivalent to each other at classical level, it may {\it not yield the same quantum theory
in the nonperturbative dynamics, depending on the parameterization, even if arriving at  the same  perturbative result.} 
\\

 Now to the case (i):  $G<G_{\rm crit} (\ne 0)\,, v\ne 0, \langle\eta(x)\rangle=0$, which exists only for $D>2$, and is more similar to the HLS Lagrangian Eq.(\ref{HLSgeneral}) or the  SM HLS Lagrangian as its scale-invariant version Eq.(\ref{SM-HLS}) which is gauge equivalent to the original SM Higgs Lagrangian, Eqs.(\ref{Higgs2}) and (\ref{Higgs1}).  In this case there remains the symmetry structure at classical level, namely both $SU(N)_{\rm global}$ and $U(1)_{\rm hidden}$ are spontaneously broken,
 in such a way that the theory is gauge equivalent to the model based on the manifold $G/H=SU(N)/[SU(N-1)\times U(1)]$ without HLS. 
  There exist $2N-2$ massless NG bosons $(\phi_1,\cdots,\phi_{N-1})$, and at classical level  the HLS gauge boson has a (bare) mass $(M_A^{(0)})^2=2 N v^2 $ 
  by the Higgs mechanism absorbing  the would-be NG boson $\hat \pi$  
  in the parametrization $\phi_N=\hat \sigma +i \hat \pi$ corresponding to in Eq.(\ref{VEVv}), where the $\hat \sigma $ 
  corresponds to $\hat \sigma$ in the SM case, 
  Eq.(\ref{Higgs1})~\footnote{Alternatively, it may be better to parameterize  $\varphi_i(x)=\sigma(x)\cdot  e^{i\theta(x)} z_i(x)$, with $\Sigma_{i=1}^N z^\dagger_i z_i=N/G$ and $\langle z_i\rangle=0 \, (i=1,\cdots, N-1)$, $z_N=1/G$, where $\theta$ is the would-be NG boson absorbed into HLS gauge boson, 
 and
  $\sigma =\sqrt{\frac{1}{N}\Sigma_{i=1}^{N} \phi_{i}^\dagger \phi_{i}}=
   v e^{\tilde \varphi/v}$, 
  with the dilaton $\tilde \varphi$ similar to $\varphi$ as the SM Higgs boson in Eq.(\ref{Higgs2}) with Eq.(\ref{NLscale}).
}.

  Yet the quantum theory also {\it develops the kinetic term
 of the HLS gauge boson precisely in the same manner as in the unbroken phase in case (ii)}:
 \begin{equation}
   \Gamma_{\mu\nu}(p) =\left( p^2 g_{\mu\nu} - p_\mu p_\nu \right)\cdot 
   f(p^2) + g_{\mu\nu}\cdot \left(2\, Nv^2\right)\, ,
    \end{equation}
up to the 
additional mass term. 
Similarly to the
 unbroken phase, it reads the kinetic term 
  $-\frac{1}{4\, g_{_{\rm HLS}}^2} F_{\mu\nu}^2$ 
  ($g_{_{\rm HLS}}^2=-f^{-1}(\mu^2)) $ 
 at the scale $\mu$. 
    Again the dynamically generated kinetic term is a dimension $4$ operator and thus breaks the scale symmetry,
similarly to $\langle \eta(x)\rangle\ne 0$ in the case (ii), 
this time by $\langle \hbox{\boldmath$\phi$}_N\rangle=\sqrt{N} v\ne 0$  
(see footnote \ref{kineticscale}).
 In this case  the HLS gauge boson would have a mass at quantum level,  
 \begin{eqnarray}
 M^2_A=-{
 f}^{-1}(M^2_A)\, (2Nv^2)= g_{_{\rm HLS}}^2 \cdot  \left(2 \cdot N v^2\right)
 \end{eqnarray}
  (now dimension 2), in precisely the standard form of the Higgs mechanism, with the factor 2 characteristic to all the
  Grassmannian models, a remnant of the KSRF II  relation (see e.g., Eq.(\ref{rhomass})). 
  The result is the same as that read from the Lagrangian at quantum level including the induced kinetic term, after it is rescaled  to the canonical form  $-\frac{1}{4 g_{_{\rm HLS}}^2} F_{\mu\nu}^2\rightarrow -\frac{1}{4} F_{\mu\nu}^2$.

   Actually, the pole
    of the HLS gauge boson is moved into the Euclidean region having imaginary part of the mass, in such a way that it decays into  
    the massless NG bosons with threshold at $p^2=(2 M_{\hbox{\boldmath$\phi$}})^2=0$, although the generation of the kinetic term
    is operative to the off-shell dynamics, such as the soliton like the skyrmion as in Ref.(\cite{Matsuzaki:2016iyq,MOY2018}). 
  \\
   
   Finally, the $D=4$ case, which is a cutoff theory, not a renormalizable theory in the usual sense
   (see Ref.~\cite{Weinberg:1997rv} for the effective theory approach). In the same sense as in the 
   D=4 NJL model discussed in Appendix \ref{NJL}, we identify  the cutoff $\grave{a}$ la Ref.\cite{Bardeen:1989ds}
   as a Landau pole, where 
   the dynamically generated kinetic term of the HLS gauge boson disappears, or the  induced 
   gauge coupling diverges. 
   
   In the case (ii)  $G>G_{\rm crit}$,  the gap equation 
   Eq.(\ref{CPN:sta3}) or Eq.(\ref{CPN:gap ep2}) with $D=4$  reads
   \begin{equation}
   \frac{1}{G} -\frac{1}{G_{\rm crit} }= -\frac{1}{(4\pi)^2} \eta \ln \frac{\Lambda^2}{\eta}\,,
   \end{equation}
   and the kinetic term of the massless HLS gauge boson reads
   \begin{equation}
   {\cal L}_{\rm HLS}^{({\rm kin})}= - \frac{1}{4 g_{_{\rm HLS}}^2} F_{\mu\nu}^2 \,, \quad  \frac{1}{g_{_{\rm HLS}}^2 }=-f(0)=\frac{N}{48 \pi^2}\,  
   \ln \frac{\Lambda^2}{\eta
   }\,. 
      \end{equation}
which is compared with the renormalizable theory Eq.(\ref{kineticD}) with $2\leq D<4$.  When the momentum integration is done from $\Lambda$ down to $\mu$ in the sense of Wilsonian renormalization group, the 
gauge coupling reads $1/g_{_{\rm HLS}}^2 = [N/(48\pi^2)] \ln (\Lambda^2/\mu^2)$. 
Now the gauge coupling $g_{_{\rm HLS}}^2$  has a Landau pole at $\mu=\Lambda$ where the
dynamically generated kinetic term does vanish: $1/g_{_{\rm HLS}}^2 \rightarrow 0$ as $\mu \rightarrow \Lambda$.  
Similarly, in the case (ii) $G<G_{\rm crit}$ (perturbative/broken phase),  the kinetic term of the massive HLS gauge boson is generated as ${\cal L}_{\rm HLS}^{({\rm kin})}=\frac{1}{4 g_{_{\rm HLS}}^2} F_{\mu\nu}^2$, with $1/g_{_{\rm HLS}}^2=-f^{-1}(\mu^2)$, and the mass reads $M_A^2= -f^{-1}(M_A^2) (2N v^2)= g_{_{\rm HLS}}^2 \cdot  (2N v^2)$.

\section{Dynamical Generation of the Kinetic Term of the Auxiliary Fields in the NJL Model}
\label{NJL}

Here we summarize the dynamical generation of the auxiliary fields in  the nonperturbative quantum theory of the $SU(2)_L\times SU(2)_R$ NJL model,
 in a way~\cite{Eguchi:1976iz,Kugo:1976tq,Bardeen:1989ds} particularly developed by Ref.\cite{Bardeen:1989ds} to reformulate the top quark condensate model of Ref.\cite{Miransky:1988xi} which has the $SU(2)_R$ violating four-fermion coupling. The quantum dynamical phenomenon discussed below is essentially the same in both cases. 
Further details including the top quark condensate case are given in, e.g., \cite{Yamawaki:2015tmu,Yamawaki:2016kdz} and references therein.

The NJL Lagrangian reads~\cite{Nambu:1961tp}:
\begin{eqnarray}
{\cal L}_{\rm NJL}^{\rm classical} &=& \bar \psi i\gamma^\mu\partial_\mu \psi + \frac{G}{4} \left[ (\bar \psi \psi)^2 +(\bar \psi i \gamma_5 \tau^a \psi)^2\right]\nonumber \\
&=& 
\bar \psi i\gamma^\mu\partial_\mu \psi + \frac{G}{4} \left[ (\bar \psi \psi)^2 +(\bar \psi i \gamma_5 \tau^a \psi)^2\right]\
-\frac{1}{2G}\left( \hat \pi^a + \frac{G}{\sqrt{2}} \bar \psi i\gamma_5 \tau^a \psi\right)^2 
- \frac{1}{2G}\left( \hat \sigma + \frac{G}{\sqrt{2}} \bar \psi \psi \right)^2
\nonumber\\
&=& \bar \psi \left(i\gamma^\mu\partial_\mu  -\frac{1}{\sqrt{2}}(\hat \sigma +i\gamma_5 \tau^a \hat \pi_a ) \right)\psi -\frac{m_0^2}{2}\left({\hat \sigma}^2 +{\hat \pi_a}^2\right)\,,  \quad 
\left(m_0^2=\frac{1}{G} >0 \right) \,,
\label{NJLeq}
\end{eqnarray}
where the equation of motion for the auxiliary fields  $\hat \sigma = - G \bar \psi \psi/\sqrt{2}$ and ${\hat \pi}_a = -G \bar \psi i\gamma_5 \tau_a\psi/\sqrt{2}$
may be plugged in the second line to get back to the original Lagrangian on the first line.

By integrating out the high frequency modes from the cutoff scale $\Lambda$ down to $\mu$ in the Wilsonian sense at leading order of $1/N_c$ s.t. $N_c \rightarrow \infty$ with $N_c\, G =$ fixed, we have a quantum theory which {\it does generate kinetic term} of $\hat \sigma/\hat \pi_a$ and the quartic coupling:
\begin{eqnarray}
{\cal L}_{\rm NJL}^{1/N_c} &=&\bar \psi \left(i\gamma^\mu\partial_\mu  -\frac{1}{\sqrt{2}}(\hat \sigma +i\gamma_5 \tau^a \hat \pi_a ) \right)\psi -\frac{m^2_0(\mu)}{2}\left({\hat \sigma}^2 +{\hat \pi_a}^2\right)\nonumber\\
&+& \frac{1}{2} Z_\phi(\mu) \left[(\partial_\mu \hat \sigma)^2 +(\partial_\mu \hat \pi_a)^2\right] -
\frac{\lambda_0(\mu)}{4}  \left[ (\hat \sigma)^2+ (\hat \pi_a^2)\right]^2\,,\nonumber \\
\lambda_0(\mu)&=&Z_\phi(\mu)= \frac{N_c}{8\pi^2} \ln \frac{\Lambda^2}{\mu^2}\,,\quad m^2_0(\mu)=\frac{1}{G} - \frac{N_c}{4 \pi^2}(\Lambda^2-\mu^2) \,,
\end{eqnarray}
with $\hat \sigma, \hat \pi_a$ now being  the dynamical tachyon $m_0^2(0) <0$ (in contrast to $m_0^2>0$) for $G>G_{\rm crit}= \frac{4\pi^2}{N_c \Lambda^2}$, the phase change from the unbroken phase into the broken one by the quantum effects
(quadratic divergence),  in accord with the gap equation for the dynamically generated fermion mass $M_F \ne 0$ (valid in the formal  limit $\mu\sim M_F\ll v (={\cal O} (\sqrt{N_c} M_F) \ll \Lambda$) where $m_0^2(M_F)\simeq m_0^2(0)$:~\cite{Nambu:1961tp}
\begin{equation}
\frac{1}{G} -\frac{1}{G_{\rm crit}} = - 2 M_F^2\, \left(\frac{N_c}{8\pi^2} \ln \frac{\Lambda^2}{M_F^2}\right)  
=-v^2<0\,, \quad\left(G> G_{\rm crit}\right)\,.
\label{NJLgap4D}
\end{equation}  
 Taking $\mu\rightarrow \Lambda$ we get back to the original classical theory, Eq.(\ref{NJLeq}), with $m^2_0(\mu) \rightarrow \frac{1}{G}$, $\lambda_0(\mu)=Z_\phi(\mu)\rightarrow 0$.

 After rescaling the kinetic term $Z^{1/2} _\phi(\mu)(\hat \sigma, \hat \pi_a) \rightarrow (\hat \sigma,\hat \pi_a)$ to the canonical one,  we have 
 \begin{eqnarray}
{\cal L}_{\rm NJL}^{1/N_c} &=&\bar \psi \left(i\gamma^\mu\partial_\mu  -\frac{g_Y}{\sqrt{2}}(\hat \sigma +i\gamma_5 \tau^a \hat \pi_a ) \right)\psi -\frac{m^2}{2}\left({\hat \sigma}^2 +{\hat \pi_a}^2\right)\nonumber\\
&+& \frac{1}{2} \left[(\partial_\mu \hat \sigma)^2 +(\partial_\mu \hat \pi_a)^2\right] -
\frac{\lambda}{4}\left[ {\hat \sigma}^2+ {\hat \pi}_a^2 \right]^2\,,\nonumber \\
\lambda&=&\lambda_0(\mu)\, Z^{-2}_\phi(\mu) =Z^{-1}_\phi(\mu)=
\frac{1}{\frac{N_c}{8\pi^2} \ln \frac{\Lambda^2}{\mu^2}} 
=g_Y^2 \,,\quad m^2= m_0^2(\mu) \cdot Z^{-1}_\phi(\mu) \,,
\label{rescaledNJL}
 \end{eqnarray} 
which is precisely the same form as the SM Higgs Lagrangian Eq.(\ref{Higgs1}),  plus the Yukawa term, 
both   having the Landau pole $\lambda=g_Y^2\rightarrow \infty$ at $\mu\rightarrow \Lambda$ (``compositeness condition'')\cite{Bardeen:1989ds} , where $v^2 =\langle \sigma^2\rangle_{m_F}=\langle  {\hat \sigma}^2+ {\hat \pi_a}^2 \rangle_{m_F} =\frac{-m^2}{\lambda}|_{m_F} = -m_0^2(M_F) $.  
  
    Note that {\it there appear extra free parameter}, $\lambda$ (and/or $g_Y$), which is absent at classical level but does exist
 in the quantum theory originating from the cutoff $\Lambda$ in $D=4$, the phenomenon also realized  in the SM case with $g_Y$ (rescaling of the kinetic term) and $\lambda$ (quartic coupling)  corresponding to an extra free parameter $g_{_{\rm HLS}}$ of the kinetic term of the SM rho.

  The result is of course the same as the
  popular (original NJL) dynamical calculation, 
the  gap equation for the fermion mass generation $M_F\ne 0$ and also the Bethe-Salpeter equation summing up the infinite geometric 
series of the one-loop bubble diagram producing the bound states, massless NG boson  $\pi$ and massive $\sigma$ (Higgs as a dilaton $\phi$) (physical modes, not $\hat \pi$ and $\hat \sigma$, see discussions below Eq.(\ref{transformation})) in the large $N_c$ limit.
 (The relation $\lambda=g_Y^2$ is a dynamical consequence of the NJL model specific to the large $N_c$ limit
 (subject to modification at higher orders), which leads to the famous NJL mass relation $M_\phi^2= 2\lambda v^2= 2 g_Y^2 v^2= 4(g_Y\, v/\sqrt{2})^2= (2\, m_F)^2$.) 
  
 \section{Direct calculation for the $\rho$ universality}
 \label{universality}
We here confirm that the $\rho-$ universality is realized independently of $a$
in the large $N$ limit, Eq.(\ref{universality-a}),  by  direct calculations~\footnote{The calculations here are
largely owe to Taichiro Kugo, private communication.}  
 \\
 
 Let us study the 
 $\rho\pi\pi$ vertex
  \begin{eqnarray}
   \Gamma^{\rho\pi\pi, \nu} (q,k, q+k)\big|^{k^2=(k+q)^2=0}_{\phi-{\rm amputated}}  
  &=&  \left[\frac{a}{2} g_{\mu\nu}
  +
 \frac{a}{2} \left\{B_{\mu\nu}(q)+ B_{\mu\lambda}(q)\cdot \left(\frac{a}{2}-1\right)\frac{G}{N} B^\lambda_\nu(q) +\cdots
  \right\}\left(\frac{a}{2}-1\right)\frac{G}{N} \right] \cdot (q+2k)^\nu\nonumber\\
  &&=\left[
  \frac{a}{2} g_{\mu\nu}
  +\frac{2}{a} \left(
  {\tilde \Gamma}^{(\rho)}_{\mu\nu}(q) -\frac{a}{2} \frac{N}{G}\cdot g_{\mu\nu}  \right)
   \left(\frac{a}{2}-1\right)\frac{G}{N} 
  \right]   \cdot (q+2k)^\nu    \nonumber\\
  &&=\left[
  1\cdot  g_{\mu\nu}
  + 
  {\tilde \Gamma}^{(\rho)}_{\mu\nu}(q)   
   \left(1-\frac{2}{a}\right)\frac{G}{N} 
  \right]   \cdot (q+2k)^\nu\,,
  \end{eqnarray}
  where  $ {\tilde \Gamma}^{(\rho)}_{\mu\nu}(q) =\frac{1}{2}  \Gamma^{(\rho)}_{\mu\nu}(q)  =-\langle \rho_\mu \rho_\nu\rangle^{-1}(q)$ is given in  Eq.(\ref{Gammarho-a}). We  then
  multiply the $\rho_\mu$ propagator  
  $\langle \rho_\mu \rho_\nu\rangle (q)$ in Eq.(\ref{rhopropagatoruniversal}) to get 
  \begin{eqnarray}
\langle \rho_\mu \rho_\nu\rangle(q) \cdot \Gamma^{\rho\pi\pi, \nu} (q,k, q+k)\big|^{k^2=(k+q)^2=0}_{\phi-{\rm amputated}}
&=& \left[\langle \rho_\mu \rho_\nu\rangle(q) - g_{\mu\nu}\cdot \left(1-\frac{2}{a}\right)\frac{G}{N}
\right]\cdot (q+2k)^\nu\nonumber\\
 &=&\frac{1}{N}\frac{- f^{-1}(q^2,\eta)}{q^2 - (- v^2 f^{-1}(q^2,\eta))} \left(g_{\mu\nu} -\frac{q_\mu q_\nu}{- v^2 f^{-1}(q^2,\eta)}
\right)\cdot (q+2k)^\nu\,, \nonumber\\
&=&2 \frac{g^2_{_{\rm HLS}}(q^2,\eta)}{q^2 - g_{_{\rm HLS}}(q^2,\eta) \cdot F_\rho^2} \left(g_{\mu\nu} -\frac{q_\mu q_\nu}{g_{_{\rm HLS}}(q^2,\eta)\cdot F_\rho^2}
\right)\cdot (q+2k)^\nu\,, 
 \label{universalitycheck}
\end{eqnarray}
which is $a-$independent, where $g^{-2}_{_{\rm HLS}}(q^2,\eta)\equiv -2 N f(q^2,\eta)$ and $F_\rho^2=N F_\pi^2=2 N v^2$. Although both $\langle \rho_\mu \rho_\nu\rangle (q)$
and   $\Gamma^{\rho\pi\pi, \nu} (q,k, q+k)\big|_{\phi-{\rm amputated}} $ have $a-$dependence, it  is cancelled out in the 
combination in the Green function. This we already have seen for the
$\pi\pi$ scattering amplitude $T_{\mu\nu}(q)$ for $a=0$ in Eq.(\ref{massivevector}), which has contact terms cancellation between the tree and the infinite bubble sum, thus
realizing the VMD. The present example clearly shows that it is also the case for
the $\rho_\mu$ vertex for arbitrary $a$.

Actually, the result is nothing but the manifestation of the Ward-Takahashi identity:
\begin{eqnarray}
0&=&\int {\cal D} \phi \frac{\delta}{\delta \rho_\mu(x)} \left(\phi(y) \phi(z)\cdot e^{i S[\phi]}\right)
=\int {\cal D} \phi \left(\frac{aN}{2G} \right) \left(\rho_\mu(x)- \alpha_{\mu,||}(x)
\right)\cdot \phi(y) \phi (z)\cdot e^{i S[\phi]}
\end{eqnarray}
 (cf. Eq.(\ref{adependence}) for the two-point functions),
 which yields
\begin{eqnarray}
\langle \rho_\mu \rho_\nu\rangle(q) \cdot \Gamma^{\rho\pi\pi, \nu} (q,k, q+k)\big|^{k^2=(k+q)^2=0}_{\phi-{\rm amputated}}
&=&\langle \rho_\mu (q) \phi (k) \phi (q+k)\rangle \big|^{k^2=(k+q)^2=0}_{\phi-{\rm amputated}}\nonumber\\
&=&\langle \alpha_{\mu,||} (q) \phi (q) \phi (q+k)\rangle\big|^{k^2=(k+q)^2=0}_{\phi-{\rm amputated}}
\label{rhopipiGreenfn}
\end{eqnarray}
where the last term is independent of the auxiliary field $\rho_\mu$ and hence is obviously independent of $a$.
So the $\rho\pi\pi$ Green function is $a-$independent as it should be.

Now we define those quantities for the  ``renormalized'' $\rho_\mu^{(R)}$ by rescaling the ``kinetic term'' to the canonical one by $g^{-2}_{_{\rm HLS}}(q^2,\eta)=-2 N f(q^2,\eta)\rightarrow 1$:
\begin{eqnarray}
\rho_\mu^{(R)}&=&g^{-1}_{_{\rm HLS}}(q^2,\eta)\cdot \rho_\mu
=\left[-2N f(q^2,\eta)\right]^{\frac{1}{2}}\cdot \rho_\mu\nonumber\\
\langle \rho^{(R)}_\mu \rho^{(R)}_\nu\rangle(q) &=& g^{-2}_{_{\rm HLS}}(q^2,\eta) \cdot \langle \rho_\mu \rho_\nu\rangle(q)\nonumber\\
\Gamma^{\rho^{(R)}\pi\pi}_\mu (q,k, q+k)\big|^{k^2=(k+q)^2=0}_{\phi-{\rm amputated}} &=&g_{_{\rm HLS}}(q^2,\eta)\cdot 
\Gamma^{\rho\pi\pi}_\mu (q,k, q+k)\big|^{k^2=(k+q)^2=0}_{\phi-{\rm amputated}}
\,,
 \end{eqnarray}
by which the renormalized $\rho\pi\pi$ Green function (intrinsically $a-$independent as mentioned above) is a product of the canonical $\rho_\mu$ propagator times
the renormalized $\rho\pi\pi$ coupling 
$g_{\rho\pi\pi}(q^2)$ defined for the renormalized $\rho^{(R)}_\mu$:
\begin{eqnarray}
\langle \rho^{(R)}_\mu(q) \phi(k) \phi(q+k)\rangle \big|^{k^2=(k+q)^2=0}_{\phi-{\rm amputated}}&=& 2\cdot \frac{
g_{\mu\nu} -
\frac{
q_\mu q_\nu}{g^2_{_{\rm HLS}}(q^2,\eta) \cdot F_\rho^2
}
}{q^2 - g^2_{_{\rm HLS}}(q^2,\eta)\cdot F_\rho^2 } 
\cdot \left[g_{\rho\pi\pi}(q^2)\cdot  (q+2k)^\nu\right]  \,, 
\end{eqnarray}
On the other hand,  Eq.(\ref{universalitycheck}) yields the same renormalized Green function as
\begin{eqnarray}
\langle \rho^{(R)}_\mu(q) \phi(k) \phi(q+k)\rangle \big|^{k^2=(k+q)^2=0}_{\phi-{\rm amputated}}&=&
\langle \rho^{(R)}_\mu \rho^{(R)}_\nu\rangle(q) \cdot\Gamma^{\rho^{(R)}\pi\pi, \nu} (q,k, q+k)\big|^{k^2=(k+q)^2=0}_{\phi-{\rm amputated}}\nonumber\\
&=& g^{-1}_{_{\rm HLS}}(q^2,\eta)\cdot \langle \rho_\mu \rho_\nu\rangle(q) \cdot \Gamma^{\rho\pi\pi, \nu} (q,k, q+k)\big|^{k^2=(k+q)^2=0}_{\phi-{\rm amputated}}\nonumber\\
&=&
2\cdot\frac{
g_{\mu\nu} -
\frac{
q_\mu q_\nu}{g^2_{_{\rm HLS}}(q^2,\eta) \cdot F_\rho^2 
}
}{q^2 -  g^2_{_{\rm HLS}}(q^2,\eta)\cdot F_\rho^2 } 
\cdot \left[g_{_{\rm HLS}}(q^2,\eta)\cdot  (q+2k)^\nu\right] \,,
\end{eqnarray}
 which implies:
\begin{eqnarray}
g_{\rho\pi\pi}(q^2) =g_{_{\rm HLS}}(q^2,\eta)\,,\quad  \left({\rm universality}\,,  \,\,a-{\rm independent}\right) .\end{eqnarray}
In particular,  the on-shell $\rho_\mu$ coupling reads:
\begin{eqnarray}
 g_{\rho\pi\pi} = g_{_{\rm HLS}}&=&\left[- 2N  f(M_\rho^2,0)\right]^{-1/2} \,, \quad \left({\rm broken} \,\, {\rm phase}\right)\nonumber\\
 &=&\left[- 2N  f(0,\eta)\right]^{-1/2}  \,, \quad \left({\rm symmtric} \,\, {\rm phase}\right)\,. 
  \end{eqnarray}

\end{document}